\documentclass{article}
\usepackage[preprint]{neurips_2025}

\usepackage[utf8]{inputenc} 
\usepackage[T1]{fontenc}    
\usepackage{hyperref}       
\usepackage{url}            
\usepackage{booktabs}       
\usepackage{amsfonts}       
\usepackage{nicefrac}       
\usepackage{microtype}      
\usepackage{graphicx}
\usepackage{caption}
\usepackage{amsmath}
\usepackage{amsthm}
\usepackage{colortbl}
\usepackage{multirow}
\usepackage{xcolor}

\usepackage{algorithm}
\usepackage{algpseudocode}
\usepackage{wrapfig}
\usepackage{needspace}
\usepackage{marginnote}

\usepackage{enumitem}

\usepackage{thmtools}

\declaretheorem[name=Remark]{remark}

\declaretheorem[
  name=Assumption,
  refname={Assumption,Assumptions},
  Refname={Assumption,Assumptions}
]{assumption}

\newcommand{\ind}{\perp\!\!\!\!\perp}

\usepackage{bbm}

\newcommand{\bX}{\mathbf{X}}
\newcommand{\bx}{\mathbf{x}}
\newcommand{\expP}{\mathbb{E}_{\mathcal{P}}}
\newcommand{\probP}{\mathbb{P}_{\mathcal{P}}}
\newcommand{\probX}{\mathbb{P}_{\bX}}
\newcommand{\expe}{\mathbb{E}}
\newcommand{\peps}{\tilde{p}_{\epsilon}}

\definecolor{boxorange}{HTML}{FDAE61}
\definecolor{boxteal}{HTML}{C5DFEC}

\title{Data Fusion for Partial Identification of Causal Effects}

\author{
Quinn Lanners \\
Duke University\\
\texttt{qml@duke.edu}\\
\And 
Cynthia Rudin \\
Duke University\\
\texttt{cynthia@cs.duke.edu}\\
\AND 
Alexander Volfovsky \\
Duke University\\
\texttt{av136@duke.edu}\\
\And 
Harsh Parikh \\
Yale University\\
\texttt{harsh.parikh@yale.edu}\\
}

\begin{document}

\maketitle

\begin{abstract}
Data fusion techniques integrate information from heterogeneous data sources to improve learning, generalization, and decision-making across data sciences. In causal inference, these methods leverage rich observational data to improve causal effect estimation, while maintaining the trustworthiness of randomized controlled trials. Existing approaches often relax the strong "no unobserved confounding" assumption by instead assuming exchangeability of counterfactual outcomes across data sources. However, when both assumptions simultaneously fail---a common scenario in practice---current methods cannot identify or estimate causal effects. We address this limitation by proposing a novel partial identification framework that enables researchers to answer key questions such as: \textit{Is the causal effect positive/negative?} and \textit{How severe must assumption violations be to overturn this conclusion?} Our approach introduces interpretable sensitivity parameters that quantify assumption violations and derives corresponding causal effect bounds. We develop doubly robust estimators for these bounds and operationalize breakdown frontier analysis to understand how causal conclusions change as assumption violations increase. We apply our framework to the Project STAR study, which investigates the effect of classroom size on students’ third-grade standardized test performance. Our analysis reveals that the Project STAR results are robust to simultaneous violations of key assumptions, both on average and across various subgroups of interest. This strengthens confidence in the study's conclusions despite potential unmeasured biases in the data.
\end{abstract}
\section{Introduction}
Modern evidence-based decision-making increasingly relies on combining information from various sources -- a practice known as \textit{data fusion}. From integrating satellite imagery across multiple spatial resolutions \citep{tang2016combined,yan2021large} to merging genetic markers with electronic health records \citep{hall2016merging,conroy2023uk,zawistowski2023michigan}, data fusion enables more robust, generalizable, and efficient analysis \citep{meng2020survey}. In causal inference, data fusion has emerged as a popular paradigm, recently recognized as one of the top ten research directions for advancing the field \citep{mitra2022future,bareinboim2016causal}. Data fusion approaches in causal inference have focused on generalizing or transporting evidence from experimental studies \citep{degtiar2023review,pearl2015generalizing,dahabreh2019generalizing,lu2019causal}, precisely estimating heterogeneous causal effects \citep{brantner2023methods,yang2023elastic}, improving efficiency \citep{rosenman2023combining,lin2025combining}, and mitigating estimation bias \citep{kallus2018removing,colnet2024causal}, among other things.

Consider the example of the Project STAR study, which investigated the impact of class size on students’ academic performance. The dataset comprises both a randomized controlled trial (RCT) -- where students were randomly assigned to different classroom sizes -- and an observational cohort where students self-selected into classrooms \citep{mosteller1995tennessee, project_star_data}. While the experimental data likely ensures internal validity, it may suffer from limited external validity or generalizability \citep{von2018does,justman2018randomized}. Conversely, the observational data better reflects real-world settings but may suffer from unobserved confounding \citep{athey2020combining,parikh2023double}. Merging the two sources can yield more precise and externally valid treatment effect estimates under milder, partially testable assumptions \citep{parikh2023double,wu2022integrative}. Specifically, the average treatment effect (ATE) becomes identifiable if \textit{either} the RCT generalizes well to the target population or the observational data satisfies no unmeasured confounding (NUC) \citep{lin2025combining,yang2023elastic}.

However, if both assumptions fail -- i.e., the RCT lacks external validity and the observational data is confounded -- then the treatment effect is no longer point-identifiable, even in the limit of infinite data. In such settings, classical estimators break down. Nevertheless, researchers may still answer important questions like: \textit{Is the treatment effect positive?} or \textit{How severe must assumption violations be to overturn this conclusion?} These are questions of \textit{partial identification}, where the goal is to estimate a plausible range or \textit{bounds} on the treatment effect rather than a single point estimate \citep{cornfield1959smoking,manski2003partial}.

While a rich literature on partial identification exists, many approaches rely on strong distributional or parametric assumptions and often ignore opportunities to tighten bounds by leveraging multiple datasets \citep{rosenbaum1983assessing,blackwell2014selection,ding2016sensitivity,bonvini2022sensitivity,nguyen2017sensitivity, nguyen2018sensitivity, nie2021covariate, colnet2022causal, dahabreh2023sensitivity,huang2022sensitivity}. This creates a critical gap in sensitivity analysis frameworks that are both flexible and informative when combining experimental and observational data.

\textbf{Contributions.} We propose a general framework for partially identifying treatment effects by integrating complementary strengths of experimental and observational studies. Our key contributions~are:
\begin{enumerate}[leftmargin=*]
    \item We introduce \textit{interpretable sensitivity parameters}, $\gamma$ and $\rho$, that quantify the extent of external validity violations and unmeasured confounding, respectively.
    \item  We develop a \textit{double machine learning estimator} based on the efficient influence function (EIF) for estimating treatment effect bounds as a function of $(\gamma, \rho)$, without relying on strong distributional or parametric assumptions.
    \item We operationalize an efficient \textit{breakdown frontier analysis}, which characterizes regions in the $(\gamma, \rho)$ space where the treatment effect remains conclusively positive (or negative) -- allowing for assessment of the robustness of causal conclusions under simultaneous assumption violations.
\end{enumerate}

Our framework enables comprehensive sensitivity analyses in data fusion settings, helping researchers transparently explore the consequences of assumption violations. By leveraging the internal validity of RCTs and the representativeness of observational studies, our approach yields tighter, more interpretable bounds on treatment effects and offers a principled way to assess robustness.

\section{Preliminaries}\label{sec:prelims}

We consider the setting where we have a sample $\mathcal{D}_n = \{1, \dots, n\}$ of $n$ units across an experimental cohort ($\mathcal{D}_e$) and an observational study ($\mathcal{D}_o$) drawn identically and independently from $\mathcal{P}$. For each unit $i \in \mathcal{D}_n = \mathcal{D}_e \cup \mathcal{D}_o$, $S_i = \mathbbm{1}[i \in \mathcal{D}_e]$ is a binary experimental cohort indicator, $T_i \in \{0,1\}$ is the binary treatment indicator, $Y_i$ is the observed outcome, and $\bX_i$ is the vector of pretreatment covariates. We assume the outcome space is bounded and \textit{positive}, but note that, without loss of generality, our approach can be applied to all bounded outcome scenarios by simply shifting the outcome domain. $Y_i(0)$ and $Y_i(1)$ denote the two potential outcomes for unit $i$. We assume the stable unit treatment value assumption (SUTVA), which ensures no interference between units and a single version of each treatment, as well as the consistency assumption, so that $Y_i = T_i Y_i(1) + (1 - T_i) Y_i(0)$.

Typically, one is interested in using these datasets to estimate the following two standard estimands, namely \textit{the average treatment effect (ATE):  } $\tau = \expP[Y(1) - Y(0)],$ and the \textit{conditional average treatment effect (CATE):  } $\tau(\bx) = \expP[Y(1) - Y(0) \mid \bX=\bx].$

We assume the following standard conditions hold:
\begin{assumption}\label{assum:pos-treatment}
(Treatment Positivity). For $s\in\{0,1\}$ and all $\bx$, $\exists c > 0$ such that\newline
$c < P(T = 1 \mid \bX = \bx, S=s) < 1-c$.
\end{assumption}

\begin{assumption}\label{assum:pos-exp}
(Study Positivity). For all $\bx$, $\exists c > 0$ such that $c < P(S = 1 \mid \bX = \bx) \leq 1$
\end{assumption}

\begin{assumption}\label{assum:exp-random}
(Internal Validity of the Experiment). $(Y(0), Y(1)) \ind T \mid \bX = \bx, S = 1$
\end{assumption}

However, we acknowledge the possibility of unobserved confounders that concurrently influence $S$, $T$, and $Y$. Due to such unobserved confounding, the following exchangeability assumptions, which are standard in the literature, may fail to hold:
\begin{assumption}\label{assum:obs-ignorability}
(No Unobserved Confounding (NUC) in the Observational Data).\newline
$(Y(0), Y(1)) \ind T \mid \bX = \bx, S = 0$
\end{assumption}

\begin{assumption}\label{assum:exchange}
(Study Exchangeability). $(Y(0), Y(1)) \ind S \mid \bX = \bx$
\end{assumption}

In this paper, we explicitly consider scenarios in which \ref{assum:obs-ignorability} and \ref{assum:exchange} assumptions are simultaneously violated, thereby challenging the point identifiability of $\tau$ and $\tau(\bx)$.

\textbf{Discussion of Assumptions}. \ref{assum:pos-treatment} is the standard treatment positivity assumption, ensuring overlap between treated and control groups. \ref{assum:exp-random} and \ref{assum:obs-ignorability} are structurally equivalent, differing only in the sample subset (experimental vs. observational units). Internal validity in RCTs is generally accepted due to randomization, whereas NUC is stronger, as treatment may depend on unobserved confounders. Combine experimental and observational samples requires the additional \ref{assum:pos-exp} and \ref{assum:exchange} assumptions. \ref{assum:pos-exp} states that each unit must have a nonzero probability of being an experimental unit and is necessary to ensure overlap between the two study cohorts. \ref{assum:exchange} is the study exchangeability assumption, which states that, conditional on covariates, potential outcomes are exchangeable across studies. Like NUC, it can be a strong assumption, as study participation may depend on unobservables.

\subsection{Quantifying Assumption Violations}\label{sec:quant-assum-viol}

We introduce two additional terms, $\rho$ and $\gamma$, that separately quantify violations of \ref{assum:obs-ignorability} and \ref{assum:exchange}, respectively. The value of $\rho \geq 0$ quantifies the level of unobserved confounding in the observational data, corresponding to a violation of \ref{assum:obs-ignorability} when $\rho > 0$. The value of $\gamma \geq 0$ quantifies the difference in potential outcomes between the RCT data and the observational data, corresponding to a violation of \ref{assum:exchange} when $\gamma > 0$. Both terms report the level of violation as relative measures of the observed outcomes. For example, $\rho=0.2$ corresponds to the setting that unobserved confounding in the observational dataset affects outcomes by 20\%. In Section~\ref{sec:partial-id}, we formally define $\rho$ and $\gamma$, expand on their interpretation, and employ them as sensitivity parameters for partial identification of $\tau$ and $\tau(\bx)$.

\subsection{Breakdown Frontiers}\label{sec:bf-intro}
\cite{masten2020inference} introduce an approach for visualizing how conclusions about a parameter of interest vary as a set of assumptions are relaxed. We leverage this framework to plot how treatment effect conclusions change as we relax \ref{assum:obs-ignorability} and \ref{assum:exchange} via our sensitivity parameters $\rho$ and $\gamma$.
\Needspace{16\baselineskip} 
\begin{wrapfigure}[16]{r}{0.4\textwidth}
    \vspace{-\intextsep}   
    \centering
    \includegraphics[width=0.4\textwidth]{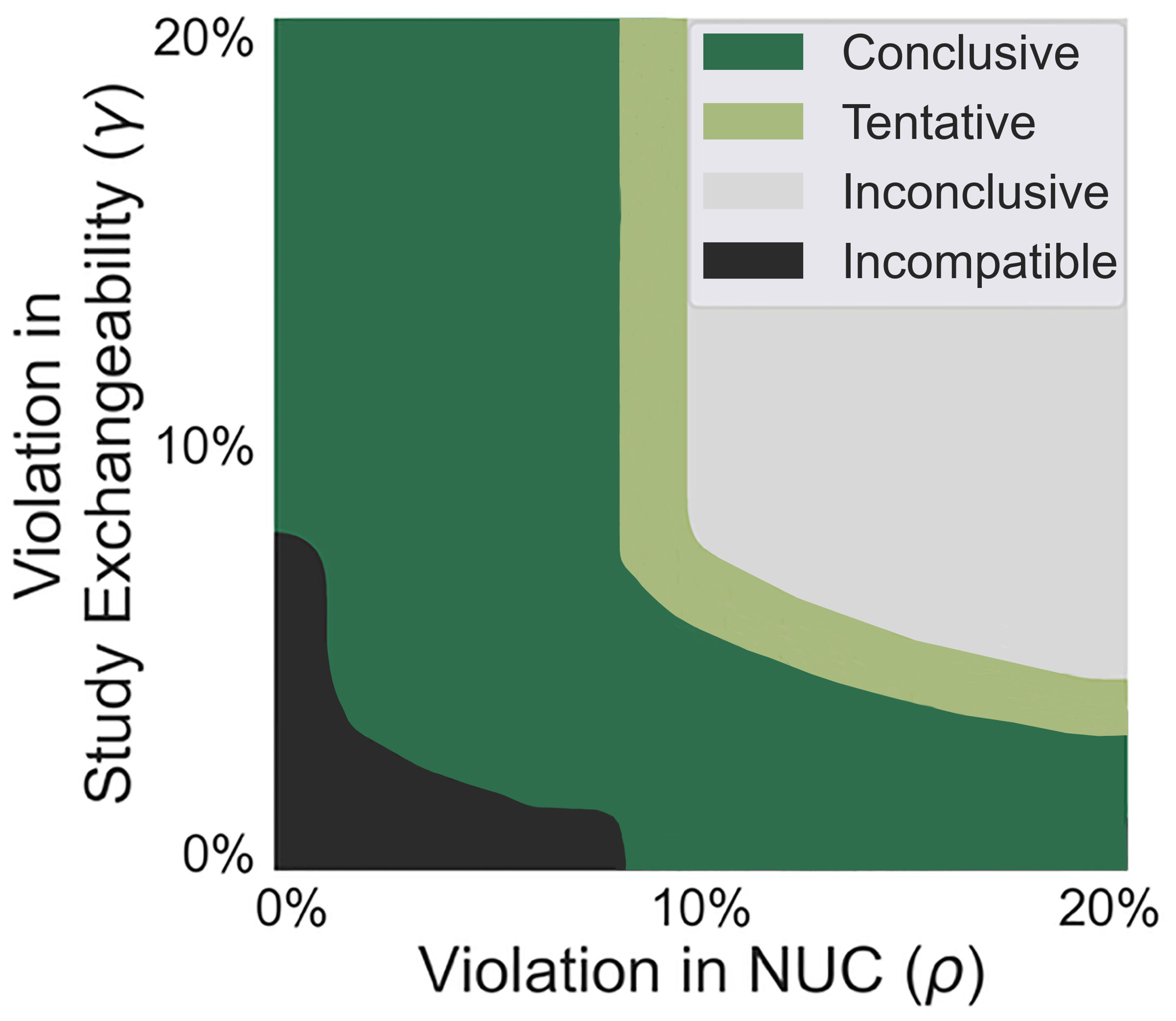}
    \caption{Example breakdown frontier plot for $\rho$ and $\gamma$.}
    \label{fig:bf-toy}
\end{wrapfigure}
Figure~\ref{fig:bf-toy} illustrates a breakdown frontier plot constructed by bounding the treatment effect for various $(\rho, \gamma)$ pairs. The x-axis represents violations of \ref{assum:obs-ignorability}, expressed as percentages corresponding to $(100\times\rho)$\%. Likewise, the y-axis represents violations of \ref{assum:exchange}, expressed as $(100\times\gamma)$\%. As a result, the bottom-left of the plot corresponds to stronger assumptions (small $\rho, \gamma$), and the top-right to weaker assumptions (large $\rho, \gamma$). For each ($\rho, \gamma$) pair, we estimate upper and lower bounds on the treatment effect. These estimates divide the plot into four regions: (i) \textbf{\textit{Conclusive}}: The point estimates of the upper and lower bounds are both positive (or both negative), and both confidence intervals exclude zero at the chosen confidence level \citep[equivalent to the robust region of][]{masten2020inference}. (ii) \textbf{\textit{Tentative}}: The point estimates of the upper and lower bounds are both positive (or both negative), but at least one of the corresponding confidence intervals includes zero. (iii) \textbf{\textit{Inconclusive}}: The point estimates of the upper and lower bounds are not the same sign. (iv) \textbf{\textit{Incompatible}}: The sensitivity parameter values lie outside the admissible range for this dataset; they imply assumptions that contradict observed discrepancies between the study groups (see Section~\ref{sec:partial-id}). We note that Figure~\ref{fig:bf-toy} is an example for illustration, and that the size and shape of the regions vary by dataset.

\section{Relevant Literature}\label{sec:lit-review}

\textbf{Data Fusion for Causal Inference.} Data fusion methods leverage randomized trials to mitigate unmeasured confounding in observational data and have become central to causal inference. Broadly, approaches differ based on whether treatments and outcomes are observed only in the trial \citep{degtiar2023review} or in both datasets \citep{brantner2023methods, lin2024data}. Our setting aligns with the latter. A more detailed review is provided in Appendix~\ref*{appdx:ext-lit-review}.

\textbf{Partial Identification.} Partial identification (ID) and sensitivity analysis frameworks are widely used to assess robustness to assumption violations in observational studies \citep{cornfield1959smoking, rosenbaum1983assessing, liu2013introduction, ding2016sensitivity, bonvini2022sensitivity}. In data fusion, most sensitivity approaches focus on violations of study exchangeability when observational treatments or outcomes are missing \citep{nguyen2017sensitivity, nguyen2018sensitivity, nie2021covariate, colnet2022causal, dahabreh2023sensitivity, huang2022sensitivity}.

In settings where treatments and outcomes are available from the observational cohort (like ours), both non-unmeasured confounding (NUC) and study exchangeability must be addressed. Most existing work assumes exchangeability and focuses on NUC violations \citep{lin2024data, lin2025combining, triantafillou2023learning, chen2021minimax, oberst2022understanding, kallus2018removing, yang2020improved, rosenman2023combining, yang2023elastic}. \cite{yang2023elastic} and \cite{parikh2023double} propose tests for assumption violations: the former attributing test failures to NUC, the latter recognizing that failures may stem from either assumption, though requiring knowledge of which one fails—a difficult task in practice.

\textbf{Partial Id w/ Data Fusion.} While partial ID and sensitivity methods are well-developed for fusion without observational treatments/outcomes, they remain sparse when these are available. Related work includes partial ID approaches in contextual bandits \citep{joshi2023towards} and structural causal modeling with qualitative knowledge \citep{zhang2022partial}. Most closely related, \cite{yu2024using} develop a two-parameter sensitivity analysis for fusion settings, addressing NUC and study positivity, but assuming study exchangeability. 

\section{Partial Identification}\label{sec:partial-id}

In this section, we present a general framework for partial identification of treatment effects under simultaneous violations of the no unmeasured confounding assumption (\ref{assum:obs-ignorability}) and study exchangeability (\ref{assum:exchange}). When either assumption fails, point identification of the average treatment effect (ATE) and conditional average treatment effect (CATE) becomes impossible.

To quantify the degree of these violations, we introduce two interpretable parameters: $\rho$, capturing the extent of unmeasured confounding in the observational cohort, and $\gamma$, capturing the extent of study exchangeability violation between the experimental and observational populations. 
We define:
\[\small
    \rho := \sup_{\bx, t}
    \left|1 -
      \frac{
        \colorbox{boxorange}{\boxed{\expP[Y(t)\mid \bX=\bx, S=0, T=1-t]}}
      }{
        \colorbox{boxteal}{\boxed{\expP[Y(t)\mid \bX=\bx, S=0, T=t]}}
      }
    \right| \text{, }
    \gamma := \sup_{\bx, t}
    \left|1 - 
      \frac{
        \colorbox{boxorange}{\boxed{\expP[Y(t)\mid \bX=\bx, S=0]}}
      }{
        \colorbox{boxteal}{\boxed{\expP[Y(t)\mid \bX=\bx, S=1]}}
      }
    \right|.
\]

The blue denominator terms represent quantities point-identifiable without assumptions \ref{assum:obs-ignorability} or \ref{assum:exchange}, whereas the orange numerator terms involve counterfactual quantities that are not directly observed. 
If \ref{assum:obs-ignorability} or \ref{assum:exchange} holds, then $\rho=0$ or $\gamma=0$, respectively, and treatment effects are point-identifiable. Conversely, dissimilarity between $\expP[Y(t)\mid \bX=\bx, S=0, T=1-t]$ and $\expP[Y(t)\mid \bX=\bx, S=0, T=t]$ reflects unobserved confounding (i.e., a violation of \ref{assum:obs-ignorability}) and leads to nonzero values of $\rho$. Similarly, dissimilarity between $\expP[Y(t)\mid \bX=\bx, S=0]$ and $\expP[Y(t)\mid \bX=\bx, S=1]$ reflects study selection bias (i.e. a violation of \ref{assum:exchange}) and corresponds to nonzero values of $\gamma$. Both $\rho$ and $\gamma$ are relative measures of dissimilarity; for example, $\rho=1$ implies a 100\% difference between the observed outcome expectation and the counterfactual counterpart. Note that while $\rho$ and $\gamma$ could be extended to functions of $\bx$ and $t$, we conservatively treat $\rho$ and $\gamma$ as scalars, taking the supremum over all covariate-treatment profiles $(\bx, t)$. This enables tractable, worst-case sensitivity analyses.

We now focus on using $\rho$ and $\gamma$ to bound $\tau$ and $\tau(\bx)$ when \ref{assum:obs-ignorability} and \ref{assum:exchange} are simultaneously violated. For parsimony, we define the following estimable quantities: (i) study selection score: $g_s(\bx) = \probP(S = s \mid \bX = \bx)$, (ii) treatment propensity score: $e_t(\bx, s) = \probP(T = t \mid \bX = \bx, S = s)$, and (iii) expected outcome: $\mu(\bx, s, t) = \expP[Y \mid \bX = \bx, S = s, T = t].$ Using the law of iterated expectation over study selection, we express:
\begin{equation}\label{eq:it-exp}
    \begin{split}
        \expP[Y(t) \mid \bX=\bx] & = g_1(\bx)\expP[Y(t) \mid \bX=\bx, S=1] + g_0(\bx)\expP[Y(t) \mid \bX=\bx, S=0] \\
        & = g_1(\bx)\mu(\bx, 1, t) + g_0(\bx)\expP[Y(t) \mid \bX=\bx, S=0],
    \end{split}    
\end{equation}
where we invoke \ref{assum:exp-random} to replace the potential outcome expectation $\expP[Y(t) \mid \bX=\bx, S=1]$ with the observed outcome expectation $\mu(\bx, 1, t)$, since treatment is randomized in the experimental study ($S=1$). In contrast, without \ref{assum:obs-ignorability} or \ref{assum:exchange}, the term $\expP[Y(t) \mid \bX=\bx, S=0]$ remains unidentifiable. However, we can leverage $\rho$ and $\gamma$ to construct sharp, identifiable bounds on $\expP[Y(t) \mid \bX=\bx, S=0]$, and thereby obtain bounds on the overall potential outcome $\expP[Y(t) \mid \bX=\bx]$.
Towards this, we define two estimable functions that upper bound $\expP[Y(t) \mid \bX=\bx, S=0]$:
\[
v(\bx,t,\gamma) := (1+\gamma)\mu(\bx,1,t), \quad
w(\bx,t,\rho) := e_t(\bx,0)\mu(\bx,0,t) + e_{1-t}(\bx,0)(1+\rho)\mu(\bx,0,t).
\]

The function $v(\bx, t, \gamma)$ is derived from experimental study data. It is based on the relative deviation of $\expP[Y(t) \mid \bX=\bx, S=0]$ from the identifiable quantity $\expP[Y(t) \mid \bX=\bx, S=1] = \mu(\bx,1,t)$, as governed by the parameter $\gamma$. Conversely, the function $w(\bx, t, \rho)$ is derived from observational study data. It combines the identifiable component $\expP[Y(t)\mid \bX=\bx, S=0, T=t] = \mu(\bx,0,t)$, weighted by the treatment propensity, $e_t(\bx,0)$, with a term that inflates the counterfactual component $\expP[Y(t)\mid \bX=\bx, S=0, T=1-t]$ according to the parameter $\gamma$.

In Lemma~\ref{lemma:pot-out-bounds}, we combine the upper bounds $v(\bx, t, \gamma)$ and $w(\bx, t, \rho)$, along with their lower bound counterparts $v(\bx, t, -\gamma)$ and $w(\bx, t, -\rho)$, to construct tight, identifiable bounds on $\expP[Y(t) \mid \bX=\bx]$. Specifically, we replace the unidentifiable term $\expP[Y(t) \mid \bX=\bx, S=0]$ in Equation~\ref{eq:it-exp} with the $\min$ of the two upper bounds and the $\max$ of the two lower bounds.

\begin{restatable}[Conditional Potential Outcome Bounds]{lemma}{potoutlemma}\label{lemma:pot-out-bounds}
Suppose \ref{assum:pos-treatment}-\ref{assum:exp-random} hold. Then for any $t \in \{0,1\}$ and given $\bx$, if $v(\bx,t,-\gamma) \leq w(\bx,t,\rho)$ and $w(\bx,t,-\rho) \leq v(\bx,t,\gamma)$,
the conditional potential outcome satisfies
\begin{gather*}
    \expP[Y(t)\mid\bX=\bx] \in [l(\bx,t,\rho,\gamma), u(\bx,t,\rho,\gamma)], \;\;\textrm{  where} \\
    l(\bx,t,\rho,\gamma) = g_1(\bx)\mu(\bx,1,t) + g_0(\bx)\max\left\{w(\bx,t,-\rho), v(\bx,t,-\gamma)\right\}, \\
    u(\bx,t,\rho,\gamma) = g_1(\bx)\mu(\bx,1,t) + g_0(\bx)\min\left\{w(\bx,t,\rho), v(\bx,t,\gamma)\right\}.
\end{gather*}
\end{restatable}

A full derivation of the results leading to Lemma~\ref{lemma:pot-out-bounds} is provided in Appendix~\ref*{appdx:pid-proofs}. Building on this results, we next derive bounds on the conditional and average treatment effects.

\begin{restatable}[Treatment Effect Bounds]{theorem}{tebounds}\label{thm:cate_bound}
Suppose \ref{assum:pos-treatment}-\ref{assum:exp-random} hold and that for each $\bx$ and $t \in \{0,1\}$, $v(\bx,t,-\gamma) \leq w(\bx,t,\rho)$ and $w(\bx,t,-\rho) \leq v(\bx,t,\gamma)$.
Then, the conditional average treatment effect satisfies:
$l(\bx,1,\rho,\gamma) - u(\bx,0,\rho,\gamma) \leq \tau(\bx) \leq u(\bx,1,\rho,\gamma) - l(\bx,0,\rho,\gamma)$.

Further, if this holds for all $\bx$ such that $\probP(\bX=\bx) > 0$, then the average treatment effect satisfies:

$\expP[l(\bX,1,\rho,\gamma) - u(\bX,0,\rho,\gamma)] \leq \tau \leq \expP[u(\bX,1,\rho,\gamma) - l(\bX,0,\rho,\gamma)]$.
\end{restatable}

We derive doubly robust estimators for the treatment effect bounds established in Theorem~\ref{thm:cate_bound} in the next section. Before proceeding, we discuss infeasibility conditions in Remark~\ref{rem:incompatible}.

\begin{remark}[\textit{(In)compatible $\rho$ and $\gamma$.}] 
Every $(\rho, \gamma)$ pair corresponds to a data-generating process that could, in principle, have produced the observed data. However, some values of $(\rho, \gamma)$ imply assumptions that conflict with what we observe. To illustrate this, consider a scenario where $\exists (\bx, t)$ such that the conditional expectations of the outcome differ across study groups, i.e. $\left| \expP[Y \mid \bX = \bx, S=1, T=t] - \expP[Y \mid \bX = \bx, S=0, T=t] \right| = \Delta(t) > 0$. \cite{parikh2023double} shows that $\Delta(t) > 0$ implies that \ref{assum:obs-ignorability} and/or \ref{assum:exchange} is violated. Therefore, setting $(\rho, \gamma) = (0,0)$--- which implies both assumptions hold---contradicts the observed difference $\Delta(t) > 0$.

More broadly, the bounds in Lemma~\ref{lemma:pot-out-bounds} and Theorem~\ref{thm:cate_bound} are valid only when both $v(\bx,t,-\gamma) \leq w(\bx,t,\rho)$ and $w(\bx,t,-\rho) \leq v(\bx,t,\gamma)$; a violation makes the parameters incompatible. Because checking the inequalities at every $(\bx, t)$ is infeasible in most settings, we test them in expectation over $\bX$ for each treatment arm. The null distribution is estimated with a resampling test that keeps the fitted propensity and outcome models fixed, as generating resamples that satisfy the null and re-estimate these models is non-trivial. This may label some pairs incompatible that a full bootstrap would not. Intuition behind and estimation of (in)compatibility are discussed further in Appendix~\ref*{appdx:incompatible}.
\label{rem:incompatible}
\end{remark}


\section{Semiparametrically Efficient Estimation}\label{sec:estimation}
We now turn to the problem of estimating the bounds identified in Section~\ref{sec:partial-id}. Our goal is to construct doubly robust estimators that offer both statistical efficiency and robustness to model misspecification \citep{chernozhukov2018double}. However, a key challenge arises: the presence of non-differentiable $\max$ and $\min$ operators in our estimands makes it intractable to directly derive the efficient influence functions (EIFs) needed for such estimators. To address this issue, in Section~\ref{sec:boltz} we introduce smooth approximations to the bounds using the Boltzmann operator in place of the $\max$ and $\min$ functions. These approximations enable the derivation of EIFs and, in turn, the construction of bias-corrected estimators in Section~\ref{sec:eff-estimators}. 
In this section, we focus on bounds for $\tau$, but note that similar steps apply to deriving efficient estimators for bounds on $\tau(\bx)$. 

\subsection{Smooth Bounds}\label{sec:boltz}
For any $x_1,x_2\in\mathbbm{R}$, the Boltzmann operator is of the form $\lambda_1 x_1 + \lambda_2 x_2$ where
\begin{equation*}
    \lambda_1 := \frac{\exp(\alpha x_1)}{\exp(\alpha x_1) + \exp(\alpha x_2)},
    \lambda_2 := \frac{\exp(\alpha x_2)}{\exp(\alpha x_1) + \exp(\alpha x_2)}.
\end{equation*}

This operator is similar to the popular softmax function in that $\lambda_1 + \lambda_2 = 1$ and their relative magnitudes are linked to $x_1$ and $x_2$. However, it differs in its incorporation of the hyperparameter $\alpha$ which causes $\lambda_1 x_1 + \lambda_2 x_2$ to approach $\max(x_1, x_2)$ as $\alpha\rightarrow \infty$ and $\min(x_1, x_2)$ as $\alpha\rightarrow -\infty$. 

In Lemma~\ref{lemma:smooth-bounds}, we show that the Boltzmann operator can be used to construct smooth approximations of our partial identification bounds. Specifically, we replace the functions $l(\bX, t, \rho, \gamma)$ and $u(\bX, t, \rho, \gamma)$ from Lemma~\ref{lemma:pot-out-bounds} with $b(\bX, t, \rho, \gamma, \alpha)$, which uses the Boltzmann operator in place of the $\max$ and $\min$ functions. We then establish treatment effect bounds using these smooth approximations---analogous to Theorem~\ref{thm:cate_bound}---and show that, as $\alpha$ increases, these estimates converge to the $\max$ and $\min$ bounds.

\begin{restatable}[name=Smooth Bounds]{lemma}{smoothboundslemma}\label{lemma:smooth-bounds}
    Consider a setting where \ref{assum:pos-treatment}-\ref{assum:exp-random} hold but \ref{assum:obs-ignorability} and \ref{assum:exchange} may not. Define
    \begin{equation*}
        b(\bX, t, \rho, \gamma, \alpha) :=
        g_1(\bX)\mu(\bX,1,t) +
        g_0(\bX)\left\{
        \lambda_1(\bX, t, \rho, \gamma, \alpha) v + \lambda_2(\bX, t, \rho, \gamma, \alpha) w
        \right\}, \textrm{ where}
    \end{equation*}
    \[
    \lambda_1(\bX, t, \rho, \gamma, \alpha) = \frac{\exp(\alpha v)}{\exp(\alpha v) + \exp(\alpha w)}, \quad
    \lambda_2(\bX, t, \rho, \gamma, \alpha) = \frac{\exp(\alpha w)}{\exp(\alpha v) + \exp(\alpha w)},
    \]
    and \( v = v(\bX, t, \gamma) \), \( w = w(\bX, t, \rho) \).
    Then for any $\alpha > 0$, $\rho$, and $\gamma$ such that $\forall t\in\{0,1\}$ and $\forall \bx$ for which $\probP(\bX = \bx) >0$, $b(\bx, t, -\rho, -\gamma, \alpha) \leq b(\bx, t, \rho, \gamma, -\alpha)$, it follows that
    \begin{equation*}
        \expP[b(\bX, 1, -\rho,-\gamma, \alpha) - b(\bX, 0, \rho,\gamma, -\alpha)]
        \leq \tau \leq 
        \expP[b(\bX,1,\rho,\gamma, -\alpha) - b(\bX,0,-\rho,-\gamma, \alpha)],    
    \end{equation*}
    and 
    \begin{align*}
        \lim_{\alpha\rightarrow\infty}\expP[b(\bX, 1, -\rho,-\gamma, \alpha) - b(\bX, 0, \rho,\gamma, -\alpha)] &= \expP[l(\bX, 1, \rho,\gamma) - u(\bX, 0, \rho,\gamma)], \\
        \lim_{\alpha\rightarrow\infty}\expP[b(\bX,1,\rho,\gamma, -\alpha) - b(\bX,0,-\rho,-\gamma, \alpha)] &= \expP[u(\bX,1,\rho,\gamma) - l(\bX,0,\rho,\gamma)].
    \end{align*}    
\end{restatable}

\subsection{Efficient Estimators}\label{sec:eff-estimators}
Having established smooth, differentiable approximations of our bounds, we now derive their corresponding EIFs, which form the basis for the bias-corrected estimators. We begin by defining $\theta(t,\rho,\gamma,\alpha) := \expP[b(\bX, t, \rho, \gamma, \alpha)]$. Using this, we can express the bounds on $\tau$ as
\begin{equation*}\label{eq:bc-ate-bounds}
    \theta(1,-\rho,-\gamma,\alpha) - \theta(0,\rho,\gamma,-\alpha)
    \leq \tau \leq 
    \theta(1,\rho,\gamma,-\alpha) - \theta(0,-\rho,-\gamma,\alpha),
\end{equation*}
where each bound is written as a difference between two instances of $\theta$ with different parameter settings. Therefore, once we establish an EIF for the general form $\theta(t,\rho,\gamma,\alpha)$, we can obtain EIFs for both the upper and lower bounds by leveraging the linearity of EIFs.

Following the approach of \cite{schuler2024moderncausalinference}, we derive the EIF for $\theta(t,\rho,\gamma,\alpha)$, denoted by $\phi(Z; t,\rho,\gamma,\alpha)$ with $Z = (\bX, S, T, Y)$. The full, centered EIF is given below with each term tagged by a superscript \((\cdot)\) for reference. For brevity, we omit the explicit arguments of $v$, $w$, $\lambda_1$, and $\lambda_2$, which match those of $\phi$.
\[
    \begin{aligned}
    \phi(Z; t,\rho,\gamma,\alpha) 
      &= \Bigg[S\mu(\bX,1,t) + (1 - S)\{\lambda_1 v + \lambda_2 w\}\Bigg]^{\textrm{(i)}}
       + \Bigg[\frac{S\mathbb{I}(T=t)}{e_t(\bX,1)}\{Y - \mu(\bX,1,t)\}\Bigg]^{\textrm{(ii)}}\\
      &\quad+ \Bigg[S(1+\gamma)\{\lambda_1 + \alpha\lambda_1\lambda_2(v - w)\}
       \frac{\mathbb{I}_t(T)g_0}{e_t(\bX,1)g_1}\{Y - \mu(\bX,1,t)\}\Bigg]^{\textrm{(iii)}}\\
      &\quad+ \Bigg[(1 - S)\{\lambda_2 + \alpha\lambda_1\lambda_2(w - v)\}
       \biggl\{\frac{\mathbb{I}_t(T)}{e_t(\bX,0)}\{Y - \mu(\bX,0,t)\}(1 + \rho e_{1-t})\\
      &\hspace{3.5cm} + \rho\,\mu(\bX,0,t)(\mathbb{I}_{1-t}(T) - e_{1-t})\biggr\}\Bigg]^{\textrm{(iv)}}
       - \theta(t,\rho,\gamma,\alpha)^{\textrm{(v)}}.
    \end{aligned}
\]

The five superscripted terms correspond to: (i) A plug-in term from the experimental and observational samples. (ii) A correction term for $g_1(\bX)\mu(\bX,1,t)$, using experimental samples. (iii) A correction term for $g_0(\bX)\lambda_1 v$, using experimental samples. (iv) A correction term for $g_0(\bX)\lambda_2 w$, using observational samples. (v) A centering term, $-\theta(t,\rho,\gamma,\alpha)$, to ensure $\expP[\phi(Z; t,\rho,\gamma,\alpha)] = 0$.

We will use the EIF for the generic $\theta(t,\rho,\gamma,\alpha)$ to obtain EIFs for the lower and upper bounds. Denoting the lower bound estimand as $\theta_{LB}(\rho,\gamma,\alpha) = \theta(1,-\rho,-\gamma, \alpha) - \theta(0, \rho, \gamma, -\alpha)$, and the upper bound as $\theta_{UB}(\rho,\gamma,\alpha) = \theta(1,\rho,\gamma,-\alpha) - \theta(0, -\rho, -\gamma, \alpha)$, their EIFs are given by
\begin{align*}
    \phi_{LB}(Z; \rho,\gamma,\alpha) & = \phi(Z; 1,-\rho,-\gamma,\alpha) - \phi(Z; 0,\rho,\gamma,-\alpha), \\
    \phi_{UB}(Z; \rho,\gamma,\alpha) & = \phi(Z; 1,\rho,\gamma,-\alpha) - \phi(Z; 0,-\rho,-\gamma,\alpha).
\end{align*}

We use these EIFs to construct bias-corrected estimators for the lower and upper bounds. Let $\hat{g}_s$, $\hat{e}_t$, and $\hat{\mu}$ denote the estimated study selection, treatment propensity, and outcome regression functions used to compute $\theta_{LB}$ and $\theta_{UB}$. These are commonly referred to as nuisance functions, as they are not themselves of interest but are necessary for estimation. We collectively denote them by $\hat{\eta} = (\hat{g}_s, \hat{e}_t, \hat{\mu})$, where the hat symbol $\hat{\cdot}$ indicates an estimated quantity. The bias-corrected estimator allows these components to be estimated with flexible machine learning models, which helps protect against model misspecification, while still enabling valid inference \citep{chernozhukov2018double}. The form of the lower bound estimator is
\begin{align*}
    \hat{\theta}_{LB}^{bc}(\rho, \gamma, \alpha; \hat{\eta}) = \hat{\theta}_{LB}^{plugin}(\rho, \gamma, \alpha; \hat{\eta}) + \frac{1}{n}\sum_{i=1}^n \hat{\phi}_{LB}(Z_i; \rho,\gamma,\alpha; \hat{\eta}),
\end{align*}
where $\hat{\theta}_{LB}^{plugin}(\rho,\gamma,\alpha; \hat{\eta})$ is the plug-in estimate and $\hat{\phi}_{LB}(Z_i; \rho,\gamma,\alpha; \hat{\eta})$ is the corresponding centered EIF evaluated at each sample $Z_i = (\bX_i, S_i, T_i, Y_i)$. The estimator for the upper bound is defined analogously using the corresponding plug-in and EIF components. 

To ensure valid inference, we employ cross-fitting and assume standard convergence conditions on nuisance functions, leading to asymptotic normality \citep{kennedy2024semiparametric,chernozhukov2018double,rudolph2024improvingefficiencytransportingaverage,schuler2024moderncausalinference}. These properties enable our estimator to adapt to complex data-generating processes while maintaining statistical validity. We present full implementation details in Algorithm~\ref{alg:bias_corrected_bounds} (Appendix~\ref*{appdx:algorithms}) and formally state the asymptotic properties of our estimators in Theorem~\ref{thm:asymptotics}.
I 
\begin{restatable}[name=Asymptotic Properties]{theorem}{asymptoticsthm}\label{thm:asymptotics} If \ref{assum:pos-treatment}-\ref{assum:exp-random} are satisfied, then for any $\alpha > 0$, $\rho$, and $\gamma$ such that $\forall t\in\{0,1\}$ and $\forall \bx$ for which $\probP(\bX = \bx) >0$, $b(\bx, t, -\rho, -\gamma, \alpha) \leq b(\bx, t, \rho, \gamma, -\alpha)$, we have that $\sqrt{n}\left(\hat{\theta}_{LB}^{bc}(\rho,\gamma,\alpha; \hat{\eta}) - \theta_{LB}(\rho,\gamma,\alpha) \right) \xrightarrow{d} \mathcal{N}(0, \sigma_{LB}^2)$ and $\sqrt{n}\left(\hat{\theta}_{UB}^{bc}(\rho,\gamma,\alpha; \hat{\eta}) - \theta_{UB}(\rho,\gamma,\alpha) \right) \xrightarrow{d} \mathcal{N}(0, \sigma_{UB}^2)$,
where $\sigma_{LB}^2 = Var[\phi_{LB}(Z; \rho,\gamma,\alpha)]$ and $\sigma_{UB}^2 = Var[\phi_{UB}(Z; \rho,\gamma,\alpha)]$.
\end{restatable}

Theorem~\ref{thm:asymptotics} leverages results on estimators derived from EIFs \citep{kennedy2024semiparametric,chernozhukov2018double} and establishes that our estimators for the partial identification bounds are asymptotically unbiased under standard regularity conditions. Variance can be estimated using the sample variance of the estimated influence functions or via resampling methods such as the bootstrap, enabling valid confidence interval construction. We note that while larger values of $\alpha$ yield closer approximations to the non-smooth bounds, they may also lead to estimator instability in small samples due to the increasingly steep gradients of the smoothed function near the $\max$/$\min$ crossover point.

\section{Experimental Results}\label{sec:experiments}
In this section, we bring together the partial identification bounds developed in Section~\ref{sec:partial-id}, the estimators derived in Section~\ref{sec:estimation}, and the breakdown frontier plots from \cite{masten2020inference} introduced in Section~\ref{sec:bf-intro} to demonstrate how our sensitivity parameters, $\rho$ and $\gamma$, enable comprehensive sensitivity analysis. We begin with synthetic data to illustrate key properties of our framework under varying data generating processes. We then return to the Project STAR study from the \hyperref[sec:intro]{Introduction}, examining the robustness of treatment effect estimates in the presence of unobserved confounding.

\subsection{Simulation Study}\label{sec:sim-study}
\begin{figure}[ht]
    \centering
    \includegraphics[width=\textwidth]{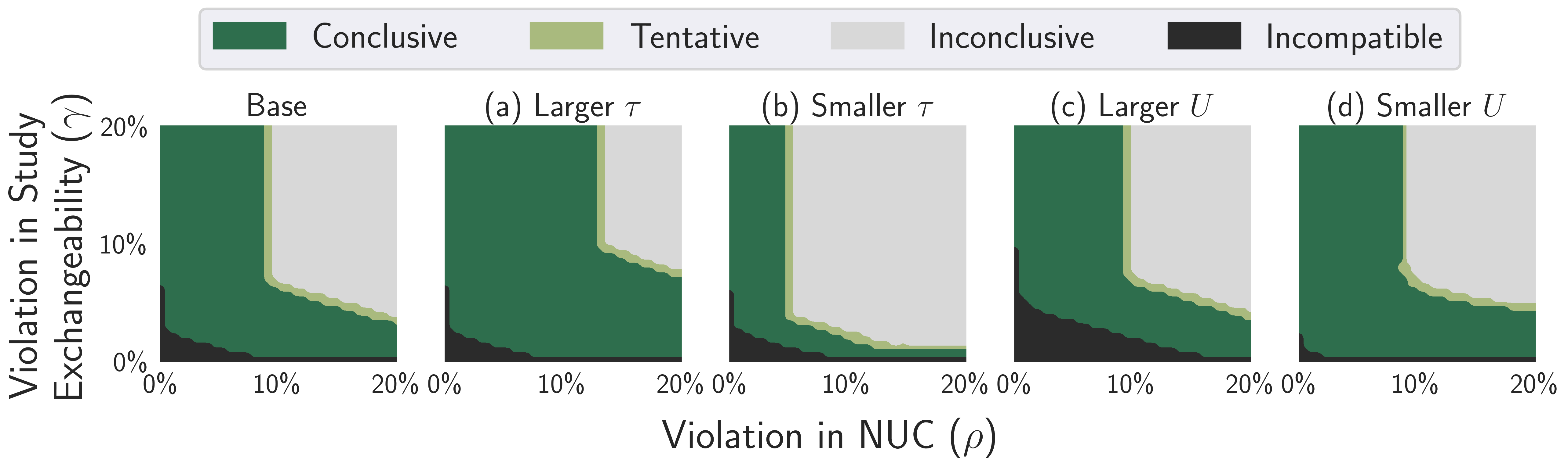}
    \caption{Breakdown frontier plots for various synthetic datasets. Figure titles indicate the relation between the data used to generate that plot to the data used to generate the \textit{Base} plot. Conclusive and tentative regions are distinguished using 95\% confidence intervals, computed from the sample variance of the efficient influence function.}
    \label{fig:sim-varieties-breakdown-frontier}
\end{figure}

We consider a data generating process with an unobserved confounder, $U$, which simultaneously affects study selection $(S)$, treatment assignment among observational units $(T | S=0)$, and outcomes $(Y)$. We generate a baseline dataset with a positive treatment effect, as well as four variants where we (a) increase the treatment effect, (b) decrease the treatment effect, (c) increase the amount of unobserved confounding, and (d) decrease the amount of unobserved confounding. We plot the breakdown frontiers for each of these datasets in Figure~\ref{fig:sim-varieties-breakdown-frontier}.

We observe that the conclusive region (dark green) expands as the magnitude of the treatment effect increases. Conversely, weaker effects lead to a smaller conclusive region. We also observe that as the amount of unobserved confounding shrinks, so does the incompatible region (black), while greater confounding enlarges it. This behavior demonstrates how the breakdown frontier plot effectively summarizes the strength of evidence for a conclusive treatment effect by incorporating both the effect size and the observed discrepancies between observational and experimental data.

Specific details of the data generating process are provided in Appendix~\ref*{appdx:sim-setup}, and an algorithm for constructing the breakdown frontier plot is included in Appendix~\ref*{appdx:bf-algo}. The procedure involves a handful of hyperparameters, including the minimum and maximum values of $\rho$ and $\gamma$, the confidence level, and the $\alpha$ scale used in the Boltzmann operator.

\subsection{Project STAR}\label{sec:star-results}

Project STAR was a large-scale study conducted in Tennessee to investigate the effect of class size on student learning outcomes \citep{mosteller1995tennessee, project_star_data}. The experimental cohort included 11,601 students randomly assigned to one of two groups: small classes (13–17 students) ($T=0$) and regular classes (22–25 students) ($T=1$). An observational cohort of 1,780 students---assigned to the same class size types but without randomization---was also available. Demographic data, including gender, race, birth year, birth month, and free lunch eligibility, were collected for both groups. Learning outcomes were measured using standardized test administered from kindergarten through third grade. Our analysis focuses on test scores from third grade.

\begin{figure}[ht] \centering \includegraphics[width=0.9\textwidth]{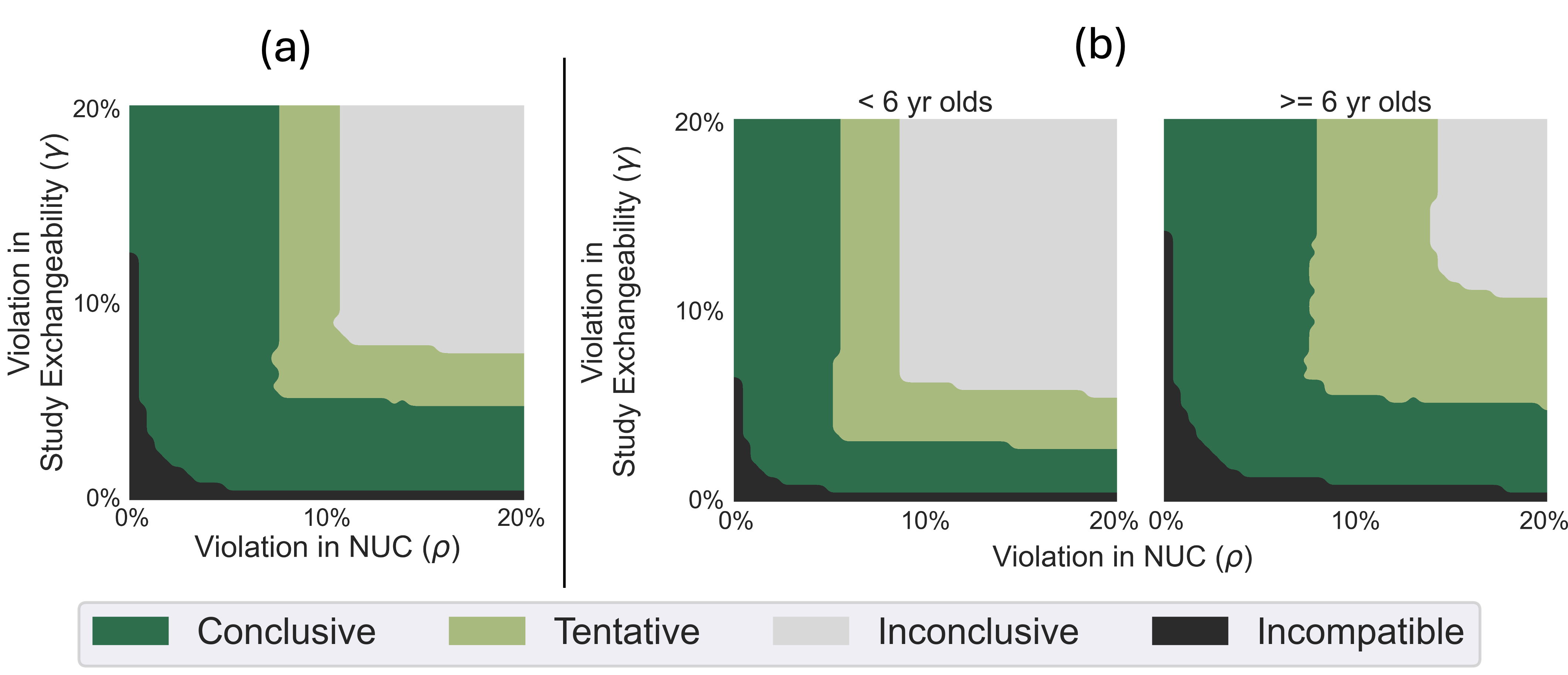} \caption{Breakdown frontier plots for Project STAR (a) population ATE and (b) subgroup-specific CATEs for students enrolled before age six (left) and at age six or older (right). Conclusive and tentative regions are based on 95\% confidence intervals computed via bootstrap resampling.} \label{fig:star} \end{figure}

Figure~\ref{fig:star}(a) presents a breakdown frontier analysis of the Project STAR ATE, varying the sensitivity parameters $\rho$ and $\gamma$. The incompatible region at small values of both parameters aligns with prior findings on unmeasured confounding in the dataset \citep{von2018does, justman2018randomized, athey2020combining, parikh2023double}. In contrast to existing estimation approaches, which require assuming either \ref{assum:exchange} or \ref{assum:obs-ignorability} holds, our framework enables investigation of causal effects under simultaneous violations of both. The analysis show that as long as study exchangeability violations remain below 5\%, there is conclusive evidence of a positive ATE---even under substantial NUC violations. Given that scores range from 486–745 (mean 618), this suggests that study selection bias would need to shift outcomes by over 30 points on average to render the results inconclusive.

Beyond the population ATE, our framework supports subgroup comparisons. Figure~\ref{fig:star}(b) shows breakdown frontier plots for students who enrolled in kindergarten before age six (left) and at six or older (right). Consistent with simulation insights, the positive treatment effect is more robust to assumption violations for the older subgroup, suggesting a larger benefit for older entrants. Developmental Psychology describes significant changes in cognitive development around the typical kindergarten entry age \citep{piaget1964cognitive}, and education research has shown that students who begin kindergarten at an older age tend to experience early learning advantages \citep{datar2006does}. While neither directly addresses class size, these findings provide context for why older students may be better positioned to benefit from the learning environment of smaller classes---a hypothesis further supported by our analysis. The older subgroup also exhibits a larger incompatible region, potentially reflecting additional unmeasured confounding related to delayed school entry (ages six to eight) and its influence on study participation, class assignment, and outcomes.

\section{Conclusion}\label{sec:conclusion}
Causal inference methods for data fusion typically assume either study exchangeability or NUC. Our work addresses settings where both assumptions may be violated, filling a gap in partial identification and sensitivity analysis. We introduce interpretable sensitivity parameters--- $\gamma$ for external validity violations and $\rho$ for unmeasured confounding---that enable transparent robustness assessments. We derive treatment effect bounds under these parameters and develop double machine learning estimators.
We use breakdown frontier plots to visualize regions where treatment effects remain conclusively positive or negative. Applications to synthetic data and Project STAR highlight our method's utility. In the Project STAR analysis, we find that the positive effect of small class sizes is robust under substantial violation of both \ref{assum:exchange} and \ref{assum:obs-ignorability}. Subgroup analysis further reveals heterogeneity in this robustness, with stronger conclusions for students who enrolled at older ages.

\textbf{Limitations \& Future Work.} Our framework supports a single binary treatment and does not handle multiple experimental or observational datasets. Extending it to continuous treatments or dynamic regimes is a promising direction. Large values of $\alpha$ can cause instability, particularly in small samples, due to steep gradients in the smoothed approximations. Our approach to identifying incompatible $(\rho, \gamma)$ values uses a simplified resampling test that does not account for uncertainty from estimating nuisance functions (Appendix~\ref*{appdx:incompatible}). While sufficient for visual diagnostics, improving this test is a direction for future work. Finally, our framework supports sensitivity analysis across $(\rho, \gamma)$ values but does not prescribe how to select them. Although our use of relative measures helps, domain expertise is needed to interpret plausible violation levels. Appendix~\ref*{appdx:bf-algo} discusses guidance and computational considerations for selecting breakdown frontier plot parameters.

\bibliographystyle{apalike}
\bibliography{biblio}

\newpage
\section*{NeurIPS Paper Checklist}

\begin{enumerate}

\item {\bf Claims}
    \item[] Question: Do the main claims made in the abstract and introduction accurately reflect the paper's contributions and scope?
    \item[] Answer: \answerYes{}
    \item[] Justification: Our contributions are clearly outlined at the end of our Introduction. We introduce \textit{interpretable sensitivity parameters} to enable partial identification of treatment effects in Section~\ref{sec:partial-id}. We proceed to develop a \textit{double machine learning estimator} for the bounds in Section~\ref{sec:estimation}. We introduce breakdown frontier plots in Section~\ref{sec:bf-intro} and operationalize them for our framework in Section~\ref{sec:experiments}.
    \item[] Guidelines:
    \begin{itemize}
        \item The answer NA means that the abstract and introduction do not include the claims made in the paper.
        \item The abstract and/or introduction should clearly state the claims made, including the contributions made in the paper and important assumptions and limitations. A No or NA answer to this question will not be perceived well by the reviewers. 
        \item The claims made should match theoretical and experimental results, and reflect how much the results can be expected to generalize to other settings. 
        \item It is fine to include aspirational goals as motivation as long as it is clear that these goals are not attained by the paper. 
    \end{itemize}

\item {\bf Limitations}
    \item[] Question: Does the paper discuss the limitations of the work performed by the authors?
    \item[] Answer: \answerYes{} 
    \item[] Justification: We include a limitations and future work section in the Conclusion (Section~\ref{sec:conclusion}). We reference the potential issue with small sample sizes and instability at the end of Section~\ref{sec:estimation}. We reference limitations with the test we use to determine incompatible sensitivity parameter values in Remark~\ref{rem:incompatible} and discuss this concept in detail in Appendix~\ref*{appdx:incompatible}. We also discuss considerations when it comes to selecting parameters for constructing breakdown frontier plots in Appendix~\ref*{appdx:bf-algo}.
    \item[] Guidelines:
    \begin{itemize}
        \item The answer NA means that the paper has no limitation while the answer No means that the paper has limitations, but those are not discussed in the paper. 
        \item The authors are encouraged to create a separate "Limitations" section in their paper.
        \item The paper should point out any strong assumptions and how robust the results are to violations of these assumptions (e.g., independence assumptions, noiseless settings, model well-specification, asymptotic approximations only holding locally). The authors should reflect on how these assumptions might be violated in practice and what the implications would be.
        \item The authors should reflect on the scope of the claims made, e.g., if the approach was only tested on a few datasets or with a few runs. In general, empirical results often depend on implicit assumptions, which should be articulated.
        \item The authors should reflect on the factors that influence the performance of the approach. For example, a facial recognition algorithm may perform poorly when image resolution is low or images are taken in low lighting. Or a speech-to-text system might not be used reliably to provide closed captions for online lectures because it fails to handle technical jargon.
        \item The authors should discuss the computational efficiency of the proposed algorithms and how they scale with dataset size.
        \item If applicable, the authors should discuss possible limitations of their approach to address problems of privacy and fairness.
        \item While the authors might fear that complete honesty about limitations might be used by reviewers as grounds for rejection, a worse outcome might be that reviewers discover limitations that aren't acknowledged in the paper. The authors should use their best judgment and recognize that individual actions in favor of transparency play an important role in developing norms that preserve the integrity of the community. Reviewers will be specifically instructed to not penalize honesty concerning limitations.
    \end{itemize}

\item {\bf Theory assumptions and proofs}
    \item[] Question: For each theoretical result, does the paper provide the full set of assumptions and a complete (and correct) proof?
    \item[] Answer: \answerYes{} 
    \item[] Justification: Each lemma and theorem includes the necessary assumptions and the full proofs are all included in Appendix~\ref*{appdx:theory-proofs}. The derivation of the efficient influence function is in Appendix~\ref*{appdx:eif}
    \item[] Guidelines:
    \begin{itemize}
        \item The answer NA means that the paper does not include theoretical results. 
        \item All the theorems, formulas, and proofs in the paper should be numbered and cross-referenced.
        \item All assumptions should be clearly stated or referenced in the statement of any theorems.
        \item The proofs can either appear in the main paper or the supplemental material, but if they appear in the supplemental material, the authors are encouraged to provide a short proof sketch to provide intuition. 
        \item Inversely, any informal proof provided in the core of the paper should be complemented by formal proofs provided in appendix or supplemental material.
        \item Theorems and Lemmas that the proof relies upon should be properly referenced. 
    \end{itemize}

    \item {\bf Experimental result reproducibility}
    \item[] Question: Does the paper fully disclose all the information needed to reproduce the main experimental results of the paper to the extent that it affects the main claims and/or conclusions of the paper (regardless of whether the code and data are provided or not)?
    \item[] Answer: \answerYes{} 
    \item[] Justification: All algorithms and experimental details are included in the Appendix. In particular, Appendix~\ref*{appdx:algorithms} has algorithms for our estimators and constructing breakdown frontier plots values. Appendix~\ref*{appdx:sim-setup} has details on the data generation process for the simulated data. Appendix~\ref*{appdx:exp-details} has implementation details for Section~\ref{sec:experiments}.
    \item[] Guidelines:
    \begin{itemize}
        \item The answer NA means that the paper does not include experiments.
        \item If the paper includes experiments, a No answer to this question will not be perceived well by the reviewers: Making the paper reproducible is important, regardless of whether the code and data are provided or not.
        \item If the contribution is a dataset and/or model, the authors should describe the steps taken to make their results reproducible or verifiable. 
        \item Depending on the contribution, reproducibility can be accomplished in various ways. For example, if the contribution is a novel architecture, describing the architecture fully might suffice, or if the contribution is a specific model and empirical evaluation, it may be necessary to either make it possible for others to replicate the model with the same dataset, or provide access to the model. In general. releasing code and data is often one good way to accomplish this, but reproducibility can also be provided via detailed instructions for how to replicate the results, access to a hosted model (e.g., in the case of a large language model), releasing of a model checkpoint, or other means that are appropriate to the research performed.
        \item While NeurIPS does not require releasing code, the conference does require all submissions to provide some reasonable avenue for reproducibility, which may depend on the nature of the contribution. For example
        \begin{enumerate}
            \item If the contribution is primarily a new algorithm, the paper should make it clear how to reproduce that algorithm.
            \item If the contribution is primarily a new model architecture, the paper should describe the architecture clearly and fully.
            \item If the contribution is a new model (e.g., a large language model), then there should either be a way to access this model for reproducing the results or a way to reproduce the model (e.g., with an open-source dataset or instructions for how to construct the dataset).
            \item We recognize that reproducibility may be tricky in some cases, in which case authors are welcome to describe the particular way they provide for reproducibility. In the case of closed-source models, it may be that access to the model is limited in some way (e.g., to registered users), but it should be possible for other researchers to have some path to reproducing or verifying the results.
        \end{enumerate}
    \end{itemize}

\item {\bf Open access to data and code}
    \item[] Question: Does the paper provide open access to the data and code, with sufficient instructions to faithfully reproduce the main experimental results, as described in supplemental material?
    \item[] Answer: \answerYes{} 
    \item[] Justification: Anonymized code and data included in submission and will be included as GitHub link in camera-ready version.
    \item[] Guidelines:
    \begin{itemize}
        \item The answer NA means that paper does not include experiments requiring code.
        \item Please see the NeurIPS code and data submission guidelines (\url{https://nips.cc/public/guides/CodeSubmissionPolicy}) for more details.
        \item While we encourage the release of code and data, we understand that this might not be possible, so “No” is an acceptable answer. Papers cannot be rejected simply for not including code, unless this is central to the contribution (e.g., for a new open-source benchmark).
        \item The instructions should contain the exact command and environment needed to run to reproduce the results. See the NeurIPS code and data submission guidelines (\url{https://nips.cc/public/guides/CodeSubmissionPolicy}) for more details.
        \item The authors should provide instructions on data access and preparation, including how to access the raw data, preprocessed data, intermediate data, and generated data, etc.
        \item The authors should provide scripts to reproduce all experimental results for the new proposed method and baselines. If only a subset of experiments are reproducible, they should state which ones are omitted from the script and why.
        \item At submission time, to preserve anonymity, the authors should release anonymized versions (if applicable).
        \item Providing as much information as possible in supplemental material (appended to the paper) is recommended, but including URLs to data and code is permitted.
    \end{itemize}

\item {\bf Experimental setting/details}
    \item[] Question: Does the paper specify all the training and test details (e.g., data splits, hyperparameters, how they were chosen, type of optimizer, etc.) necessary to understand the results?
    \item[] Answer: \answerYes{} 
    \item[] Justification: Included in Appendix~\ref*{appdx:exp-details}.
    \item[] Guidelines:
    \begin{itemize}
        \item The answer NA means that the paper does not include experiments.
        \item The experimental setting should be presented in the core of the paper to a level of detail that is necessary to appreciate the results and make sense of them.
        \item The full details can be provided either with the code, in appendix, or as supplemental material.
    \end{itemize}

\item {\bf Experiment statistical significance}
    \item[] Question: Does the paper report error bars suitably and correctly defined or other appropriate information about the statistical significance of the experiments?
    \item[] Answer: \answerYes{} 
    \item[] Justification: While we do not use traditional error bars, we quantify uncertainty through confidence regions in the breakdown frontier plots. We clearly explain how variance of our estimators can be estimated in Section~\ref{sec:estimation} and  Appendix~\ref*{appdx:exp-details} includes specific details on variance estimation for the breakdown frontier plots in Section~\ref{sec:experiments}.
    \item[] Guidelines:
    \begin{itemize}
        \item The answer NA means that the paper does not include experiments.
        \item The authors should answer "Yes" if the results are accompanied by error bars, confidence intervals, or statistical significance tests, at least for the experiments that support the main claims of the paper.
        \item The factors of variability that the error bars are capturing should be clearly stated (for example, train/test split, initialization, random drawing of some parameter, or overall run with given experimental conditions).
        \item The method for calculating the error bars should be explained (closed form formula, call to a library function, bootstrap, etc.)
        \item The assumptions made should be given (e.g., Normally distributed errors).
        \item It should be clear whether the error bar is the standard deviation or the standard error of the mean.
        \item It is OK to report 1-sigma error bars, but one should state it. The authors should preferably report a 2-sigma error bar than state that they have a 96\% CI, if the hypothesis of Normality of errors is not verified.
        \item For asymmetric distributions, the authors should be careful not to show in tables or figures symmetric error bars that would yield results that are out of range (e.g. negative error rates).
        \item If error bars are reported in tables or plots, The authors should explain in the text how they were calculated and reference the corresponding figures or tables in the text.
    \end{itemize}

\item {\bf Experiments compute resources}
    \item[] Question: For each experiment, does the paper provide sufficient information on the computer resources (type of compute workers, memory, time of execution) needed to reproduce the experiments?
    \item[] Answer: \answerYes{} 
    \item[] Justification: Information on compute resources used is included in Appendix~\ref*{appdx:exp-details}.
    \item[] Guidelines:
    \begin{itemize}
        \item The answer NA means that the paper does not include experiments.
        \item The paper should indicate the type of compute workers CPU or GPU, internal cluster, or cloud provider, including relevant memory and storage.
        \item The paper should provide the amount of compute required for each of the individual experimental runs as well as estimate the total compute. 
        \item The paper should disclose whether the full research project required more compute than the experiments reported in the paper (e.g., preliminary or failed experiments that didn't make it into the paper). 
    \end{itemize}
    
\item {\bf Code of ethics}
    \item[] Question: Does the research conducted in the paper conform, in every respect, with the NeurIPS Code of Ethics \url{https://neurips.cc/public/EthicsGuidelines}?
    \item[] Answer: \answerYes{} 
    \item[] Justification: This research does not involve human subjects or confidential data. As a sensitivity analysis framework, our method poses minimal risk of societal harm. We have taken care to ensure that our results are clear, reproducible and in full alignment with the NeurIPS Code of Ethics.
    \item[] Guidelines:
    \begin{itemize}
        \item The answer NA means that the authors have not reviewed the NeurIPS Code of Ethics.
        \item If the authors answer No, they should explain the special circumstances that require a deviation from the Code of Ethics.
        \item The authors should make sure to preserve anonymity (e.g., if there is a special consideration due to laws or regulations in their jurisdiction).
    \end{itemize}

\item {\bf Broader impacts}
    \item[] Question: Does the paper discuss both potential positive societal impacts and negative societal impacts of the work performed?
    \item[] Answer: \answerNo{} 
    \item[] Justification: We discuss the potential positive societal impacts of encouraging greater exploration of the robustness of causal conclusions to standard assumptions. However, we do not explicitly discuss potential negative societal impacts. As a methodological contribution in partial identification and sensitivity analysis, we view the risk of harm to be minimal.
    \item[] Guidelines:
    \begin{itemize}
        \item The answer NA means that there is no societal impact of the work performed.
        \item If the authors answer NA or No, they should explain why their work has no societal impact or why the paper does not address societal impact.
        \item Examples of negative societal impacts include potential malicious or unintended uses (e.g., disinformation, generating fake profiles, surveillance), fairness considerations (e.g., deployment of technologies that could make decisions that unfairly impact specific groups), privacy considerations, and security considerations.
        \item The conference expects that many papers will be foundational research and not tied to particular applications, let alone deployments. However, if there is a direct path to any negative applications, the authors should point it out. For example, it is legitimate to point out that an improvement in the quality of generative models could be used to generate deepfakes for disinformation. On the other hand, it is not needed to point out that a generic algorithm for optimizing neural networks could enable people to train models that generate Deepfakes faster.
        \item The authors should consider possible harms that could arise when the technology is being used as intended and functioning correctly, harms that could arise when the technology is being used as intended but gives incorrect results, and harms following from (intentional or unintentional) misuse of the technology.
        \item If there are negative societal impacts, the authors could also discuss possible mitigation strategies (e.g., gated release of models, providing defenses in addition to attacks, mechanisms for monitoring misuse, mechanisms to monitor how a system learns from feedback over time, improving the efficiency and accessibility of ML).
    \end{itemize}
    
\item {\bf Safeguards}
    \item[] Question: Does the paper describe safeguards that have been put in place for responsible release of data or models that have a high risk for misuse (e.g., pretrained language models, image generators, or scraped datasets)?
    \item[] Answer: \answerNA{} 
    \item[] Justification: As a methodological contribution in partial identification and sensitivity analysis, our paper does not pose such a risk.
    \item[] Guidelines:
    \begin{itemize}
        \item The answer NA means that the paper poses no such risks.
        \item Released models that have a high risk for misuse or dual-use should be released with necessary safeguards to allow for controlled use of the model, for example by requiring that users adhere to usage guidelines or restrictions to access the model or implementing safety filters. 
        \item Datasets that have been scraped from the Internet could pose safety risks. The authors should describe how they avoided releasing unsafe images.
        \item We recognize that providing effective safeguards is challenging, and many papers do not require this, but we encourage authors to take this into account and make a best faith effort.
    \end{itemize}

\item {\bf Licenses for existing assets}
    \item[] Question: Are the creators or original owners of assets (e.g., code, data, models), used in the paper, properly credited and are the license and terms of use explicitly mentioned and properly respected?
    \item[] Answer: \answerYes{} 
    \item[] Justification: The Project STAR study and dataset are properly cited \citep{mosteller1995tennessee, project_star_data} and no license is needed to use.
    \item[] Guidelines:
    \begin{itemize}
        \item The answer NA means that the paper does not use existing assets.
        \item The authors should cite the original paper that produced the code package or dataset.
        \item The authors should state which version of the asset is used and, if possible, include a URL.
        \item The name of the license (e.g., CC-BY 4.0) should be included for each asset.
        \item For scraped data from a particular source (e.g., website), the copyright and terms of service of that source should be provided.
        \item If assets are released, the license, copyright information, and terms of use in the package should be provided. For popular datasets, \url{paperswithcode.com/datasets} has curated licenses for some datasets. Their licensing guide can help determine the license of a dataset.
        \item For existing datasets that are re-packaged, both the original license and the license of the derived asset (if it has changed) should be provided.
        \item If this information is not available online, the authors are encouraged to reach out to the asset's creators.
    \end{itemize}

\item {\bf New assets}
    \item[] Question: Are new assets introduced in the paper well documented and is the documentation provided alongside the assets?
    \item[] Answer: \answerYes{} 
    \item[] Justification: We release code implementing our estimation procedure and experiments. We include experimentation details in Appendix~\ref*{appdx:exp-details} and provide documentation of the code repository.
    \item[] Guidelines:
    \begin{itemize}
        \item The answer NA means that the paper does not release new assets.
        \item Researchers should communicate the details of the dataset/code/model as part of their submissions via structured templates. This includes details about training, license, limitations, etc. 
        \item The paper should discuss whether and how consent was obtained from people whose asset is used.
        \item At submission time, remember to anonymize your assets (if applicable). You can either create an anonymized URL or include an anonymized zip file.
    \end{itemize}

\item {\bf Crowdsourcing and research with human subjects}
    \item[] Question: For crowdsourcing experiments and research with human subjects, does the paper include the full text of instructions given to participants and screenshots, if applicable, as well as details about compensation (if any)? 
    \item[] Answer: \answerNA{} 
    \item[] Justification: The paper does not involve crowdsourcing nor research with human subjects.
    \item[] Guidelines:
    \begin{itemize}
        \item The answer NA means that the paper does not involve crowdsourcing nor research with human subjects.
        \item Including this information in the supplemental material is fine, but if the main contribution of the paper involves human subjects, then as much detail as possible should be included in the main paper. 
        \item According to the NeurIPS Code of Ethics, workers involved in data collection, curation, or other labor should be paid at least the minimum wage in the country of the data collector. 
    \end{itemize}

\item {\bf Institutional review board (IRB) approvals or equivalent for research with human subjects}
    \item[] Question: Does the paper describe potential risks incurred by study participants, whether such risks were disclosed to the subjects, and whether Institutional Review Board (IRB) approvals (or an equivalent approval/review based on the requirements of your country or institution) were obtained?
    \item[] Answer: \answerNA{} 
    \item[] Justification: The paper does not involve crowdsourcing nor research with human subjects.
    \item[] Guidelines:
    \begin{itemize}
        \item The answer NA means that the paper does not involve crowdsourcing nor research with human subjects.
        \item Depending on the country in which research is conducted, IRB approval (or equivalent) may be required for any human subjects research. If you obtained IRB approval, you should clearly state this in the paper. 
        \item We recognize that the procedures for this may vary significantly between institutions and locations, and we expect authors to adhere to the NeurIPS Code of Ethics and the guidelines for their institution. 
        \item For initial submissions, do not include any information that would break anonymity (if applicable), such as the institution conducting the review.
    \end{itemize}

\item {\bf Declaration of LLM usage}
    \item[] Question: Does the paper describe the usage of LLMs if it is an important, original, or non-standard component of the core methods in this research? Note that if the LLM is used only for writing, editing, or formatting purposes and does not impact the core methodology, scientific rigorousness, or originality of the research, declaration is not required.
    \item[] Answer: \answerNA{} 
    \item[] Justification: The core method development in this research does not involve LLMs as any important, original, or non-standard components
    \item[] Guidelines:
    \begin{itemize}
        \item The answer NA means that the core method development in this research does not involve LLMs as any important, original, or non-standard components.
        \item Please refer to our LLM policy (\url{https://neurips.cc/Conferences/2025/LLM}) for what should or should not be described.
    \end{itemize}

\end{enumerate}

\newpage
\appendix
\section{Extended Literature Review on Data Fusion Methods for Causal Inference}\label{appdx:ext-lit-review}
In this section, we expand on our literature review of data fusion methods for causal inference. We note that this review is not meant to serve as an exhaustive summary, but rather provide further detail for interested readers. We direct readers to the following literature reviews for more comprehensive discussions on various topics within data fusion for causal inference:
\begin{itemize}
    \item \cite{colnet2024causal}: Combining RCTs and observational studies.
    \item \cite{degtiar2023review}: Generalizing and transporting treatment effects.
    \item \cite{brantner2023methods}: Estimates heterogeneous treatment effects.
    \item \cite{lin2024data}: Efficiency gains in ATE estimation.
\end{itemize}

\paragraph{\textbf{No treatments or outcomes in RWD.}} 
Oftentimes, it is the case that we have treatment and outcome data from an RCT but only demographic information from an observational dataset (e.g. census data). In this setting, data fusion methods are focused on generalizing or transporting the experimental results from the RCT to a different target population. Taking on the terminology of \cite{dahabreh2019extending}, this task is defined as generalization when the RCT sample is drawn from the population of interest and transportation when the population of interest extends beyond the trial-eligible cohort. Typically, these approaches do not require treatment or outcome information to be available for the observational data. Therefore, they have no notion of NUC in the observational dataset (Assumption~\ref*{assum:obs-ignorability}) and only rely on the assumptions that the randomized trial is internally valid (Assumptions~\ref*{assum:exp-random}) and that its findings can be generalized to the target population (Assumptions~\ref*{assum:pos-exp} and \ref*{assum:exchange}).

\cite{degtiar2023review} categorized generalization and transportation approaches into the familiar buckets of matching and weighting, outcome regression based, or doubly robust methods that combine the two techniques.

Matching and weighting techniques utilize a balancing score to account for different distributions of effect modifiers between the experimental and observational sample. \cite{cole2010generalizing} and \cite{stuart2011use} were among the first to propose the use of a propensity score that models the likelihood of inclusion in the experimental study as a way to assess the difference between the two populations and extrapolate the treatment effect to the target population of interest. The most popular use of these scores is in the form of inverse probability of sampling weighted (IPSW) methods that closely resemble the widely used inverse probability weighting (IPW) methods \citep{wooldridge2002inverse, ding2010estimating, seaman2013review}. 

There has been substantial work extending and supplementing the IPSW approach. \citep{kern2016assessing, lesko2017generalizing} explored the use of non-parametric methods and \citep{colnet2022reweighting}  established finite sample error results and assessed the effect of variable selection. Furthermore, \citep{buchanan2018generalizing} established consistency guarantees for parametric estimators and \citep{colnet2024causal} did the same for non-parametric estimators. Finally, under parametric assumptions, \citep{buchanan2018generalizing} derived the asymptotic variance for the tasks of generalization and \citep{zivich2022delicatessen} did so for transportation, while \citep{dahabreh2019generalizing} proposed a bootstrap variance estimator to use with non-parametric models.

While the majority of work on weighting methods has focused on IPSW, \cite{hartman2015sample} used a maximum entropy weighting approach similar to what \cite{hainmueller2012entropy} proposed for achieving covariate balance in observational studies. And in the context of stratification, \cite{stuart2011use, o2014generalizing, tipton2013improving} have proposed propensity based stratification techniques which \cite{buchanan2018generalizing} compared to IPSW methods in regards to finite sample performance.

A separate approach to generalization and transportation utilizes outcome regression to extrapolate the treatment effect to the population of interest. These methods, often referred to as plug-in g-formula estimators, use randomized data to fit an outcome regression model that estimates the conditional means and then marginalize over the distribution of the target population. This estimator can be a simple ordinary least squares model \citep{kern2016assessing} or more complex machine learning models \citep{kern2016assessing, lesko2017generalizing}. \cite{dahabreh2019extending} established consistency results for the parametric estimators while \cite{colnet2024causal} propose more general consistency results that apply to non-parametric estimators.

Doubly robust methods constitute a third approach for generalizing and transporting treatment effects. Arguably the the most popular such approach is Augmented IPSW. \cite{colnet2024causal} describes this estimator, which adapts the standard Augmented IPW estimator of \cite{glynn2010introduction} for generalizability and transportability. The Augmented IPSW combines outcome and inclusion propensity regressions in an estimator that is asymptotically unbiased when at least one of the regression functions is consistent and is asymptotically efficient when both are consistent. \cite{dahabreh2020extending} and \cite{colnet2022causal} provide consistency results for the Augmented IPSW estimator in these tasks while \cite{dahabreh2019extending} and \cite{li2023note} establish conditions under which the estimator is asymptotically normal.

\paragraph{\textbf{Treatments and outcomes for RWD.}}
In other cases, we do have treatment and outcome information in the RCT and the observational data. While this is sometimes done strategically, by running studies with randomized and non-randomized portions, growing access to detailed individual level data, from sources such as electronic health records, makes the procurement of RWD with treatments and outcomes that match those measured in an RCT a growing application. Methods developed for this setting tend to focus primarily on estimating heterogenous treatment effect and utilizing the additional observational data to improve efficiency \citep{brantner2023methods,lin2024data}.

As summarized by \cite{brantner2023methods}, these methods can primarily be broken up into approaches that either combine separate estimates from the RCT and observational datasets \citep{rosenman2023combining,triantafillou2023learning, yang2023elastic, cheng2021adaptive,chen2021minimax,oberst2022understanding}, or use the experimental data to account for confounding bias in the estimate from the observational data \citep{kallus2018removing, hatt2022combining, joshi2023towards, yang2020improved}. We further posit the existence of a third category of methods that use observational and RCT data together to train a consistent and efficient estimator \citep{wu2022integrative, parikh2023double, lu2019causal,lin2025combining}

There are a several approaches that use various techniques to combine estimators derived from the experimental and observational datasets. \cite{rosenman2022propensity} propose an approach to estimate CATEs with respect to a known stratification scheme. They use the propensity score to assign the experimental and observational data to strata and then combine the estimates from the two data sources in each strata. 
\cite{rosenman2023combining} propose another method for estimating CATEs within known strata that uses shrinkage estimators. They derive optimality results and discuss variance estimation and confidence interval construction using their estimators. \cite{cheng2021adaptive} propose a similar approach to the shrinkage estimators that allows for CATE estimation on individual covariates. Their doubly robust estimator aggressively weights towards the estimator for the randomized data when the estimator for the observational data is heavily biased.

\cite{triantafillou2023learning} use Bayesian methods and a weighted combination of estimators from the experimental and observational data. Similar to other work, they use the probability that an adjustment set exists (i.e. Assumption~\ref*{assum:obs-ignorability} holds) to determine the weight for the observational estimator. \cite{yang2023elastic} propose a semiparametric efficient strategy for combining estimates from experimental and observational data when the CATE function is consistent across the two datasets and there is no unobserved confounding in the observational data. They posit a test to determine if these assumptions hold, and derive the asymptotic distribution of their estimator under null, fixed, and local alternatives.

Another subset of approaches focus on using the (often larger) observational dataset to estimate a CATE function. These methods subsequently leverage the RCT data to estimate confounding bias, which is then combined with the CATE estimator to yield a final, unbiased estimate. Such "bias correction" methods have been implemented in a variety of ways in \cite{kallus2018removing, hatt2022combining, joshi2023towards, yang2020improved}.

Several recent works have derived estimators that use the experimental and observational data together, rather than combining the estimates derived from each dataset separately. These methods fully utilize the observational data for increased efficiency and account for the bias when unobserved confounding is present to ensure asymptotic consistency. \cite{wu2022integrative} put forward an integrative R-learner that uses both datasets to train a consistent estimator that is at least as efficient as the estimator trained only on the experimental data. \cite{lin2025combining} proposed a power likelihood approach that incorporates the observational data into the likelihood via a learning rate to regulate the amount of information drawn from the potentially confounded data. Similar to \cite{yang2023elastic}, \cite{parikh2023double} propose another approach to test for a violation in key assumptions. They posit an Augmented IPSW approach for efficiently estimating the ATE under the violation of one of these two assumptions. However, importantly, they require the user to know which of the two assumptions is violated.

We conclude this section by noting that there are several other ways to categorize data fusion methods for causal inference. For instance, \cite{lin2024data} classified these methods into Bayesian borrowing \citep{lin2025combining, triantafillou2023learning}, weighted estimator combinations \citep{chen2021minimax,oberst2022understanding}, bias correction \citep{kallus2018removing, yang2020improved}, and shrinkage estimators \citep{rosenman2023combining}. Additionally, the boundaries between methodological subfields are often unclear. For example, the topic of data fusion has also been extensively studied through the lens of structural causal modeling \citep{pearl2015generalizing}; however, we do not cover this subfield in detail, and such methods are largely omitted from the reviews referenced above.

\section{Theoretical Results}\label{appdx:theory-proofs}

\subsection{Section~\ref*{sec:partial-id} Proofs: Lemma~\ref*{lemma:pot-out-bounds} \& Theorem~\ref*{thm:cate_bound}}\label{appdx:pid-proofs}

\potoutlemma*
\begin{proof}
    Start by rewriting
    \begin{align*}
        \expP[Y(t) \mid \bX=\bx] &= \expP[Y(t) \mid \bX=\bx, S=1]\probP(S=1|\bX=\bx) \nonumber \\
        &\quad + \expP[Y(t) \mid \bX=\bx, S=0](1 - \probP(S=1|\bX=\bx))
    \end{align*}
    using the law of iterated expectation over study selection. By the internal calidity of the RCT treatment randomization (Assumption~\ref*{assum:exp-random}), we have that
    \begin{equation*}
        \expP[Y(t) \mid \bX=\bx, S=1] = \expP[Y(t) \mid \bX=\bx, S=1, T=t] =
        \expP[Y \mid \bX=\bx, S=1, T=t].
    \end{equation*}
    Plugging this in, we get
    \begin{align*}
        \expP[Y(t) \mid \bX=\bx] &= \mathbb{E}[Y \mid \bX=\bx, S=1, T=t]\probP(S=1|\bX=\bx) \nonumber \\
        &\quad + \expP[Y(t) \mid \bX=\bx, S=0](1 - \probP(S=1|\bX=\bx)) \\
        &= \mu(\bx, 1, t)g_1(\bx) + \expP[Y(t) \mid \bX=\bx, S=0](1 - g_1(\bx)),
    \end{align*}
    
    where $\mu(\bx, 1, t)$ and $g_1(\bx)$ are the shorthand notation for these identifiable quantities. Our attention now turns to bounding $\expP[Y(t) \mid \bX=\bx, S=0]$. Recall that $\gamma$ is defined as
    
    \begin{equation*}
        \gamma = \sup_{\bx, t}\left|\frac{\expP[Y(t) | \bX=\bx, S=1] - \expP[Y(t) | \bX=\bx, S=0]}{\expP[Y(t) | \bX=\bx, S=1]}\right|.
    \end{equation*}
    
     We again use Assumption~\ref*{assum:exp-random} to replace $\expP[Y(t) | \bX=\bx, S=1]$ with its identifiable shorthand notation $\mu(\bx, 1, t)$. Then, for any $\bx$ and $t\in\{0,1\}$ we have that

    \begin{align*}
    \gamma &\geq \left| 
    \frac{ \mu(\bx, 1, t) 
           - \expP[Y(t) \mid \bX = \bx, S = 0] }
         { \mu(\bx, 1, t) } 
    \right| \\
    \gamma \cdot \mu(\bx, 1, t) 
    &\geq 
    \left| \mu(\bx, 1, t) 
           - \expP[Y(t) \mid \bX = \bx, S = 0] \right|
    \end{align*}
     
     where we use the fact that the outcome space is positive to multiply each side by $\mu(\bx, 1, t)$. We can bound the quantity inside the absolute value using the positive and negative versions of the left-hand side, and then rearrange terms to establish identifiable lower and upper bounds on $\expP[Y(t) \mid \bX = \bx, S = 0]$, as shown below:
    
    \begin{align*}
        -\gamma \cdot \mu(\bx, 1, t) 
        &\leq \mu(\bx, 1, t) - \expP[Y(t) \mid \bX = \bx, S = 0] 
        \leq \gamma \cdot \mu(\bx, 1, t) \\
        (1 - \gamma) \mu(\bx, 1, t) 
        &\leq \expP[Y(t) \mid \bX = \bx, S = 0] 
        \leq (1 + \gamma) \mu(\bx, 1, t)
    \end{align*}

    We then define $v(\bx, t, \gamma) = (1 + \gamma) \mu(\bx, 1, t)$ and have that
    \begin{equation*}
        \expP[Y(t) \mid \bX = \bx, S = 0]\in[v(\bx, t, -\gamma), v(\bx, t, \gamma)].
    \end{equation*}

    Now, we turn our objective to using $\rho$ to bound $\expP[Y(t) | \bX=\bx, S=0]$. Recall that $\rho$ is defined as
    \begin{equation*}
        \begin{aligned}
            \rho &= \sup_{\bx, t}\left|\frac{\expP[Y(t) | \bX=\bx, S=0, T=t] - \expP[Y(t) | \bX=\bx, S=0, T=1-t]}{\expP[Y(t) | \bX=\bx, S=0, T=t]}\right| \\
            &= \sup_{\bx, t}\left|\frac{\mu(\bx, 0, t) - \expP[Y(t) | \bX=\bx, S=0, T=1-t]}{\mu(\bx, 0, t)}\right|,
        \end{aligned}
    \end{equation*}
    where $\mu(\bx, 0, t)$ is the shorthand notation for the identifiable quantity $\expP[Y(t) | \bX=\bx, S=0, T=t]$. Then, for any $\bx$ and $t\in\{0,1\}$, using the fact that the outcome space is positive, we have that

    \begin{align*}
    \rho &\geq \left| 
    \frac{ \mu(\bx, 0, t) 
           - \expP[Y(t) | \bX=\bx, S=0, T=1-t] }
         { \mu(\bx, 0, t) } 
    \right| \\
    \rho \cdot \mu(\bx, 0, t) 
    &\geq 
    \left| \mu(\bx, 0, t) 
           - \expP[Y(t) | \bX=\bx, S=0, T=1-t] \right|.
    \end{align*}    

    Just as before, we bound the quantity inside the absolute value using the positive and negative versions of the left-hand side and rearrange terms to establish identifiable lower and upper bounds on $\expP[Y(t) | \bX=\bx, S=0, T=1-t]$.

    \begin{align*}
        -\rho \cdot \mu(\bx, 0, t) 
        &\leq \mu(\bx, 0, t) - \expP[Y(t) | \bX=\bx, S=0, T=1-t] 
        \leq \rho \cdot \mu(\bx, 0, t) \\
        (1 - \rho) \mu(\bx, 0, t) 
        &\leq \expP[Y(t) | \bX=\bx, S=0, T=1-t] 
        \leq (1 + \rho) \mu(\bx, 0, t).
    \end{align*}

    From here, we note that the bound established from $\rho$ is for the incorrect quantity. Namely, we have thus-far used $\rho$ to bound $\expP[Y(t) | \bX=\bx, S=0, T=1-t]$, not $\expP[Y(t) | \bX=\bx, S=0]$. To reconcile this, first observe that we can use the iterated expectation over treatment selection in the RWD to decompose $\expP[Y(t) | \bX=\bx, S=0]$ as
    \begin{equation*}
    \begin{aligned}
         \expP[Y(t) | \bX=\bx, S=0] &= \expP[Y(t) \mid \bX=\bx, S=0, T=t]P(T=t \mid \bX=\bx, S=0) 
         \\& \quad + \expP[Y(t) \mid \bX=\bx, S=0, T=1-t]P(T=1-t \mid \bX=\bx, S=0) \\
         &= \expP[Y \mid \bX=\bx, S=0, T=t]P(T=t \mid \bX=\bx, S=0) 
         \\& \quad + \expP[Y(t) \mid \bX=\bx, S=0, T=1-t]P(T=1-t \mid \bX=\bx, S=0) \\
         &= \mu(\bx, 0, t)e_t(\bx, 0) + \expP[Y(t) \mid \bX=\bx, S=0, T=1-t](1-e_t(\bx, 0)),
    \end{aligned}
    \end{equation*}
    where $e_t(\bx, 0)$ are the shorthand notation for the propensity score in the RWD. Then, we can simply plug in our above bound for $\expP[Y(t) \mid \bX=\bx, S=0, T=1-t]$ to bound $\expP[Y(t) | \bX=\bx, S=0]$:
    \begin{align*}
        &\mu(\bx, 0, t)e_t(\bx, 0) 
        + (1 - \rho)\mu(\bx, 0, t)(1 - e_t(\bx, 0)) \\
        &\quad \leq \expP[Y(t) \mid \bX = \bx, S = 0] \\
        &\quad \leq \mu(\bx, 0, t)e_t(\bx, 0) 
        + (1 + \rho)\mu(\bx, 0, t)(1 - e_t(\bx, 0)).
    \end{align*}

    We then define $w(\bx, t, \rho) = (1 + \rho)\mu(\bx, 0, t)(1 - e_t(\bx, 0))$ and have that
    \begin{equation*}
        \expP[Y(t) \mid \bX = \bx, S = 0]\in[w(\bx, t, -\rho), w(\bx, t, \rho)].
    \end{equation*}    

    Now, we have used our sensitivity parameters $\gamma$ and $\rho$ to construct two intervals around $\expP[Y(t) \mid \bX = \bx, S = 0]$. Namely,
    \begin{gather*}
        \expP[Y(t) \mid \bX = \bx, S = 0]\in[v(\bx, t, -\gamma), v(\bx, t, \gamma)], \textrm{ and } \\
        \expP[Y(t) \mid \bX = \bx, S = 0]\in[w(\bx, t, -\rho), w(\bx, t, \rho)].
    \end{gather*}

    We can use these two separate interval to construct a tightest interval by simply taking the maximum of the two lower bounds and the minimum of the two upper bounds. In particular, we have that
    \begin{equation*}
        \expP[Y(t) \mid \bX = \bx, S = 0]\in[\max\{v(\bx, t, -\gamma), w(\bx, t, -\rho)\}, \min\{v(\bx, t, \gamma), w(\bx, t, \rho) \}].
    \end{equation*}        

    From here, we note that this interval is valid if and only if $\max\{v(\bx, t, -\gamma), w(\bx, t, -\rho)\} \leq \min\{v(\bx, t, \gamma), w(\bx, t, \rho)\}$. Since $v(\bx, t, -\gamma) \leq v(\bx, t, \gamma)$ and $ w(\bx, t, -\rho) \leq w(\bx, t, \rho)$ (because the outcome space is strictly positive), $\max\{v(\bx, t, -\gamma), w(\bx, t, -\rho)\} \leq \min\{v(\bx, t, \gamma), w(\bx, t, \rho)\}$ is equivalent to the condition in the lemma that $v(\bx,t,-\gamma) \leq w(\bx,t,\rho)$ and $w(\bx,t,-\rho) \leq v(\bx,t,\gamma)$.

    We can then use these bounds on $\expP[Y(t) \mid \bX = \bx, S = 0]$ to establish that $\expP[Y(t) \mid \bX=\bx]\in[l(\bx, t, \rho, \gamma), u(\bx, t, \rho, \gamma)]$ where
    \begin{align*}
        l(\bx,t,\rho,\gamma) &= g_1(\bx)\mu(\bx,1,t) + g_0(\bx)\max\left\{w(\bx,t,-\rho), v(\bx,t,-\gamma)\right\}, \\
        u(\bx,t,\rho,\gamma) &= g_1(\bx)\mu(\bx,1,t) + g_0(\bx)\min\left\{w(\bx,t,\rho), v(\bx,t,\gamma)\right\}.
    \end{align*}
\end{proof}

\tebounds*
\begin{proof}
    We first provide a proof for the bound on the CATE. Start by noting that 
    \begin{equation*}
        \tau(\bx) = \expP[Y(1) - Y(0) \mid \bX=\bx] = \expP[Y(1) \mid \bX=\bx] - \expP[Y(0) \mid \bX=\bx]
    \end{equation*} 
    by the linearity of expectation. Given that for both $t=0$ and $t=1$, $v(\bx,t,-\gamma) \leq w(\bx,t,\rho)$ and $w(\bx,t,-\rho) \leq v(\bx,t,\gamma)$, we have from Lemma~\ref*{lemma:pot-out-bounds} that
    \begin{gather*}
        \expP[Y(1) \mid \bX=\bx] \in[l(\bx, 1, \rho, \gamma), u(\bx, 1, \rho, \gamma)], \textrm{ and } \\
        \expP[Y(0) \mid \bX=\bx] \in[l(\bx, 0, \rho, \gamma), u(\bx, 0, \rho, \gamma)].
    \end{gather*}
    From here, we note that the maximum possible value of $\tau(\bx)$ is the difference between the maximum value of $\expP[Y(1) \mid \bX=\bx]$ (i.e. $u(\bx, 1, \rho, \gamma)$) and the minimum value of $\expP[Y(0) \mid \bX=\bx]$ (i.e. $l(\bx, 0, \rho, \gamma)$). And coversely, the minimum possible value of $\tau(\bx)$ is the difference between the minimum value of $\expP[Y(1) \mid \bX=\bx]$ (i.e. $l(\bx, 1, \rho, \gamma)$) and the maximum value of $\expP[Y(0) \mid \bX=\bx]$ (i.e. $u(\bx, 0, \rho, \gamma)$). Summarizing, we can bound the conditional average treatment effect at $\bx$ as
    \begin{equation*}
        l(\bx, 1, \rho,\gamma) - u(\bx, 0, \rho,\gamma)
        \leq \tau(\bx) \leq 
        u(\bx, 1, \rho,\gamma) - l(\bx, 0, \rho,\gamma).
    \end{equation*}

    We now turn to proving the bound on the ATE. We start by using the law of iterated expectation to write
    \begin{align*}
        \tau &= \expP[Y(1) - Y(0)] \\
        & = \int \expP[Y(1) - Y(0) | \bX=\bx]d\probX(\bx) \\
        & = \int \tau(\bx)d\probX(\bx)
    \end{align*}     
    where $\probX$ is the probability measure induced by $\bX$ over $\mathcal{P}$.

    Then, for any $\bx$ where $\probX(\bx) > 0$, because for both $t=0$ and $t=1$, $v(\bx,t,-\gamma) \leq w(\bx,t,\rho)$ and $w(\bx,t,-\rho) \leq v(\bx,t,\gamma)$, we have from the above result that 
    \begin{equation*}
        l(\bx, 1, \rho,\gamma) - u(\bx, 0, \rho,\gamma)
        \leq \tau(\bx) \leq 
        u(\bx, 1, \rho,\gamma) - l(\bx, 0, \rho,\gamma).
    \end{equation*}
    Thus,
    \begin{align*}
        \int \left(l(\bx, 1, \rho,\gamma) - u(\bx, 0, \rho,\gamma)\right)d\probX(\bx)
        \leq \tau \leq 
        \int \left(u(\bx, 1, \rho,\gamma) - l(\bx, 0, \rho,\gamma)\right)d\probX(\bx).
    \end{align*}         
    Since, for $t\in\{0,1\}$, $l(\bx, t, \rho,\gamma)$ and $u(\bx, t, \rho,\gamma)$ depend on $\bx$, we can express the bounds in expectation notation as
    \begin{equation*}
        \begin{gathered}
            \expP[l(\bX, 1, \rho,\gamma) - u(\bX, 0, \rho,\gamma)]
            \leq \tau \leq 
            \expP[u(\bX,1,\rho,\gamma) - l(\bX,0,\rho,\gamma)].
        \end{gathered}                
    \end{equation*}    
\end{proof}

\subsection{Section~\ref*{sec:estimation} Proofs: Lemma~\ref*{lemma:smooth-bounds} \& Theorem~\ref*{thm:asymptotics}}\label{appdx:est-proofs}

\smoothboundslemma*
\begin{proof}
    As we did using the hard $\max$ and $\min$ functions in Lemma~\ref*{lemma:pot-out-bounds}, we can use the weighted Boltzmann operator quantities to bound 
    \begin{equation*}
    \begin{aligned}
    \expP[Y(t) \mid \bX = \bx, S = 0] \in \bigl[\, &\lambda_1(\bx,t,-\rho,-\gamma,\alpha)\,v(\bx,t,\rho) + \lambda_2(\bx,t,-\rho,-\gamma,\alpha)\,w(\bx,t,\gamma),\\
    &\lambda_1(\bx,t,\rho,\gamma,-\alpha)\,v(\bx,t,\rho) + \lambda_2(\bx,t,\rho,\gamma,-\alpha)\,w(\bx,t,\gamma)
    \,\bigr].
    \end{aligned}
    \end{equation*}
    
    so long as 
    \begin{equation*}
    \begin{aligned}
        \lambda_1(\bx, t, -\rho, -\gamma, \alpha)\,v(\bx,t,\rho) 
        + \lambda_2(\bx, t, -\rho, -\gamma, \alpha)\,w(\bx,t,\gamma) 
        \leq\;& \lambda_1(\bx, t, \rho, \gamma, -\alpha)\,v(\bx,t,\rho) \\
        &+ \lambda_2(\bx, t, \rho, \gamma, -\alpha)\,w(\bx,t,\gamma).
    \end{aligned}
    \end{equation*}
    
    Because $\bx$ and $t$ are constant in this inequality, this is equivalent to the condition that
    \begin{equation*}
        b(\bx, t, -\rho, -\gamma, \alpha) \leq b(\bx, t, \rho, \gamma, -\alpha).
    \end{equation*}
    Then, under these conditions we have that
    \begin{equation*}
        \expP[Y(t) \mid \bX = \bx]\in [b(\bx, t, -\rho, -\gamma, \alpha), b(\bx, t, \rho, \gamma, -\alpha)].
    \end{equation*}
    Then, we can proceed as we did in the proof for Theorem~\ref*{thm:cate_bound} to bound $\tau(\bx)$, and subsequently, $\tau$. Namely, we note that we can bound $\tau(\bx)$ between (i) the difference of the lower bound on $\expP[Y(1) \mid \bX = \bx]$ and the upper bound on $\expP[Y(0) \mid \bX = \bx]$ and (ii) the difference between the upper bound on $\expP[Y(1) \mid \bX = \bx]$ and the lower bound on $\expP[Y(0) \mid \bX = \bx]$. Concretely, we have that
    \begin{equation*}
        b(\bx, 1, -\rho, -\gamma, \alpha) - b(\bx, 0, \rho, \gamma, -\alpha) \leq \tau(\bx) \leq b(\bx, 1, \rho, \gamma, -\alpha) - b(\bx, 0, -\rho, -\gamma, \alpha).
    \end{equation*}
    Then, since for any $\bx$ where $\probX(\bx) > 0$ and $t\in\{0,1\}$, $b(\bx, t, -\rho, -\gamma, \alpha) \leq b(\bx, t, \rho, \gamma, -\alpha)$, we can take the expectation over the upper and lower $\tau(\bX)$ bounds as we did in the proof for Theorem~\ref*{thm:cate_bound} to conclude that 
    \begin{equation*}
        \begin{gathered}
            \expP[b(\bX, 1, -\rho, -\gamma, \alpha) - b(\bX, 0, \rho, \gamma, -\alpha)]
            \leq \tau \leq 
            \expP[b(\bX, 1, \rho, \gamma, -\alpha) - b(\bX, 0, -\rho, -\gamma, \alpha)].
        \end{gathered}                
    \end{equation*}        
    
    We now prove the convergence of these smooth bounds to their sharp counterparts as $\alpha\rightarrow\infty$. By the properties of the Boltzmann operator, we have that for any $\rho$, $\gamma$, $\bx$, and $t\in\{0,1\}$ 
    \begin{gather*}
        \lambda_1(\bx, t, \rho, \gamma, \alpha) v(\bx,t,\rho) + \lambda_2(\bx, t, \rho, \gamma, \alpha)w(\bx,t,\gamma) \rightarrow \max\{v(\bx,t,\rho),w(\bx,t,\gamma)\} \textrm{ as } \alpha\rightarrow\infty, \\
        \lambda_1(\bx, t, \rho, \gamma, \alpha) v(\bx,t,\rho) + \lambda_2(\bx, t, \rho, \gamma, \alpha)w(\bx,t,\gamma) \rightarrow \min\{v(\bx,t,\rho),w(\bx,t,\gamma)\} \textrm{ as } \alpha\rightarrow -\infty.
    \end{gather*}
    Then, note that
    \begin{equation*}
        \begin{aligned}
            g_0(\bX)\min\{v(\bx,t,\rho), w(\bx,t,\gamma)\} 
            \;\leq\;& g_0(\bX)\big[\lambda_1(\bX,t,\rho,\gamma,\alpha)\,v(\bX,t,\rho) \\
            &+ \lambda_2(\bX,t,\rho,\gamma,\alpha)\,w(\bX,t,\gamma)\big] \\
            \;\leq\;& g_0(\bX)\max\{v(\bx,t,\rho), w(\bx,t,\gamma)\}
        \end{aligned}
    \end{equation*}
    and 
    \begin{equation*}
        -\infty < E[g_0(\bX)\min\{v(\bx,t,\rho),w(\bx,t,\gamma)\}] \leq E[g_0(\bX)\max\{v(\bx,t,\rho),w(\bx,t,\gamma)\}] < \infty
    \end{equation*} 
    because the outcome space is bounded. We then have by linearity of expectation and the dominated convergence theorem that 
    \begin{align*}
    \lim_{\alpha\rightarrow\infty} 
    &\expP\big[
        b(\bX, 1, -\rho, -\gamma, \alpha) 
        - b(\bX, 0, \rho, \gamma, -\alpha)
    \big] \\
    = \;\;
    \lim_{\alpha\rightarrow\infty} 
    &\expP\big[
        g_1(\bX)\mu(\bX,1,1) 
        + g_0(\bX)\bigl\{
            \lambda_1(\bX,1,-\rho,-\gamma,\alpha)\,v(\bX,1,-\rho) \\
    &\hspace{4.2em}
            + \lambda_2(\bX,1,-\rho,-\gamma,\alpha)\,w(\bX,1,-\gamma)
        \bigr\} \\
    &\hspace{3.2em}
        - g_1(\bX)\mu(\bX,1,0) 
        - g_0(\bX)\bigl\{
            \lambda_1(\bX,0,\rho,\gamma,-\alpha)\,v(\bX,0,\rho) \\
    &\hspace{4.2em}
            + \lambda_2(\bX,0,\rho,\gamma,-\alpha)\,w(\bX,0,\gamma)
        \bigr\}
    \big] \\
    = \;\;
    &\expP[g_1(\bX)\mu(\bX,1,1)]
        + \expP\big[
            \lim_{\alpha\to\infty} g_0(\bX)\bigl\{
                \lambda_1(\bX,1,-\rho,-\gamma,\alpha)\,v(\bX,1,-\rho) \\
    &\hspace{6.2em}
                + \lambda_2(\bX,1,-\rho,-\gamma,\alpha)\,w(\bX,1,-\gamma)
            \bigr\}
        \big] \\
    &\quad - \expP[g_1(\bX)\mu(\bX,1,0)]
        - \expP\big[
            \lim_{\alpha\to\infty} g_0(\bX)\bigl\{
                \lambda_1(\bX,0,\rho,\gamma,-\alpha)\,v(\bX,0,\rho) \\
    &\hspace{6.2em}
                + \lambda_2(\bX,0,\rho,\gamma,-\alpha)\,w(\bX,0,\gamma)
            \bigr\}
        \big] \\
    = \;\;
    &\expP[g_1(\bX)\mu(\bX,1,1)] 
        + \expP\big[g_0(\bX)\max\{v(\bx,1,-\rho), w(\bx,1,-\gamma)\}\big] \\
    &\quad - \expP[g_1(\bX)\mu(\bX,1,0)] 
        - \expP\big[g_0(\bX)\min\{v(\bx,0,\rho), w(\bx,0,\gamma)\}\big] \\
    = \;\;
    &\expP\big[l(\bX, 1, \rho, \gamma)\big] 
        - \expP\big[u(\bX, 0, \rho, \gamma)\big] \\
    = \;\;
    &\expP\big[
        l(\bX, 1, \rho, \gamma) 
        - u(\bX, 0, \rho, \gamma)
    \big].
    \end{align*}
    
    The same steps can be done to show that
    \begin{gather*}
        \lim_{\alpha\rightarrow\infty}\expP[b(\bX,1,\rho,\gamma, -\alpha) - b(\bX,0,-\rho,-\gamma, \alpha)] = \expP[u(\bX,1,\rho,\gamma) - l(\bX,0,\rho,\gamma)].
    \end{gather*}
    
\end{proof}

\asymptoticsthm*
\begin{proof}
    Our estimator is a one-step, bias-corrected estimator built from the efficient influence function. To establish asymptotic normality, it is sufficient to show that both the empirical process term and second-order remainder term are $o_p(n^{-1/2})$. The empirical process term can be controlled either by imposing Donsker conditions or by using cross-fitting; we adopt the latter of the two. 
    
    To control the second-order remainder term, we assume that all the nuisance functions--namely, the study selection score ($g_s(\bx)$), the treatment propensity score ($e_t(\bx, s)$), and the expected outcome ($\mu(\bx, s, t)$)---are estimated at $n^{-1/4}$ convergence rates (or faster). 
    
    Under these assumptions, the $\sqrt{n}$ convergence of our bias-corrected estimators, $\hat{\theta}_{LB}^{bc}(\rho,\gamma,\alpha; \hat{\eta})$ and $\hat{\theta}_{UB}^{bc}(\rho,\gamma,\alpha; \hat{\eta})$, follows directly from previously established results by \cite{kennedy2024semiparametric} (Proposition 2) and \cite{chernozhukov2018double} (Theorem 3.1). 
\end{proof}
\section{EIF Derivation}\label{appdx:eif}
\subsection{Setup Recap}

We start by reiterating our setup and relevant notation. Recall that we have the following variables,
\begin{itemize}
    \item $\bX$: A vector of pretreatment covariates.
    \item $S \in \{0, 1\}$: A binary variable indicating study assignment (RWD vs RCT data).
    \item $T \in \{0, 1\}$: A binary treatment indicator.
    \item $Y \in \mathbb{R}^{+}$: The outcome of interest.
\end{itemize}

The distribution of the population is denoted by $\mathcal{P}$ over $(\bX, S, T, Y)$, which we assume admits a probability density function, denoted by $p(\bX, S, T, Y)$.

We then defined the quantities (i) study selection score: $g_s(\bx) = \probP(S = s \mid \bX = \bx)$, (ii) treatment propensity score: $e_t(\bx, s) = \probP(T = t \mid \bX = \bx, S = s)$, and (iii) expected outcome: $\mu(\bx, s, t) = \expP[Y \mid \bX = \bx, S = s, T = t].$

And, using our sensitivity parameters $\gamma$ and $\rho$ we constructed the following terms:

\begin{align*}
    w(\bX, t, \rho) &:= e_t(\bX, 0) \mu(\bX, 0, t) + \big(1 - e_t(\bX, 0)\big)(1 + \rho) \mu(\bX, 0, t), \\
    v(\bX, t, \gamma) &:= (1 + \gamma)\mu(\bX, 1, t), \\
    \lambda_1(\bX, t, \rho, \gamma, \alpha) &:= \frac{\exp(\alpha v(\bX, t, \gamma))}{\exp(\alpha v(\bX, t, \gamma)) + \exp(\alpha w(\bX, t, \rho))}, \\
    \lambda_2(\bX, t, \rho, \gamma, \alpha) &:= \frac{\exp(\alpha w(\bX, t, \rho))}{\exp(\alpha v(\bX, t, \gamma)) + \exp(\alpha w(\bX, t, \rho))}.
\end{align*}

Our goal is to derive an efficient influence function for $\theta(t,\rho, \gamma, \alpha)$, where 

\begin{gather*}
    \theta(t,\rho,\gamma,\alpha) = \\
    \expP\left[g_1(\bX)\mu(\bX,t,1) + g_0(\bX)\left\{\lambda_1(\bX, t, \rho, \gamma, \alpha)v(\bX,t, \gamma) + \lambda_2(\bX, t, \rho, \gamma, \alpha)w(\bX, t,\rho) \right\}\right].
\end{gather*}

Note that we can write
\[
\theta(t,\rho,\gamma,\alpha) = \theta_1(t) + \theta_2(t,\rho,\gamma,\alpha),
\]
where:
\[
\theta_1(t) = \mathbb{E}\Big[g_1(\bX)\mu(\bX, 1, t)\Big], \textrm{ and }
\]
\[
\theta_2(t,\rho,\gamma,\alpha) = \mathbb{E}\Big[g_0(\bX)\Big\{\lambda_1(\bX, t, \rho, \gamma, \alpha)v(\bX, t, \gamma)
+ \lambda_2(\bX, t, \rho, \gamma, \alpha)w(\bX, t, \rho)\Big\}\Big].
\]

\subsection{Efficient Influence Function}
We use point mass contamination to derive a candidate EIF, $\phi(Z; t,\rho,\gamma,\alpha)$, for our estimand, $\theta(t,\rho,\gamma,\alpha)$. We proceed to verify that the candidate is indeed a valid influence function, and subsequently the \textit{efficient} influence function because we are in a fully-saturated model space, by confirming that for any generic score, $h$, we have
\[\nabla_{h}\theta(t,\rho,\gamma,\alpha) = \expP[\phi(Z; t,\rho,\gamma,\alpha) h].\]

We start by letting $\Phi$ be an EIF operator which "takes a parameter and returns its efficient influence function" \citep{schuler2024moderncausalinference}. In our case, $\Phi(\theta(t,\rho,\gamma,\alpha)) = \phi(Z; t,\rho,\gamma,\alpha)$.

Using this notation, we can break our estimand of interest into separate components for easier derivation. We start by noting that thanks to the linearity property of the EIF
\[
\Phi(\theta(t,\rho,\gamma,\alpha)) = \Phi(\theta_1(t)) + \Phi(\theta_2(t,\rho,\gamma,\alpha)).
\]

Before proceeding to derive $\Phi(\theta_1(t))$ and $\Phi(\theta_2(t,\rho,\gamma,\alpha))$, we first compute $\Phi(\cdot)$ for various subcomponents that make up our estimand.

\begin{align*}
    \Phi(p(\bx)) &= \mathbb{I}_{\bx}(\bX) - p(\bx), \\
    \Phi(g_s(\bx)) &= \frac{\mathbb{I}_{\bx}(\bX)}{p(\bx)}\left[\mathbb{I}_s(S)-g_s(\bx)\right], \\
    \Phi(e_t(\bx, s)) &= \frac{\mathbb{I}_{\bx}(\bX)\mathbb{I}_s(S)}{p(\bx,s)}\left[\mathbb{I}_t(T)-e_t(\bx,s)\right], \\
    \Phi(\mu(\bx, s, t)) &= \frac{\mathbb{I}_{\bx}(\bX)\mathbb{I}_s(S)\mathbb{I}_{t}(T)}{p(\bx,s,t)}\left[Y-\mu(\bx, s, t)\right].
\end{align*}

In the equations above, and throughout the remainder of this section, $\mathbb{I}_{a}(A)$ denotes an indicator function that equals $1$ when the random variable $A$ takes the value $a$, and 0 otherwise.

\subsubsection{Candidate EIF for $\theta_1$}
Taking the point mass contamination approach, we first rewrite $\theta_1(t)$ assuming all of our covariates are discrete. Namely,

\[\theta_1(t) = \sum_{\bx} g_1(\bx)\mu(\bx,1,t)p(\bx).\]

We apply the product rule and plug in values to calculate $\Phi(\theta_1(t))$ below.

\begin{equation*}
    \begin{split}
        \Phi(\theta_1(t)) = & \sum_{\bx} \Phi(g_1(\bx))\mu(\bx,1,t)p(\bx) + g_1(\bx)\Phi(\mu(\bx,1,t))p(\bx) + g_1(\bx)\mu(\bx,1,t)\Phi(p(\bx)) \\
        = & \sum_{\bx} \frac{\mathbb{I}_{\bx}(\bX)}{p(\bx)}\left[S-g_1(\bx)\right]\mu(\bx,1,t)p(\bx) \\
        & + \sum_{\bx} \frac{S\mathbb{I}_{\bx}(\bX)\mathbb{I}_{t}(T)}{p(\bx,1,t)}\left[Y-\mu(\bx, 1, t)\right]g_1(\bx)p(\bx) + \left[\mathbb{I}_{\bx}(\bX) - p(\bx)\right]g_1(\bx)\mu(\bx,1,t) \\
        = & \sum_{\bx} \frac{S\mathbb{I}_{\bx}(\bX)\mathbb{I}_{t}(T)}{e_{t}(\bx,1)}\left[Y-\mu(\bx, 1, t)\right] + \mathbb{I}_{\bx}(\bX)S\mu(\bx,1,t) - g_1(\bx)\mu(\bx,1,t)p(\bx) \\
        = & \frac{S\mathbb{I}_{t}(T)}{e_{t}(\bX,1)}\left[Y-\mu(\bX, 1, t)\right] + S\mu(\bX,1,t) - \theta_1(t).
    \end{split}
\end{equation*}

In summary, the candidate EIF for $\theta_1(t)$ is:
\[
\phi_{\theta_1} = \frac{S\mathbb{I}(T=t)}{e_t(\bX, 1)} \Big[Y - \mu(\bX, 1, t)\Big]
+ S\mu(\bX, 1, t) - \theta_1(t).
\]

We drop the arguments for $\phi$ and simply denote the portion of EIF for $\theta_1(t)$ as $\phi_{\theta_1}$ for simplicity.

\subsubsection{Candidate EIF for $\theta_2$}
Since $\gamma$, $\rho$, and $\alpha$ are predefined hyperparameters, while deriving the candidate EIF for $\theta_2$ we omit them as arguments to $w, v, \lambda_1$ and $\lambda_2$ for brevity. Importantly, these hyperparameters will become relevant when deriving the EIF, so they are still present in each function - just omitted for notational brevity. 

We again start by rewriting $\theta_2$ assuming all of our covariates are discrete.

\[\theta_2 = \sum_{\bx} g_0(\bx)\left\{\lambda_1(\bx)v(\bx) + \lambda_2(\bx)w(\bx) \right\}p(\bx) = \sum_{\bx} g_0(\bx)\left\{v(\bx) + \lambda_2(\bx)\left[w(\bx) - v(\bx)\right] \right\}p(\bx).\]

Then, we can apply the product rule to see that
\begin{equation*}
    \begin{split}
        \Phi(\theta_2) = & \sum_{\bx} \Phi(g_0(\bx))p(\bx)\left\{v(\bx) + \lambda_2(\bx)\left[w(\bx) - v(\bx)\right] \right\} \\
        & + \sum_{\bx} \Phi(v(\bx))g_0(\bx)p(\bx)\left\{1 - \lambda_2(\bx)\right\} \\
        & + \sum_{\bx} \Phi(w(\bx))g_0(\bx)p(\bx)\lambda_2(\bx) \\
        & + \sum_{\bx} \Phi(\lambda_2(\bx))g_0(\bx)p(\bx)\left\{w(\bx) - v(\bx) \right\} \\
        & + \sum_{\bx} \Phi(p(\bx))g_0(\bx)\left\{v(\bx) + \lambda_2(\bx)\left[w(\bx) - v(\bx)\right] \right\}.
    \end{split}
\end{equation*}

We already know $\Phi(g_0(\bx))$ and $\Phi(p(\bx))$. But we need to calculate $\Phi(v(\bx))$, $\Phi(w(\bx))$, and $\Phi(\lambda_2(\bx))$. First, note that we can rewrite

\begin{equation*}
    w(\bx) = w(\bx, t, \rho) = e_t(\bx, 0) \mu(\bx, 0, t) + e_{1-t}(\bx, 0)(1 + \rho) \mu(\bx, 0, t) = \mu(\bx, 0, t)\left[1 + \rho e_{1-t}(\bx, 0)\right].
\end{equation*}

Then, 
\begin{equation*}
    \begin{split}
        \Phi(v(\bx)) = & \Phi(v(\bx, t, \gamma)) \\
        = & \Phi\left((1 + \gamma)\mu(\bx,1,t) \right) \\
        = & (1 + \gamma)\Phi\left(\mu(\bx,1,t) \right) \\
        = & (1 + \gamma)\frac{S\mathbb{I}_{\bx}(\bX)\mathbb{I}_t(T)}{p(\bx,1,t)}\left[Y - \mu(\bx,1,t) \right],
    \end{split}
\end{equation*}

\begin{equation*}
    \begin{split}
        \Phi(w(\bx)) = & w(\bx, t, \rho) \\
        = & \Phi\left(\mu(\bx, 0, t)\left[1 + \rho e_{1-t}(\bx, 0)\right]\right) \\
        = & \Phi\left(\mu(\bx, 0, t)\right)\left[1 + \rho e_{1-t}(\bx, 0)\right] + \rho \mu(\bx, 0, t) \Phi(e_{1-t}(\bx, 0)) \\
        = & \left[1 + \rho e_{1-t}(\bx, 0)\right]\left(\frac{(1-S)\mathbb{I}_{\bx}(\bX)\mathbb{I}_t(T)}{p(\bx,0,t)}\left[ Y - \mu(\bx, 0, t)\right] \right) \\
        & + \rho\mu(\bx,0,t)\left(\frac{(1-S)\mathbb{I}_{\bx}(\bX)}{p(\bx,0)}\left[\mathbb{I}_{1-t}(T) - e_{1-t}(\bx, 0) \right] \right),
    \end{split}
\end{equation*}

and

\begin{equation*}
    \begin{split}
        \Phi(\lambda_2(\bx)) = & \Phi\left(\frac{\exp(\alpha w(\bx))}{\exp(\alpha v(\bx)) + \exp(\alpha w(\bx))} \right) \\
        = & \alpha\exp(\alpha w(\bx)) \left\{\exp(\alpha v(\bx)) + \exp(\alpha w(\bx)) \right\}^{-2} \\
        & \times \left\{\Phi(w(\bx))\left[\exp(\alpha v(\bx)) + \exp(\alpha w(\bx))\right] - \right. \\
        & \quad \left. \Phi(v(\bx))\exp(\alpha v(\bx)) - \Phi(w(\bx))\exp(\alpha w(\bx)) \right\} \\
        = & \alpha\exp(\alpha w(\bx)) \left\{\exp(\alpha v(\bx)) + \exp(\alpha w(\bx)) \right\}^{-2} \\
        & \times \left\{\Phi(w(\bx))\exp(\alpha v(\bx)) - \Phi(v(\bx))\exp(\alpha v(\bx)) \right\} \\
        = & \alpha\lambda_2(\bx)\left(1 - \lambda_2(\bx)\right)\left[\Phi(w(\bx)) - \Phi(v(\bx)) \right] \\
        = & \alpha\lambda_1(\bx)\lambda_2(\bx)\left[\Phi(w(\bx)) - \Phi(v(\bx)) \right].
    \end{split}
\end{equation*}

Now, we plug $\Phi(\lambda_2(\bx))$ in to the full $\Phi(\theta_2)$ and, because $\Phi(\lambda_2(\bx))$ is composed of $\Phi(v(\bx))$ and $\Phi(w(\bx))$, combine like terms.

\begin{align*}
    \Phi(\theta_2) = & \sum_{\bx} \Phi(g_0(\bx))p(\bx)\left\{v(\bx) + \lambda_2(\bx)\left[w(\bx) - v(\bx)\right] \right\} \\
    & + \sum_{\bx} \Phi(v(\bx))g_0(\bx)p(\bx)\left\{1 - \lambda_2(\bx)\right\} \\
    & + \sum_{\bx} \Phi(w(\bx))g_0(\bx)p(\bx)\lambda_2(\bx) \\
    & + \sum_{\bx} \left\{\alpha\lambda_1(\bx)\lambda_2(\bx)\left[\Phi(w(\bx)) - \Phi(v(\bx)) \right]\right\}g_0(\bx)p(\bx)\left\{w(\bx) - v(\bx) \right\} \\
    & + \sum_{\bx} \Phi(p(\bx))g_0(\bx)\left\{v(\bx) + \lambda_2(\bx)\left[w(\bx) - v(\bx)\right] \right\} \\
    = & \sum_{\bx} \Phi(g_0(\bx))p(\bx)\left\{v(\bx) + \lambda_2(\bx)\left[w(\bx) - v(\bx)\right] \right\} \\
    & + \sum_{\bx} \Phi(v(\bx))g_0(\bx)p(\bx)\left\{\lambda_1(\bx) + \alpha\lambda_1(\bx)\lambda_2(\bx)\left[v(\bx) - w(\bx) \right] \right\} \\
    & + \sum_{\bx} \Phi(w(\bx))g_0(\bx)p(\bx)\left\{\lambda_2(\bx) + \alpha\lambda_1(\bx)\lambda_2(\bx)\left[w(\bx) - v(\bx) \right] \right\} \\
    & + \sum_{\bx} \Phi(p(\bx))g_0(\bx)\left\{v(\bx) + \lambda_2(\bx)\left[w(\bx) - v(\bx)\right] \right\}.
\end{align*}

We now plug in the corresponding values for the remaining $\Phi(\cdot)$'s in $\Phi(\theta_2)$.

\begin{align*}
    \Phi(\theta_2) = & \sum_{\bx} \left\{\frac{\mathbb{I}_{\bx}(\bX)}{p(\bx)}\left[1 - S - g_0(\bx)\right] \right\}p(\bx)\left\{v(\bx) + \lambda_2(\bx)\left[w(\bx) - v(\bx)\right] \right\} \\
    & + \sum_{\bx} \left\{(1 + \gamma)\frac{S\mathbb{I}_{\bx}(\bX)\mathbb{I}_t(T)}{p(\bx,1,t)}\left[Y - \mu(\bx,1,t) \right] \right\}g_0(\bx)p(\bx) \\
    & \hspace{1.2cm} \times \left\{\lambda_1(\bx) + \alpha\lambda_1(\bx)\lambda_2(\bx)\left[v(\bx) - w(\bx) \right] \right\} \\
    & + \sum_{\bx} \left\{\left[1 + \rho e_{1-t}(\bx, 0)\right]\left(\frac{(1-S)\mathbb{I}_{\bx}(\bX)\mathbb{I}_t(T)}{p(\bx,0,t)}\left[ Y - \mu(\bx, 0, t)\right] \right) \right. \\
    & \left. \hspace{1.2cm} + \rho\mu(\bx,0,t)\left(\frac{(1-S)\mathbb{I}_{\bx}(\bX)}{p(\bx,0)}\left[\mathbb{I}_{1-t}(T) - e_{1-t}(\bx, 0) \right] \right) \right\} \\
    & \hspace{1.2cm} \times g_0(\bx)p(\bx)\left\{\lambda_2(\bx) + \alpha\lambda_1(\bx)\lambda_2(\bx)\left[w(\bx) - v(\bx) \right] \right\} \\
    & + \sum_{\bx} \left\{\mathbb{I}_{\bx}(\bX) - p(\bx) \right\}g_0(\bx)\left\{v(\bx) + \lambda_2(\bx)\left[w(\bx) - v(\bx)\right] \right\}.
\end{align*}

By applying the indicator function and simplifying, we get

\begin{align*}
    \Phi(\theta_2) = & \left[1 - S - g_0(\bX)\right] \left\{v(\bX) + \lambda_2(\bX)\left[w(\bX) - v(\bX)\right] \right\} \\
    & +(1 + \gamma)\frac{S\mathbb{I}_t(T)}{e_t(\bX, 1)g_1(\bX)}\left[Y - \mu(\bX,1,t) \right] g_0(\bX) \\
    & \hspace{1.2cm} \times \left\{\lambda_1(\bX) + \alpha\lambda_1(\bX)\lambda_2(\bX)\left[v(\bX) - w(\bX) \right] \right\} \\
    & + (1-S)\left\{\lambda_2(\bX) + \alpha\lambda_1(\bX)\lambda_2(\bX)\left[w(\bX) - v(\bX) \right] \right\} \\
    & \hspace{0.5cm} \times \left\{\frac{\mathbb{I}_t(T)}{e_t(\bX, 0)}\left[ Y - \mu(\bX, 0, t)\right]\left[1 + \rho e_{1-t}(\bX, 0)\right]  \right. \\
    & \left. \hspace{1.2cm} + \rho\mu(\bX,0,t)\left[\mathbb{I}_{1-t}(T) - e_{1-t}(\bX, 0) \right] \right\} \\
    & + g_0(\bX)\left\{v(\bX) + \lambda_2(\bX)\left[w(\bX) - v(\bX)\right] \right\} \\
    & - \theta_2.
\end{align*}

Then, reorganizing the order of terms and canceling, we arrive at the final form of a candidate EIF for $\theta_2$:

\begin{align*}
    \Phi(\theta_2) = & S(1 + \gamma)\left\{\lambda_1(\bX) + \alpha\lambda_1(\bX)\lambda_2(\bX)\left[v(\bX) - w(\bX) \right] \right\} \\
    & \hspace{0.5cm} \times \left\{\frac{\mathbb{I}_t(T)}{e_t(\bX, 1)g_1(\bX)}\left[Y - \mu(\bX,1,t) \right] g_0(\bX)\right\} \\        
    & + \left[1 - S\right] \left\{\lambda_1(\bX)v(\bX) + \lambda_2(\bX)w(\bX) \right\} \\
    & + [1-S]\left\{\lambda_2(\bX) + \alpha\lambda_1(\bX)\lambda_2(\bX)\left[w(\bX) - v(\bX) \right] \right\} \\
    & \hspace{0.5cm} \times \Bigg\{\frac{\mathbb{I}_t(T)}{e_t(\bX, 0)}\left[ Y - \mu(\bX, 0, t)\right]\left[1 + \rho e_{1-t}(\bX, 0)\right] \\
    & \hspace{1.5cm} + \rho\mu(\bX,0,t)\left[\mathbb{I}_{1-t}(T) - e_{1-t}(\bX, 0) \right] \Bigg\} \\        
    & - \theta_2.
\end{align*}

As we did for $\phi_{\theta_1}$, we drop the arguments and simply denote the EIF for $\theta_2$ as $\phi_{\theta_2}$ for simplicity.

\subsubsection{Checking Candidate EIF for $\theta_1$}
We start by rewriting 
\begin{equation*}
    \begin{aligned}
        \theta_1 = & \int_{\bx}g_1(\bx)\mu(\bx,1,t)p(\bx) dx \\
        = & \int_{\bx}\int_{y} y p(y|\bx, 1, t) dy \hspace{0.1cm} p(1|\bx)p(\bx) dx,
    \end{aligned}
\end{equation*}
by replacing $\mu(\bx,1,t)$ with $\int_{y} y p(y|\bx, 1, t) dy$ and $g_1(\bx)$ with $p(1|\bx)$.

Recall that we need to show that $\nabla_{h}\theta_1 = \expP[\phi_{\theta_1}h]$ for any generic score, $h$. To compute the directional derivative for $\theta_1$ and $h$, $\nabla_{h}\theta_1$,  we introduce the notation
$\peps = (1 + \epsilon h) p$. Then, for any generic $h$ we have that

\begin{equation*}
    \begin{aligned}
        \nabla_{h}\theta_1 = & \left.\frac{\partial}{\partial\epsilon}\int_{\bx}\int_{y} y \peps(y|\bx, 1, t) dy \hspace{0.1cm} \peps(1|\bx)\peps(\bx) dx \right|_{\epsilon=0}.
    \end{aligned}
\end{equation*}

For the expectation $\expP[\phi_{\theta_1}h]$ we forego denoting the $\mathcal{P}$ for the remainder of this section for brevity. Then, we take the following steps to rewrite this expectation.

\begin{equation*}
    \begin{split}
        \expe[\phi_{\theta_1}h] = & \expe[(\phi_{\theta_1} + \theta_1)h] - \expe[\theta_1h] \\
        = & \int_{\bx}\sum_{s}\sum_{t'}\int_{y} \left[\frac{S\mathbb{I}(t'=t)}{e_t(\bx, 1)} \Big[y - \mu(\bx, 1, t)\Big] + S\mu(\bx, 1, t) \right] \\
        & \hspace{2.4cm} \times h p(y|\bx,s,t')dy \hspace{0.1cm}p(t'|\bx,s)p(s|\bx)p(\bx)dx \\
        = & \int_{\bx}\int_{y}\left[y - \mu(\bx,1,t) \right]h p(y|\bx,1,t)dy \hspace{0.1cm}p(1|\bx)p(\bx)dx
        \\ & + \int_{\bx} \mu(\bx, 1, t) \sum_{t'}\int_y hp(y|\bx,1,t')dy \hspace{0.1cm}p(t'|\bx,1)p(1|\bx)p(\bx)dx.
    \end{split}
\end{equation*}
In the above, we leveraged the fact that $\expe[\theta_1h] = 0$ for any valid score function $h$.

It is easier to prove this equality by decomposing $h = h_{Y|\bX, S, T} + h_{T|\bX, S} + h_{S|\bX} + h_{\bX}$ and showing that
\begin{gather*}
    \nabla_{h_{Y|\bX, S, T}}\theta_1 + \nabla_{h_{T|\bX, S}}\theta_1 + \nabla_{h_{S|\bX}}\theta_1 + \nabla_{h_{\bX}}\theta_1 = \\
    \expe[\phi_{\theta_1}h_{Y|\bX, S, T}] + \expe[\phi_{\theta_1}h_{T|\bX, S}] + \expe[\phi_{\theta_1}h_{S|\bX}] + \expe[\phi_{\theta_1}h_{\bX}].
\end{gather*}

In particular, we will show that
\begin{gather*}
    \nabla_{h_{\bX}}\theta_1 = \expe[\phi_{\theta_1}h_{\bX}], \\
    \nabla_{h_{S|\bX}}\theta_1 = \expe[\phi_{\theta_1}h_{S|\bX}], \\
    \nabla_{h_{T|\bX, S}}\theta_1 = \expe[\phi_{\theta_1}h_{T|\bX, S}], \\
    \nabla_{h_{Y|\bX, S, T}}\theta_1 = \expe[\phi_{\theta_1}h_{Y|\bX, S, T}].
\end{gather*}

We start with $\nabla_{h_{\bX}}\theta_1 = \expe[\phi_{\theta_1}h_{\bX}]$. For $\nabla_{h_{\bX}}\theta_1$ we replace the corresponding term in the factorized distribution function, $\peps(\bx) = (1 + \epsilon h_{\bX}(\bx))p(\bx)$, and set the other conditional probability density functions to their normal $p$ form.

\begin{equation*}
    \begin{split}
        \nabla_{h_{\bX}}\theta_1 = & \left.\frac{\partial}{\partial\epsilon}\int_{\bx}\int_{y} y p(y|\bx, 1, t)dy \hspace{0.1cm}p(1|\bx)(1 + \epsilon h_{\bX}(\bx))p(\bx) dx \right|_{\epsilon=0} \\
        = & \int_{\bx}\int_{y} y p(y|\bx, 1, t)dy \hspace{0.1cm}p(1|\bx)h_{\bX}(\bx)p(\bx) dx \\
        = & \int_{\bx}\mu(\bx,1,t) \hspace{0.1cm}p(1|\bx)h_{\bX}(\bx)p(\bx) dx \\
        = & \expe[\mu(\bX,1,t)g_1(\bX)h_{\bX}(\bX)].
    \end{split}
\end{equation*}

Then we are left to show that $\expe[\phi_{\theta_1}h_{\bX}] = \expe[\mu(\bX,1,t)g_1(\bX)h_{\bX}(\bX)]$:

\begin{equation*}
    \begin{split}
        \expe[\phi_{\theta_1}h_{\bX}] = & \int_{\bx}\int_{y}\left[y - \mu(\bx,1,t) \right]h_{\bX}(\bx) p(y|\bx,1,t)dy \hspace{0.1cm}p(1|\bx)p(\bx)dx
        \\ & + \int_{\bx} \mu(\bx, 1, t) \sum_{t'}\int_y h_{\bX}(\bx)p(y|\bx,1,t')dy \hspace{0.1cm}p(t'|\bx,1)p(1|\bx)p(\bx)dx \\
        = & \int_{\bx}h_{\bX}(\bx) \Bigg(\int_{y}\left[y - \mu(\bx,1,t) \right] p(y|\bx,1,t)dy \Bigg) \hspace{0.1cm}p(1|\bx)p(\bx)dx
        \\ & + \int_{\bx} h_{\bX}(\bx)\mu(\bx, 1, t) \Bigg(\sum_{t'}\int_y p(y|\bx,1,t')dy \hspace{0.1cm}p(t'|\bx,1) \Bigg)p(1|\bx)p(\bx)dx \\
        = & 0 + \int_{\bx} h_{\bX}(\bx)\mu(\bx, 1, t) \big(1 \big)p(1|\bx)p(\bx)dx \\
        = & \expe[\mu(\bX,1,t)g_1(\bX)h_{\bX}(\bX)]
    \end{split}
\end{equation*}

Similarly, for $\nabla_{h_{S|\bX}}\theta_1 = \expe[\phi_{\theta_1}h_{S|\bX}]$, we first simplify $\nabla_{h_{S|\bX}}\theta_1$ by replacing the corresponding term in the factorized distribution function, $\peps(1|\bx) = (1 + \epsilon h_{S|\bX}(1|\bx))p(1|\bx)$, and set the other conditional probability density functions to their normal $p$ form.

\begin{equation*}
    \begin{split}
        \nabla_{h_{S|\bX}}\theta_1 = & \left.\frac{\partial}{\partial\epsilon}\int_{\bx}\int_{y} y p(y|\bx, 1, t)dy \hspace{0.1cm}(1 + \epsilon h_{S|\bX}(1|\bx))p(1|\bx)p(\bx) dx \right|_{\epsilon=0} \\
        = & \int_{\bx}\int_{y} y p(y|\bx, 1, t)dy \hspace{0.1cm} h_{S|\bX}(1 | \bx)p(1|\bx)p(\bx) dx \\
        = & \int_{\bx}\mu(\bx,1,t) h_{S|\bX}(1 | \bx)p(1|\bx)p(\bx) dx \\
        = & \expe[\mu(\bX,1,t)g_1(\bX)h_{S|\bX}(1|\bX)].
    \end{split}
\end{equation*}

We then show that $\expe[\phi_{\theta_1}h_{S|\bX}] = \expe[\mu(\bX,1,t)g_1(\bX)h_{S|\bX}(1|\bX)]$:

\begin{equation*}
    \begin{split}
        \expe[\phi_{\theta_1}h_{S|\bX}] = & \int_{\bx}\int_{y}\left[y - \mu(\bx,1,t) \right]h_{S|\bX}(s|\bx) p(y|\bx,1,t)dy \hspace{0.1cm}p(1|\bx)p(\bx)dx
        \\ & + \int_{\bx} \mu(\bx, 1, t) \sum_{t'}\int_y h_{S|\bX}(s|\bx)p(y|\bx,1,t')dy \hspace{0.1cm}p(t'|\bx,1)p(1|\bx)p(\bx)dx \\
        = & \int_{\bx}h_{S|\bX}(s|\bx) \Bigg(\int_{y}\left[y - \mu(\bx,1,t) \right] p(y|\bx,1,t)dy \Bigg) \hspace{0.1cm}p(1|\bx)p(\bx)dx
        \\ & + \int_{\bx} h_{S|\bX}(s|\bx)\mu(\bx, 1, t) \Bigg(\sum_{t'}\int_y p(y|\bx,1,t')dy \hspace{0.1cm}p(t'|\bx,1) \Bigg)p(1|\bx)p(\bx)dx \\
        = & 0 + \int_{\bx} h_{S|\bX}(s|\bx)\mu(\bx, 1, t) \big(1 \big)p(1|\bx)p(\bx)dx \\
        = & \int_{\bx} h_{S|\bX}(1|\bx)\mu(\bx, 1, t) p(1|\bx)p(\bx)dx \\
        = & \expe[\mu(\bX,1,t)g_1(\bX)h_{S|\bX}(1|\bX)]
    \end{split}
\end{equation*}

Moving on to $\nabla_{h_{T|\bX, S}}\theta_1 = \expe[\phi_{\theta_1}h_{T|\bX, S}]$, we start by simplifying $\nabla_{h_{T|\bX, S}}\theta_1$ by replacing the corresponding term in the factorized distribution function and setting the other conditional probability density functions to their normal $p$ form. We note here that there is no $\peps(t|\bx,s)$ in the integral, so this term simplifies to zero.

\begin{equation*}
    \begin{split}
        \nabla_{h_{T|\bX, S}}\theta_1 = & \left. \frac{\partial}{\partial\epsilon}\int_{\bx}\int_{y} y p(y|\bx, 1, t)dy \hspace{0.1cm} p(1|\bx)p(\bx) dx\right|_{\epsilon=0} \\
        = & 0.
    \end{split}
\end{equation*}

We show too that $\expe[\phi_{\theta_1}h_{T|\bX, S}]$ is equal to zero.

\begin{equation*}
    \begin{split}
        \expe[\phi_{\theta_1}h_{T|\bX, S}] = & \int_{\bx}\int_{y}\left[y - \mu(\bx,1,t) \right]h_{T|\bX, S}(t| \bx, s) p(y|\bx,1,t)dy \hspace{0.1cm}p(1|\bx)p(\bx)dx
        \\ & + \int_{\bx} \mu(\bx, 1, t) \sum_{t'}\int_y h_{T|\bX, S}(t'| \bx, s)p(y|\bx,1,t')dy \hspace{0.1cm}p(t'|\bx,1)p(1|\bx)p(\bx)dx \\
        = & \int_{\bx}h_{T|\bX, S}(t| \bx, s) \Bigg(\int_{y}\left[y - \mu(\bx,1,t) \right] p(y|\bx,1,t)dy \Bigg) p(1|\bx)p(\bx)dx
        \\ & + \int_{\bx} \mu(\bx, 1, t) \sum_{t'} h_{T|\bX, S}(t'| \bx, s) \Bigg(\int_y p(y|\bx,1,t')dy \Bigg) p(t'|\bx,1)p(1|\bx)p(\bx)dx \\
        = & 0 + \int_{\bx} \mu(\bx, 1, t) \Bigg(\sum_{t'} h_{T|\bX, S}(t'| \bx, s) p(t'|\bx,1)\Bigg)p(1|\bx)p(\bx)dx \\
        = & 0 + \int_{\bx} \mu(\bx, 1, t) \big(0\big)p(1|\bx)p(\bx)dx \\
        = & 0.
    \end{split}
\end{equation*}

Similarly, for $\nabla_{h_{Y|\bX, S, T}}\theta_1 = \expe[\phi_{\theta_1}h_{Y|\bX, S, T}]$, we simplify $\nabla_{h_{Y|\bX, S, T}}\theta_1$ by replacing the corresponding term in the factorized distribution function, $\peps(y|\bx, 1, t) = (1 + \epsilon h_{Y|\bX, S, T}(y|\bx, 1, t))p(y|\bx, 1, t)$, and set the other conditional probability density functions to their normal $p$ form.

\begin{equation*}
    \begin{split}
        \nabla_{h_{Y|\bX, S, T}}\theta_1 = & \left.\frac{\partial}{\partial\epsilon}\int_{\bx}\int_{y} y (1 + \epsilon h_{Y|\bX, S, T}(y|\bx, 1, t))p(y|\bx, 1, t)dy \hspace{0.1cm} p(1|\bx)p(\bx) dx\right|_{\epsilon=0} \\
        = & \int_{\bx}\int_{y} y h_{Y|\bX, S, T}(y|\bx, 1, t)p(y|\bx, 1, t)dy \hspace{0.1cm} p(1|\bx)p(\bx) dx \\
        = & \expe[g_1(\bX)\expe[Yh_{Y|\bX, S, T}(Y|\bX, 1, t) | \bX, S = 1, T = t]]
    \end{split}
\end{equation*}

We finish validating the EIF for $\theta_1$ by showing that \newline$\expe[\phi_{\theta_1}h_{Y|\bX, S, T}] = \expe[g_1(\bX)\expe[Yh_{Y|\bX, S, T}(Y|\bX, 1, t) | \bX, S = 1, T = t]]$.

\begin{equation*}
    \begin{split}
        \expe[\phi_{\theta_1}h_{Y|\bX, S, T}] = & \int_{\bx}\int_{y}\left[y - \mu(\bx,1,t) \right]h_{Y|\bX, S, T}(y|\bx, s, t) p(y|\bx,1,t)dy \hspace{0.1cm}p(1|\bx)p(\bx)dx
        \\ & + \int_{\bx} \mu(\bx, 1, t) \sum_{t'}\int_y h_{Y|\bX, S, T}(y|\bx, s, t')p(y|\bx,1,t')dy p(t'|\bx,1)p(1|\bx)p(\bx)dx \\
        = & \expe\big[\mathbb{I}(S=1)\expe[(Y - \mu(\bX, 1, t))h_{Y|\bX, S, T}(Y | \bX, 1, t) | \bX, S = 1, T = t] \big]\\
        & + \expe\big[\mu(\bX, 1, t)\mathbb{I}(S=1) \expe[h_{Y|\bX, S, T}(Y | \bX, 1, T) | \bX, S = 1] \big] \\
        = & \expe\big[\mathbb{I}(S=1)\expe[Yh_{Y|\bX, S, T}(Y | \bX, 1, t) | \bX, S = 1, T = t] \big] \\
        & - \expe\big[\mathbb{I}(S=1)\mu(\bX, 1, t)\expe[h_{Y|\bX, S, T}(Y | \bX, 1, t) | \bX, S = 1, T = t] \big] \\
        & + \expe\big[\mu(\bX, 1, t)\mathbb{I}(S=1) \expe[h_{Y|\bX, S, T}(Y | \bX, 1, T) | \bX, S = 1] \big] \\
        = & \expe\big[\mathbb{I}(S=1)\expe[Yh_{Y|\bX, S, T}(Y | \bX, 1, t) | \bX, S = 1, T = t] \big] \\
        & - \expe\big[\mathbb{I}(S=1)\mu(\bX, 1, t)\times 0 \big] \\
        & + \expe\big[\mu(\bX, 1, t)\mathbb{I}(S=1)\times 0 \big] \\
        = & \expe\big[g_1(\bX)\expe[Yh_{Y|\bX, S, T}(Y | \bX, 1, t) | \bX, S = 1, T = t] \big]
    \end{split}
\end{equation*}

\subsubsection{Checking Candidate EIF for $\theta_2$}
We perform the same steps to check the candidate EIF for $\theta_2$. We first write $\theta_2$ as shown below:

\begin{equation*}
    \begin{split}
        \theta_2 = & \int_{\bx} g_0(\bx)\left\{\lambda_1(\bx)v(\bx) + \lambda_2(\bx)w(\bx)\right\}p(x)dx \\
        = & \int_{\bx} \Bigg[\frac{\exp\left(\alpha(1+\gamma)\int_y y p(y|\bx,1,t)dy \right)}{\exp\left(\alpha(1 + \gamma)\int_y y p(y|\bx,1,t)dy \right) + \exp\left(\alpha\int_y y p(y|\bx,0,t)dy[1 + \rho p(1-t|\bx,0)] \right)} \\
        & \hspace{1cm} \times (1 + \gamma)\int_y y p(y|\bx,1,t)dy\Bigg] p(0|\bx) p(x)dx \\
        & + \int_{\bx} \Bigg[\frac{\exp\left(\alpha\int_y y p(y|\bx,0,t)dy[1 + \rho p(1-t|\bx,0)] \right)}{\exp\left(\alpha(1 + \gamma)\int_y y p(y|\bx,1,t)dy \right) + \exp\left(\alpha\int_y y p(y|\bx,0,t)dy[1 + \rho p(1-t|\bx,0)] \right)} \\
        & \hspace{1cm} \times \int_y y p(y|\bx,0,t)dy[1 + \rho p(1-t|\bx,0)]\Bigg] p(0|\bx)p(x)dx,
    \end{split}
\end{equation*}

where we replace $g_0(\bx)$ with $p(0|\bx)$, $\mu(\bx,1,t)$ with $\int_{y} y p(y|\bx, 1, t) dy$, $\mu(\bx,0,t)$ with $\int_{y} y p(y|\bx, 0, t) dy$, $e_t(\bx, 0)$ with $p(t | \bx, 0)$ and $e_{1-t}(\bx, 0)$ with $p(1-t | \bx, 0)$.

We then have that for any generic $h$ that $\nabla_{h}\theta_2$ can be written as

\begin{equation*}
    \begin{gathered}
        \nabla_{h}\theta_2 = \\ \frac{\partial}{\partial\epsilon} \int_{\bx} \Bigg[\frac{\exp\left(\alpha(1 + \gamma)\int_y y \peps(y|\bx,1,t)dy \right)}{\exp\left(\alpha(1 + \gamma)\int_y y \peps(y|\bx,1,t)dy \right) + \exp\left(\alpha\int_y y \peps(y|\bx,0,t)dy[1 + \rho \peps(1-t|\bx,0)] \right)} \\
         \left.\times (1 + \gamma)\int_y y \peps(y|\bx,1,t)dy\Bigg] \peps(0|\bx)\peps(x)dx \right|_{\epsilon=0} \\
         + \frac{\partial}{\partial\epsilon} \int_{\bx} \Bigg[\frac{\exp\left(\alpha\int_y y \peps(y|\bx,0,t)dy[1 + \rho \peps(1-t|\bx,0)] \right)}{\exp\left(\alpha(1 + \gamma)\int_y y \peps(y|\bx,1,t)dy \right) + \exp\left(\alpha\int_y y \peps(y|\bx,0,t)dy[1 + \rho \peps(1-t|\bx,0)] \right)} \\
        \left.\times \int_y y \peps(y|\bx,0,t)dy[1 + \rho \peps(1-t|\bx,0)]\Bigg] \peps(0|\bx)\peps(x)dx \right|_{\epsilon=0}
    \end{gathered}
\end{equation*}

and $\expe[\phi_{\theta_2}h]$ can be written as

\begin{equation*}
    \begin{split}
        \expe[\phi_{\theta_2}h] = & \expe[(\phi_{\theta_2} + \theta_2)h] - \expe[\theta_2h] \\
        = & \int_{\bx}\sum_{s}\sum_{t'}\int_{y} \Bigg\{ \\
        & s(1 + \gamma)\left\{\lambda_1(\bx) + \alpha\lambda_1(\bx)\lambda_2(\bx)\left[v(\bx) - w(\bx) \right] \right\} \\
        & \hspace{0.5cm} \times \left\{\frac{\mathbb{I}_t(t')}{e_t(\bx, 1)g_1(\bx)}\left[y - \mu(\bx,1,t) \right] g_0(\bx)\right\} \\        
        & + \left[1 - s\right] \left\{\lambda_1(\bx)v(\bx) + \lambda_2(\bx)w(\bx) \right\} \\
        & + [1-s]\left\{\lambda_2(\bx) + \alpha\lambda_1(\bx)\lambda_2(\bx)\left[w(\bx) - v(\bx) \right] \right\} \\
        & \hspace{0.5cm} \times \left\{\frac{\mathbb{I}_t(t')}{e_t(\bx, 0)}\left[ y - \mu(\bx, 0, t)\right]\left[1 + \rho e_{1-t}(\bx, 0)\right] + \rho\mu(\bx,0,t)\left[\mathbb{I}_{1-t}(t') - e_{1-t}(\bx, 0) \right] \right\} \\
        & \Bigg\}\times h p(y|\bx,s,t')p(t'|\bx,s)p(s|\bx)p(\bx) dy dx.
    \end{split}
\end{equation*}

We can distribute terms to write this as
\begin{equation*}
    \begin{split}
        \expe[\phi_{\theta_2}h] = & \int_{\bx}\int_{y} (1 + \gamma)\left\{\lambda_1(\bx) + \alpha\lambda_1(\bx)\lambda_2(\bx)\left[v(\bx) - w(\bx) \right] \right\} \\
        & \hspace{1cm} \times \left[y - \mu(\bx,1,t) \right] h p(y|\bx,1,t)p(0|\bx)p(\bx) dy dx \\
        & + \int_{\bx}\sum_{t'}\int_{y} \left\{\lambda_1(\bx)v(\bx) + \lambda_2(\bx)w(\bx) \right\}h p(y|\bx,0,t')p(t'|\bx,0)p(0|\bx)p(\bx) dy dx \\
        & + \int_{\bx}\int_{y} \left\{\lambda_2(\bx) + \alpha\lambda_1(\bx)\lambda_2(\bx)\left[w(\bx) - v(\bx) \right] \right\}\left[ y - \mu(\bx, 0, t)\right]\left[1 + \rho e_{1-t}(\bx, 0)\right] \\
        & \hspace{1cm} \times h p(y|\bx,0,t)p(0|\bx)p(\bx) dy dx \\
        & + \int_{\bx}\sum_{t'}\int_{y} \left\{\lambda_2(\bx) + \alpha\lambda_1(\bx)\lambda_2(\bx)\left[w(\bx) - v(\bx) \right] \right\}\rho\mu(\bx,0,t)\left[\mathbb{I}_{1-t}(t') - e_{1-t}(\bx, 0) \right] \\
        & \hspace{2cm} \times h p(y|\bx,0,t')p(t'|\bx,0)p(0|\bx)p(\bx) dy dx.
    \end{split}
\end{equation*}

And then write in expectation notation as
\begin{equation*}
    \begin{split}        
        \expe[\phi_{\theta_2}h] = & (1 + \gamma)\expe\Big[g_0(\bX)\left\{\lambda_1(\bX) + \alpha\lambda_1(\bX)\lambda_2(\bX)\left[v(\bX) - w(\bX) \right]\right\} \\ 
        & \hspace{1.5cm} \times \expe[(Y - \mu(\bX,1,t))h | \bX, S=1, T = t]\Big] \\
        & + \expe\Big[g_0(\bX)\left\{\lambda_1(\bX)v(\bX) + \lambda_2(\bX)w(\bX)\right\}\expe[h | \bX, S=0]\Big] \\
        & + \expe\Big[g_0(\bX)\left[1 + \rho e_{1-t}(\bX, 0)\right]\left\{\lambda_2(\bX) + \alpha\lambda_1(\bX)\lambda_2(\bX)\left[w(\bX) - v(\bX) \right] \right\} \\
        & \hspace{1cm} \times \expe[(Y - \mu(\bX,0,t))h | \bX, S=0, T=t]\Big] \\
        & + \expe\Big[g_0(\bX)\rho\mu(\bX,0,t)\left\{\lambda_2(\bX) + \alpha\lambda_1(\bX)\lambda_2(\bX)\left[w(\bX) - v(\bX) \right] \right\} \\
        & \hspace{1cm} \times \expe[\left[\mathbb{I}_{1-t}(T) - e_{1-t}(\bX, 0) \right]h | \bX, S = 0]\Big].
    \end{split}
\end{equation*}

We now show equality of the separate decomposed components of the generic score. In particular, just like we did for $\theta_1$, we will show that
\begin{gather*}
    \nabla_{h_{\bX}}\theta_2 = \expe[\phi_{\theta_2}h_{\bX}], \\
    \nabla_{h_{S|\bX}}\theta_2 = \expe[\phi_{\theta_2}h_{S|\bX}], \\
    \nabla_{h_{T|\bX, S}}\theta_2 = \expe[\phi_{\theta_2}h_{T|\bX, S}], \\
    \nabla_{h_{Y|\bX, S, T}}\theta_2 = \expe[\phi_{\theta_2}h_{Y|\bX, S, T}].
\end{gather*}

Starting with $\nabla_{h_{\bX}}\theta_2 = \expe[\phi_{\theta_2}h_{\bX}]$, we first simplify $\nabla_{h_{\bX}}\theta_2$ by replacing the corresponding term in the factorized distribution function, $\peps(\bx) = (1 + \epsilon h_{\bX}(\bx))p(\bx)$, and set the other conditional probability density functions to their normal $p$ form.

\begin{equation*}
    \begin{split}
        \nabla_{h_{\bX}}\theta_2 = & \frac{\partial}{\partial\epsilon} \int_{\bx} \Bigg[\frac{\exp\left(\alpha(1+\gamma)\int_y y p(y|\bx,1,t)dy \right)}{\exp\left(\alpha(1+\gamma)\int_y y p(y|\bx,1,t)dy \right) + \exp\left(\alpha\int_y y p(y|\bx,0,t)dy[1+\rho p(1-t|\bx,0)] \right)} \\
        & \hspace{1cm} \left. \times (1+\gamma)\int_y y p(y|\bx,1,t)dy\Bigg] p(0|\bx)(1 + \epsilon h_{\bX}(\bx))p(x)dx \right|_{\epsilon=0} \\
        & + \frac{\partial}{\partial\epsilon} \int_{\bx} \Bigg[\frac{\exp\left(\alpha\int_y y p(y|\bx,0,t)dy[1+\rho p(1-t|\bx,0)] \right)}{\exp\left(\alpha(1+\gamma)\int_y y p(y|\bx,1,t)dy \right) + \exp\left(\alpha\int_y y p(y|\bx,0,t)dy[1+\rho p(1-t|\bx,0)] \right)} \\
        & \hspace{1cm} \left. \times \int_y y p(y|\bx,0,t)dy[1+\rho p(1-t|\bx,0)]\Bigg] p(0|\bx)(1 + \epsilon h_{\bX}(\bx))p(x)dx \right|_{\epsilon=0} \\
        = & \int_{\bx} \left\{\lambda_1(\bX)v(\bX) + \lambda_2(\bX)w(\bX)\right\} p(0|\bx)h_{\bX}(\bx)p(x)dx \\
        = & \expe\Big[g_0(\bX)\left\{\lambda_1(\bX)v(\bX) + \lambda_2(\bX)w(\bX)\right\}h_{\bX}(\bX)\Big].
    \end{split}
\end{equation*}

We now show that $\expe[\phi_{\theta_2}h_{\bX}] = \expe\Big[g_0(\bX)\left\{\lambda_1(\bX)v(\bX) + \lambda_2(\bX)w(\bX)\right\}h_{\bX}(\bX)\Big]$:

\begin{equation*}
    \begin{split}
        \expe[\phi_{\theta_2}h_{\bX}] = & (1 + \gamma)\expe\Big[g_0(\bX)\left\{\lambda_1(\bX) + \alpha\lambda_1(\bX)\lambda_2(\bX)\left[v(\bX) - w(\bX) \right]\right\} \\
        & \hspace{1.5cm} \times \expe[(Y - \mu(\bX,1,t))h_{\bX}(\bX) | \bX, S=1, T = t]\Big] \\
        & + \expe\Big[g_0(\bX)\left\{\lambda_1(\bX)v(\bX) + \lambda_2(\bX)w(\bX)\right\}\expe[h_{\bX}(\bX) | \bX, S = 0]\Big] \\
        & + \expe\Big[g_0(\bX)\left[1 + \rho e_{1-t}(\bX, 0)\right]\left\{\lambda_2(\bX) + \alpha\lambda_1(\bX)\lambda_2(\bX)\left[w(\bX) - v(\bX) \right] \right\} \\
        & \hspace{1cm} \times \expe[(Y - \mu(\bX,0,t))h_{\bX}(\bX) | \bX, S=0, T=t]\Big] \\
        & + \expe\Big[g_0(\bX)\rho\mu(\bX,0,t)\left\{\lambda_2(\bX) + \alpha\lambda_1(\bX)\lambda_2(\bX)\left[w(\bX) - v(\bX) \right] \right\} \\
        & \hspace{1cm} \times \expe[\left[\mathbb{I}_{1-t}(T) - e_{1-t}(\bX, 0) \right]h_{\bX}(\bX) | \bX, S = 0]\Big].
    \end{split}
\end{equation*}

We can pull the $h_{\bX}$ outside of all of the inner expectations to write as

\begin{equation*}
    \begin{split}
        \expe[\phi_{\theta_2}h_{\bX}] = & (1 + \gamma)\expe\Big[g_0(\bX)\left\{\lambda_1(\bX) + \alpha\lambda_1(\bX)\lambda_2(\bX)\left[v(\bX) - w(\bX) \right]\right\} \\
        & \hspace{1.5cm} \times h_{\bX}(\bX)\expe[(Y - \mu(\bX,1,t)) | \bX, S=1, T = t]\Big] \\
        & + \expe\Big[g_0(\bX)\left\{\lambda_1(\bX)v(\bX) + \lambda_2(\bX)w(\bX)\right\}h_{\bX}(\bX)\Big] \\
        & + \expe\Big[g_0(\bX)\left[1 + \rho e_{1-t}(\bX, 0)\right]\left\{\lambda_2(\bX) + \alpha\lambda_1(\bX)\lambda_2(\bX)\left[w(\bX) - v(\bX) \right] \right\} \\
        & \hspace{1cm} \times h_{\bX}(\bX)\expe[(Y - \mu(\bX,0,t)) | \bX, S=0, T=t]\Big] \\
        & + \expe\Big[g_0(\bX)\rho\mu(\bX,0,t)\left\{\lambda_2(\bX) + \alpha\lambda_1(\bX)\lambda_2(\bX)\left[w(\bX) - v(\bX) \right] \right\} \\
        &\hspace{1cm} \times h_{\bX}(\bX)\expe[\left[\mathbb{I}_{1-t}(T) - e_{1-t}(\bX, 0) \right] | \bX, S = 0]\Big].
    \end{split}
\end{equation*}

And then, applying the inner conditional expectations and cancelling like terms, we get

\begin{equation*}
    \begin{split}        
        \expe[\phi_{\theta_2}h_{\bX}] = & (1 + \gamma)\expe\Big[g_0(\bX)\left\{\lambda_1(\bX) + \alpha\lambda_1(\bX)\lambda_2(\bX)\left[v(\bX) - w(\bX) \right]\right\} \\
        & \hspace{1.5cm} \times h_{\bX}(\bX)(\mu(\bX,1,t) - \mu(\bX,1,t))\Big] \\
        & + \expe\Big[g_0(\bX)\left\{\lambda_1(\bX)v(\bX) + \lambda_2(\bX)w(\bX)\right\}h_{\bX}(\bX)\Big] \\
        & + \expe\Big[g_0(\bX)\left[1 + \rho e_{1-t}(\bX, 0)\right]\left\{\lambda_2(\bX) + \alpha\lambda_1(\bX)\lambda_2(\bX)\left[w(\bX) - v(\bX) \right] \right\} \\
        & \hspace{1cm} \times h_{\bX}(\bX)(\mu(\bX,0,t) - \mu(\bX,0,t))\Big] \\
        & + \expe\Big[g_0(\bX)\rho\mu(\bX,0,t)\left\{\lambda_2(\bX) + \alpha\lambda_1(\bX)\lambda_2(\bX)\left[w(\bX) - v(\bX) \right] \right\} \\
        & \hspace{1cm} \times h_{\bX}(\bX)\left[e_{1-t}(\bX, 0) - e_{1-t}(\bX, 0) \right]\Big] \\
        = & \expe\Big[g_0(\bX)\left\{\lambda_1(\bX)v(\bX) + \lambda_2(\bX)w(\bX)\right\}h_{\bX}(\bX)\Big].
    \end{split}
\end{equation*}

Next, we show $\nabla_{h_{S|\bX}}\theta_2 = \expe[\phi_{\theta_2}h_{S|\bX}]$. We start this by simplifying $\nabla_{h_{S|\bX}}\theta_2$ by replacing the corresponding term in the factorized distribution function, $\peps(0|\bx) = (1 + \epsilon h_{S|\bX}(0|\bx))p(0|\bx)$, and set the other conditional probability density functions to their normal $p$ form.

\begin{equation*}
    \begin{split}
        \nabla_{h_{S|\bX}}\theta_2 = & \frac{\partial}{\partial\epsilon} \int_{\bx} \Bigg[ \frac{\exp\left(\alpha(1+\gamma)\int_y y p(y|\bx,1,t)dy \right)}{\exp\left(\alpha(1+\gamma)\int_y y p(y|\bx,1,t)dy \right) + \exp\left(\alpha\int_y y p(y|\bx,0,t)dy[1+\rho p(1-t|\bx,0)] \right)} \\
        & \hspace{1cm} \left. \times (1+\gamma)\int_y y p(y|\bx,1,t)dy\Bigg] (1 + \epsilon h_{S|\bX}(0|\bx))p(0|\bx) p(x)dx \right|_{\epsilon=0} \\
        & + \frac{\partial}{\partial\epsilon} \int_{\bx} \Bigg[\frac{\exp\left(\alpha\int_y y p(y|\bx,0,t)dy[1+\rho p(1-t|\bx,0)] \right)}{\exp\left(\alpha(1+\gamma)\int_y y p(y|\bx,1,t)dy \right) + \exp\left(\alpha\int_y y p(y|\bx,0,t)dy[1+\rho p(1-t|\bx,0)] \right)} \\
        & \hspace{1cm} \left. \times \int_y y p(y|\bx,0,t)dy[1+\rho p(1-t|\bx,0)]\Bigg] (1 + \epsilon h_{S|\bX}(0|\bx))p(0|\bx) p(x)dx \right|_{\epsilon=0} \\
        = & \int_{\bx} \left\{\lambda_1(\bX)v(\bX) + \lambda_2(\bX)w(\bX)\right\} h_{S|\bX}(0|\bx)p(0|\bx)p(x)dx \\
        = & \expe\Big[g_0(\bX)\left\{\lambda_1(\bX)v(\bX) + \lambda_2(\bX)w(\bX)\right\}h_{S|\bX}(0|\bX)\Big].
    \end{split}
\end{equation*}

We proceed to show that $\expe[\phi_{\theta_2}h_{S|\bX}] = \expe\Big[g_0(\bX)\left\{\lambda_1(\bX)v(\bX) + \lambda_2(\bX)w(\bX)\right\}h_{S|\bX}(0|\bX)\Big]$. First, plugging in $h_{S|\bX}$,

\begin{equation*}
    \begin{split}
        \expe[\phi_{\theta_2}h_{S|\bX}] = & (1 + \gamma)\expe\Big[g_0(\bX)\left\{\lambda_1(\bX) + \alpha\lambda_1(\bX)\lambda_2(\bX)\left[v(\bX) - w(\bX) \right]\right\} \\
        & \hspace{1.5cm} \times \expe[(Y - \mu(\bX,1,t))h_{S|\bX}(S|\bX) | \bX, S=1, T = t]\Big] \\
        & + \expe\Big[g_0(\bX)\left\{\lambda_1(\bX)v(\bX) + \lambda_2(\bX)w(\bX)\right\}\expe[h_{S|\bX}(S|\bX) | \bX, S=0]\Big] \\
        & + \expe\Big[g_0(\bX)\left[1 + \rho e_{1-t}(\bX, 0)\right]\left\{\lambda_2(\bX) + \alpha\lambda_1(\bX)\lambda_2(\bX)\left[w(\bX) - v(\bX) \right] \right\} \\
        & \hspace{1cm} \times \expe[(Y - \mu(\bX,0,t))h_{S|\bX}(S|\bX) | \bX, S=0, T=t]\Big] \\
        & + \expe\Big[g_0(\bX)\rho\mu(\bX,0,t)\left\{\lambda_2(\bX) + \alpha\lambda_1(\bX)\lambda_2(\bX)\left[w(\bX) - v(\bX) \right] \right\} \\
        & \hspace{1cm} \times \expe[\left[\mathbb{I}_{1-t}(T) - e_{1-t}(\bX, 0) \right]h_{S|\bX}(S|\bX) | \bX, S = 0]\Big].
    \end{split}
\end{equation*}

Then, pulling $h_{S|\bX}$ out of the inner expectation where possible
\begin{equation*}
    \begin{split}
        \expe[\phi_{\theta_2}h_{S|\bX}] = & (1 + \gamma)\expe\Big[g_0(\bX)\left\{\lambda_1(\bX) + \alpha\lambda_1(\bX)\lambda_2(\bX)\left[v(\bX) - w(\bX) \right]\right\} \\
        & \hspace{1.5cm} \times h_{S|\bX}(1|\bX)\expe[(Y - \mu(\bX,1,t)) | \bX, S=1, T = t]\Big] \\
        & + \expe\Big[g_0(\bX)\left\{\lambda_1(\bX)v(\bX) + \lambda_2(\bX)w(\bX)\right\}h_{S|\bX}(0|\bX)\Big] \\
        & + \expe\Big[g_0(\bX)\left[1 + \rho e_{1-t}(\bX, 0)\right]\left\{\lambda_2(\bX) + \alpha\lambda_1(\bX)\lambda_2(\bX)\left[w(\bX) - v(\bX) \right] \right\} \\
        & \hspace{1cm} \times h_{S|\bX}(0|\bX)\expe[(Y - \mu(\bX,0,t)) | \bX, S=0, T=t]\Big] \\
        & + \expe\Big[g_0(\bX)\rho\mu(\bX,0,t)\left\{\lambda_2(\bX) + \alpha\lambda_1(\bX)\lambda_2(\bX)\left[w(\bX) - v(\bX) \right] \right\} \\
        & \hspace{1cm} \times h_{S|\bX}(0|\bX) \expe[\left[\mathbb{I}_{1-t}(T) - e_{1-t}(\bX, 0) \right] | \bX, S = 0]\Big].
    \end{split}
\end{equation*}

And then applying the inner expectations, cancelling terms, and simplifying,

\begin{equation*}
    \begin{split}
        \expe[\phi_{\theta_2}h_{S|\bX}] = & (1 + \gamma)\expe\Big[g_0(\bX)\left\{\lambda_1(\bX) + \alpha\lambda_1(\bX)\lambda_2(\bX)\left[v(\bX) - w(\bX) \right]\right\} \\
        & \hspace{1.5cm} \times h_{S|\bX}(1|\bX)(\mu(\bX,1,t) - \mu(\bX,1,t))\Big] \\
        & + \expe\Big[g_0(\bX)\left\{\lambda_1(\bX)v(\bX) + \lambda_2(\bX)w(\bX)\right\}h_{S|\bX}(0|\bX) \Big] \\
        & + \expe\Big[g_0(\bX)\left[1 + \rho e_{1-t}(\bX, 0)\right]\left\{\lambda_2(\bX) + \alpha\lambda_1(\bX)\lambda_2(\bX)\left[w(\bX) - v(\bX) \right] \right\} \\
        & \hspace{1cm} \times h_{S|\bX}(0|\bX)(\mu(\bX,0,t) - \mu(\bX,0,t))\Big] \\
        & + \expe\Big[g_0(\bX)\rho\mu(\bX,0,t)\left\{\lambda_2(\bX) + \alpha\lambda_1(\bX)\lambda_2(\bX)\left[w(\bX) - v(\bX) \right] \right\} \\
        & \hspace{1cm} \times h_{S|\bX}(0|\bX) \left[e_{1-t}(\bX, 0) - e_{1-t}(\bX, 0) \right]\Big] \\
        = & \expe\Big[g_0(\bX)\left\{\lambda_1(\bX)v(\bX) + \lambda_2(\bX)w(\bX)\right\} h_{S|\bX}(0|\bX)\Big].
    \end{split}
\end{equation*}

Nearly there, we now show $\nabla_{h_{T|\bX, S}}\theta_2 = \expe[\phi_{\theta_2}h_{T|\bX, S}]$. We simplify $\nabla_{h_{T|\bX, S}}\theta_2$ by replacing the corresponding term in the factorized distribution function, $\peps(1-t|\bx,0) = (1 + \epsilon h_{T|\bX,S}(1-t|\bx, 0))p(1-t|\bx,0)$, and set the other conditional probability density functions to their normal $p$ form. We have to format $\nabla_{h_{T|\bX, S}}\theta_2$ slightly differently to allow it to stay within the page margins.

\begin{equation*}
    \begin{split}
        \nabla_{h_{T|\bX, S}}\theta_2 = & \frac{\partial}{\partial\epsilon} \int_{\bx} \exp\left(\alpha(1+\gamma)\int_y y p(y|\bx,1,t)dy \right) \\
        & \hspace{0.6cm} \times \Bigg\{\exp\left(\alpha\int_y y p(y|\bx,0,t)dy[1+\rho (1 + \epsilon h_{T|\bX,S}(1-t|\bx, 0))p(1-t|\bx,0)] \right)  \\
        & \hspace{1.2cm} + \exp\left(\alpha(1+\gamma)\int_y y p(y|\bx,1,t)dy \right) \Bigg\}^{-1} \\
        & \hspace{0.6cm} \left. \times \Bigg((1+\gamma)\int_y y p(y|\bx,1,t)dy\Bigg) p(0|\bx) p(x)dx \right|_{\epsilon=0} \\
        & + \frac{\partial}{\partial\epsilon} \int_{\bx} \exp\left(\alpha\int_y y p(y|\bx,0,t)dy[1+\rho (1 + \epsilon h_{T|\bX,S}(1-t|\bx, 0))p(1-t|\bx,0)] \right) \\
        & \hspace{0.6cm} \times \Bigg\{\exp\left(\alpha\int_y y p(y|\bx,0,t)dy[1+\rho (1 + \epsilon h_{T|\bX,S}(1-t|\bx, 0))p(1-t|\bx,0)] \right)  \\
        & \hspace{1.2cm} + \exp\left(\alpha(1+\gamma)\int_y y p(y|\bx,1,t)dy \right) \Bigg\}^{-1} \\
        & \hspace{0.6cm} \times \Bigg(\int_y y p(y|\bx,0,t)dy[1+\rho (1 + \epsilon h_{T|\bX,S}(1-t|\bx, 0))p(1-t|\bx,0)]\Bigg) \\
        & \hspace{0.6cm} \times p(0|\bx)p(x)dx \Bigg|_{\epsilon=0}
    \end{split}
\end{equation*}

To simplify, we plug in $\exp\left(\alpha v(\bx) \right)$ for $\exp\left(\alpha(1+\gamma)\int_y y p(y|\bx,1,t)dy \right)$ and $\mu(\bx, 0, t)$$\int_y y p(y|\bx,0,t)dy$. We also replace $p(1-t|\bx,0)$ with $e_{1-t}(\bx, 0)$ simply for clarity in the following steps. Then, $\nabla_{h_{T|\bX, S}}\theta_2$ is equal to

\begin{equation*}
    \begin{split}        
        & \frac{\partial}{\partial\epsilon} \int_{\bx} \frac{\exp\left(\alpha v(\bx) \right)}{\exp\left(\alpha v(\bx) \right) + \exp\left(\alpha \mu(\bx, 0, t)[1+\rho (1 + \epsilon h_{T|\bX, S}(1-t|\bx, 0))e_{1-t}(\bx, 0)] \right)} \\
        & \hspace{1cm} \left. \times v(\bx) p(0|\bx)p(x)dx \right|_{\epsilon=0} \\
        & + \frac{\partial}{\partial\epsilon} \int_{\bx} \frac{\exp\left(\alpha \mu(\bx, 0, t) [1+\rho (1 + \epsilon h_{T|\bX, S}(1-t|\bx, 0)) e_{1-t}(\bx, 0)] \right)}{\exp\left(\alpha v(\bx) \right) + \exp\left(\alpha \mu(\bx, 0, t) [1+\rho (1 + \epsilon h_{T|\bX, S}(1-t|\bx, 0)) e_{1-t}(\bx, 0)] \right)} \\
        & \hspace{1cm} \left. \times \mu(\bx, 0, t) [1+\rho (1 + \epsilon h_{T|\bX, S}(1-t|\bx, 0))e_{1-t}(\bx, 0)] p(0|\bx)p(x)dx \right|_{\epsilon=0}.
    \end{split}
\end{equation*}
This is a rather complex partial derivative. To proceed, we will let 
\begin{equation*}
    w_{\epsilon_{T}}(\bx) = \mu(\bx, 0, t) [1+\rho (1 + \epsilon h_{T|\bX, S}(1-t|\bx, 0))e_{1-t}(\bx, 0)]
\end{equation*}
and rewrite the above as
\begin{equation*}
    \begin{split}
        = &  \int_{\bx} \frac{\partial}{\partial\epsilon} \left. \Big[\frac{v(\bx)\exp\left(\alpha v(\bx) \right)}{\exp\left(\alpha v(\bx) \right) + \exp\left(\alpha w_{\epsilon_{T}}(\bx) \right)} + \frac{w_{\epsilon_{T}}(\bx)\exp\left(\alpha w_{\epsilon_{T}}(\bx) \right)}{\exp\left(\alpha v(\bx) \right) + \exp\left(\alpha w_{\epsilon_{T}}(\bx) \right)} \Big] \right|_{\epsilon=0} p(0|\bx)p(x)dx .     
    \end{split}
\end{equation*}
Note that 
\begin{equation*}
    \frac{\partial}{\partial\epsilon} w_{\epsilon_{T}}(\bx) = h_{T|\bX, S}(1-t|\bx, 0)\rho\mu(\bx, 0, t)e_{1-t}(\bx, 0)
\end{equation*}
and $\left.w_{\epsilon_{T}}(\bx)\right|_{\epsilon=0} = w(\bx)$. From here, we evaluate the partial derivative piece by piece. First the left fraction:
\begin{equation*}
    \begin{split}
        \frac{\partial}{\partial\epsilon} \left. \frac{v(\bx)\exp\left(\alpha v(\bx) \right)}{\exp\left(\alpha v(\bx) \right) + \exp\left(\alpha w_{\epsilon_{T}}(\bx) \right)} \right|_{\epsilon=0} = &
        \left. \frac{-v(\bx)\exp\left(\alpha v(\bx) \right)\exp\left(\alpha w_{\epsilon_{T}}(\bx) \right)\times \alpha \frac{\partial}{\partial\epsilon} w_{\epsilon_{T}}(\bx)}{\Big[\exp\left(\alpha v(\bx) \right) + \exp\left(\alpha w_{\epsilon_{T}}(\bx) \right)\Big]^2} \right|_{\epsilon=0} \\
        = & \frac{-\alpha v(\bx)\frac{\partial}{\partial\epsilon} w_{\epsilon_{T}}(\bx)\exp\left(\alpha v(\bx) \right)\exp\left(\alpha w(\bx) \right)}{\Big[\exp\left(\alpha v(\bx) \right) + \exp\left(\alpha w(\bx) \right)\Big]^2}
    \end{split}
\end{equation*}
Second the right fraction:
\begin{align*}
    & \left. 
    \frac{\partial}{\partial\epsilon} 
    \left( 
        \frac{w_{\epsilon_{T}}(\bx)\exp\left(\alpha w_{\epsilon_{T}}(\bx) \right)}{
            \exp\left(\alpha v(\bx) \right) + \exp\left(\alpha w_{\epsilon_{T}}(\bx) \right)
        }
    \right) 
    \right|_{\epsilon=0} \\
    & = \Bigg\{\Big[\exp\left(\alpha v(\bx) \right) + \exp\left(\alpha w_{\epsilon_{T}}(\bx) \right)\Big] \\
    & \hspace{0.6cm} \times 
        \Big[
            \frac{\partial w_{\epsilon_{T}}(\bx)}{\partial \epsilon} \exp\left(\alpha w_{\epsilon_{T}}(\bx)\right)
            + 
            w_{\epsilon_{T}}(\bx) \exp\left(\alpha w_{\epsilon_{T}}(\bx)\right) \times \alpha \frac{\partial w_{\epsilon_{T}}(\bx)}{\partial \epsilon}
        \Big] \\
    &\hspace{0.6cm} - w_{\epsilon_{T}}(\bx) \exp\left(2\alpha w_{\epsilon_{T}}(\bx)\right) \times \alpha \frac{\partial w_{\epsilon_{T}}(\bx)}{\partial \epsilon}\Bigg\}
    \left[
        \exp\left(\alpha v(\bx)\right) + \exp\left(\alpha w_{\epsilon_{T}}(\bx)\right)
    \right]^{-2} \Bigg|_{\epsilon=0} \\
    &=
    \frac{\frac{\partial w_{\epsilon_{T}}(\bx)}{\partial \epsilon}\times\Big[\exp\left(\alpha v(\bx) \right) \exp\left(\alpha w(\bx) \right) + \alpha w(\bx) \exp\left(\alpha v(\bx) \right) \exp\left(\alpha w(\bx) \right) + \exp\left(2\alpha w(\bx) \right)\Big]}{\left[\exp\left(\alpha v(\bx)\right) + \exp\left(\alpha w(\bx)\right)\right]^{2}}
\end{align*}

Putting these two terms together, we have
\begin{equation*}
    \begin{gathered}
        \frac{\partial w_{\epsilon_{T}}(\bx)}{\partial \epsilon} \times \Bigg[\frac{\exp\left(\alpha v(\bx) \right) \exp\left(\alpha w(\bx) \right) + \exp\left(2\alpha w(\bx) \right)}{\left[\exp\left(\alpha v(\bx)\right) + \exp\left(\alpha w(\bx)\right)\right]^{2}} + \\
        \frac{\alpha w(\bx) \exp\left(\alpha v(\bx) \right) \exp\left(\alpha w(\bx) \right) -\alpha v(\bx) \exp\left(\alpha v(\bx) \right)\exp\left(\alpha w(\bx) \right)}{\left[\exp\left(\alpha v(\bx)\right) + \exp\left(\alpha w(\bx)\right)\right]^{2}}\Bigg]
        \\
        =
        \\
        \frac{\partial w_{\epsilon_{T}}(\bx)}{\partial \epsilon} \times \Bigg[\lambda_2(\bx) + \alpha\lambda_1(\bx)\lambda_2(\bx)\left[w(\bx) - v(\bx) \right] \Bigg]
        \\
        =
        \\
        h_{T|\bX, S}(1-t|\bx, 0)\rho\mu(\bx, 0, t)e_{1-t}(\bx, 0)\Bigg[\lambda_2(\bx) + \alpha\lambda_1(\bx)\lambda_2(\bx)\left[w(\bx) - v(\bx) \right] \Bigg]
    \end{gathered}
\end{equation*}

Now, we can plug this partial derivative evaluated at $\epsilon=0$ back into our integral, to get that 

\begin{equation*}
    \begin{split}
        \nabla_{h_{T|\bX, S}}\theta_2 = & \int_{\bx} h_{T|\bX, S}(1-t|\bx, 0)\rho\mu(\bx, 0, t)e_{1-t}(\bx, 0) \\
        & \hspace{0.6cm} \times \Big[\lambda_2(\bx) + \alpha\lambda_1(\bx)\lambda_2(\bx)\left[w(\bx) - v(\bx) \right] \Big] p(0|\bx)p(x)dx \\
        = & \int_{\bx} p(0|\bx) \rho\mu(\bx, 0, t) \left\{\lambda_2(\bx) + \alpha\lambda_1(\bx)\lambda_2(\bx)\left[w(\bx) - v(\bx) \right] \right\} \\
        & \hspace{0.6cm} \times e_{1-t}(\bx, 0)h_{T|\bX, S}(1-t|\bx, 0) p(x)dx \\
        = & \expe\Big[g_0(\bX)\rho\mu(\bX,0,t)\left\{\lambda_2(\bX) + \alpha\lambda_1(\bX)\lambda_2(\bX)\left[w(\bX) - v(\bX) \right] \right\} \\
        & \hspace{0.6cm} \times e_{1-t}(\bX, 0) h_{T|\bX, S}(1-t|\bX, 0)\Big].
    \end{split}
\end{equation*}

We proceed to show that $\expe[\phi_{\theta_2}h_{T|\bX, S}]$ equals the above value. First, plugging in $h_{T|\bX,S}$,

\begin{equation*}
    \begin{split}
        \expe[\phi_{\theta_2}h_{T|\bX, S}] = & (1 + \gamma)\expe[g_0(\bX)\left\{\lambda_1(\bX) + \alpha\lambda_1(\bX)\lambda_2(\bX)\left[v(\bX) - w(\bX) \right]\right\} \\
        & \hspace{1.5cm} \times \expe[(Y - \mu(\bX,1,t))h_{T|\bX, S}(T | \bX, S) | \bX, S=1, T = t]] \\
        & + \expe[g_0(\bX)\left\{\lambda_1(\bX)v(\bX) + \lambda_2(\bX)w(\bX)\right\}\expe[h_{T|\bX, S}(T | \bX, S) | \bX, S=0]] \\
        & + \expe[g_0(\bX)\left[1 + \rho e_{1-t}(\bX, 0)\right]\left\{\lambda_2(\bX) + \alpha\lambda_1(\bX)\lambda_2(\bX)\left[w(\bX) - v(\bX) \right] \right\} \\
        & \hspace{1cm} \times \expe[(Y - \mu(\bX,0,t))h_{T|\bX, S}(T | \bX, S) | \bX, S=0, T=t]] \\
        & + \expe[g_0(\bX)\rho\mu(\bX,0,t)\left\{\lambda_2(\bX) + \alpha\lambda_1(\bX)\lambda_2(\bX)\left[w(\bX) - v(\bX) \right] \right\} \\
        & \hspace{1cm} \times \expe[\left[\mathbb{I}_{1-t}(T) - e_{1-t}(\bX, 0) \right]h_{T|\bX, S}(T | \bX, S) | \bX, S = 0]].
    \end{split}
\end{equation*}

Then we pull $h_{T|\bX,S}$ out of the inner expectation where possible
\begin{equation*}
    \begin{split}        
        \expe[\phi_{\theta_2}h_{T|\bX, S}] = & (1 + \gamma)\expe[g_0(\bX)\left\{\lambda_1(\bX) + \alpha\lambda_1(\bX)\lambda_2(\bX)\left[v(\bX) - w(\bX) \right]\right\} \\
        & \hspace{1.5cm} \times h_{T|\bX, S}(t|\bX, 1) \expe[(Y - \mu(\bX,1,t)) | \bX, S=1, T = t]] \\
        & + \expe[g_0(\bX)\left\{\lambda_1(\bX)v(\bX) + \lambda_2(\bX)w(\bX)\right\}\expe[h_{T|\bX, S}(T|\bX, S)| \bX, S = 0]] \\
        & + \expe[g_0(\bX)\left[1 + \rho e_{1-t}(\bX, 0)\right]\left\{\lambda_2(\bX) + \alpha\lambda_1(\bX)\lambda_2(\bX)\left[w(\bX) - v(\bX) \right] \right\} \\
        & \hspace{1cm} \times h_{T|\bX, S}(t|\bX, 0)\expe[(Y - \mu(\bX,0,t)) | \bX, S=0, T=t]] \\
        & + \expe\bigg[g_0(\bX)\rho\mu(\bX,0,t)\left\{\lambda_2(\bX) + \alpha\lambda_1(\bX)\lambda_2(\bX)\left[w(\bX) - v(\bX) \right] \right\} \\
        & \hspace{1cm} \times \Big(\expe[\mathbb{I}_{1-t}(T) h_{T|\bX, S}(T|\bX, S)| \bX, S = 0] \\
        & \hspace{1.6cm} - e_{1-t}(\bX, 0)\expe[h_{T|\bX, S}(T|\bX, S)| \bX, S = 0]\Big)\bigg],
    \end{split}
\end{equation*}

and use the mean zero property of the score function to set all terms multiplied by $\expe[h_{T|\bX, S}(T|\bX, S)| \bX, S = 0]$ equal to zero:

\begin{equation*}
    \begin{split}        
        \expe[\phi_{\theta_2}h_{T|\bX, S}] = & (1 + \gamma)\expe[g_0(\bX)\left\{\lambda_1(\bX) + \alpha\lambda_1(\bX)\lambda_2(\bX)\left[v(\bX) - w(\bX) \right]\right\} \\
        & \hspace{1.5cm} \times h_{T|\bX, S}(t|\bX, 1) \expe[(Y - \mu(\bX,1,t)) | \bX, S=1, T = t]] \\
        & + \expe[g_0(\bX)\left[1 + \rho e_{1-t}(\bX, 0)\right]\left\{\lambda_2(\bX) + \alpha\lambda_1(\bX)\lambda_2(\bX)\left[w(\bX) - v(\bX) \right] \right\} \\
        & \hspace{1cm} \times h_{T|\bX, S}(t|\bX, 0)\expe[(Y - \mu(\bX,0,t)) | \bX, S=0, T=t]] \\
        & + \expe[g_0(\bX)\rho\mu(\bX,0,t)\left\{\lambda_2(\bX) + \alpha\lambda_1(\bX)\lambda_2(\bX)\left[w(\bX) - v(\bX) \right] \right\} \\
        & \hspace{1cm} \times \expe[\mathbb{I}_{1-t}(T) h_{T|\bX, S}(T|\bX, S)| \bX, S = 0]].
    \end{split}
\end{equation*}

Then applying the inner expectations and canceling like terms we get

\begin{equation*}
    \begin{split}        
        \expe[\phi_{\theta_2}h_{T|\bX, S}] = & (1 + \gamma)\expe[g_0(\bX)\left\{\lambda_1(\bX) + \alpha\lambda_1(\bX)\lambda_2(\bX)\left[v(\bX) - w(\bX) \right]\right\} \\
        & \hspace{1.5cm} \times h_{T|\bX, S}(t|\bX, 1) (\mu(\bX,1,t) - \mu(\bX,1,t))] \\
        & + \expe[g_0(\bX)\left[1 + \rho e_{1-t}(\bX, 0)\right]\left\{\lambda_2(\bX) + \alpha\lambda_1(\bX)\lambda_2(\bX)\left[w(\bX) - v(\bX) \right] \right\} \\
        & \hspace{1cm} \times h_{T|\bX, S}(t|\bX, 0)(\mu(\bX,0,t) - \mu(\bX,0,t))] \\
        & + \expe[g_0(\bX)\rho\mu(\bX,0,t)\left\{\lambda_2(\bX) + \alpha\lambda_1(\bX)\lambda_2(\bX)\left[w(\bX) - v(\bX) \right] \right\} \\
        & \hspace{1cm} \times \expe[\mathbb{I}_{1-t}(T) h_{T|\bX, S}(T|\bX, S)| \bX, S = 0]] \\
        = & \expe[g_0(\bX)\rho\mu(\bX,0,t)\left\{\lambda_2(\bX) + \alpha\lambda_1(\bX)\lambda_2(\bX)\left[w(\bX) - v(\bX) \right] \right\} \\
        & \hspace{1cm} \times \expe[\mathbb{I}_{1-t}(T) h_{T|\bX, S}(T|\bX, S)| \bX, S = 0]].
    \end{split}
\end{equation*}

Now, notice that

\begin{equation*}
    \begin{split}
        \expe[\mathbb{I}_{1-t}(T) h_{T|\bX, S}(T|\bX, S)| \bX, S = 0] = & \expe\Bigg[\sum_{t'\in\{t,1-t\}} \mathbb{I}_{1-t}(t') h_{T|\bX, S}(t'|\bX, 0) e_{t'}(\bX, 0)\Bigg] \\
        = & \expe\big[e_{1-t}(\bX, 0) h_{T|\bX, S}(1-t|\bX, 0)\big].
    \end{split}
\end{equation*}

Therefore,

\begin{equation*}
    \begin{split}        
        \expe[\phi_{\theta_2}h_{T|\bX, S}] = & \expe\Big[g_0(\bX)\rho\mu(\bX,0,t)\left\{\lambda_2(\bX) + \alpha\lambda_1(\bX)\lambda_2(\bX)\left[w(\bX) - v(\bX) \right] \right\} \\
        & \hspace{0.6cm} \times e_{1-t}(\bX, 0) h_{T|\bX, S}(1-t|\bX, 0)\Big],
    \end{split}
\end{equation*}

and we have shown that $\nabla_{h_{T|\bX, S}}\theta_2 = \expe[\phi_{\theta_2}h_{T|\bX, S}]$

Finally, we show that $\nabla_{h_{Y|\bX, S, T}}\theta_2 = \expe[\phi_{\theta_2}h_{Y|\bX, S, T}]$. As always, we start by simplifying $\nabla_{h_{Y|\bX, S, T}}\theta_2$ by replacing the corresponding terms in the factorized distribution function. This time we set $\peps(y|\bx,1,t) = (1 + \epsilon h_{Y|\bX, S, T}(y|\bx, 1, t))p(y|\bx,1,t)$ and $\peps(y|\bx,0,t) = (1 + \epsilon h_{Y|\bX, S, T}(y|\bx, 0, t))p(y|\bx,0,t)$. We set the other conditional probability density functions to their normal $p$ form. We again have to format $\nabla_{h_{Y|\bX, S, T}}\theta_2$ slightly differently to allow it to stay within the page margins.

\begin{equation*}
    \begin{split}
        \nabla_{h_{T|\bX, S}}\theta_2 = & \frac{\partial}{\partial\epsilon} \int_{\bx} \exp\left(\alpha(1+\gamma)\int_y y (1 + \epsilon h_{Y|\bX, S, T}(y|\bx, 1, t))p(y|\bx,1,t)dy \right) \\
        & \hspace{0.6cm} \times \Bigg\{\exp\left(\alpha\int_y y (1 + \epsilon h_{Y|\bX, S, T}(y|\bx, 0, t))p(y|\bx,0,t)dy[1+\rho p(1-t|\bx,0)] \right)  \\
        & \hspace{1.2cm} + \exp\left(\alpha(1+\gamma)\int_y y (1 + \epsilon h_{Y|\bX, S, T}(y|\bx, 1, t))p(y|\bx,1,t)dy \right) \Bigg\}^{-1} \\
        & \hspace{0.6cm} \left. \times \Bigg((1+\gamma)\int_y y (1 + \epsilon h_{Y|\bX, S, T}(y|\bx, 1, t))p(y|\bx,1,t)dy\Bigg) p(0|\bx) p(x)dx \right|_{\epsilon=0} \\
        & + \frac{\partial}{\partial\epsilon} \int_{\bx} \exp\left(\alpha\int_y y (1 + \epsilon h_{Y|\bX, S, T}(y|\bx, 0, t))p(y|\bx,0,t)dy[1+\rho p(1-t|\bx,0)] \right) \\
        & \hspace{0.6cm} \times \Bigg\{\exp\left(\alpha\int_y y (1 + \epsilon h_{Y|\bX, S, T}(y|\bx, 0, t))p(y|\bx,0,t)dy[1+\rho p(1-t|\bx,0)] \right)  \\
        & \hspace{1.2cm} + \exp\left(\alpha(1+\gamma)\int_y y (1 + \epsilon h_{Y|\bX, S, T}(y|\bx, 1, t))p(y|\bx,1,t)dy \right) \Bigg\}^{-1} \\
        & \hspace{0.6cm} \times \Bigg(\int_y y (1 + \epsilon h_{Y|\bX, S, T}(y|\bx, 0, t))p(y|\bx,0,t)dy[1+\rho p(1-t|\bx,0)]\Bigg) \\
        & \hspace{0.6cm} \times p(0|\bx)p(x)dx \Bigg|_{\epsilon=0}.
    \end{split}
\end{equation*}

We define the following terms to substitute in above:
\begin{equation*}
    \begin{split}
        v_{\epsilon_{Y}}(\bx) & = (1+\gamma)\int_y y (1 + \epsilon h_{Y|\bX, S, T}(y|\bx, 1, t))p(y|\bx,1,t)dy \\
        &= v(\bx) + \epsilon(1+\gamma)\int_y y h_{Y|\bX, S, T}(y|\bx, 1, t)p(y|\bx,1,t)dy, \textrm{ and } \\
        w_{\epsilon_{Y}}(\bx) &= \int_y y (1 + \epsilon h_{Y|\bX, S, T}(y|\bx, 0, t))p(y|\bx,0,t)dy[1+\rho p(1-t|\bx,0)] \\
        &= w(x) + \epsilon[1+\rho e_{1-t}(\bx, 0)]\int_y y h_{Y|\bX, S, T}(y|\bx, 0, t)p(y|\bx,0,t)dy.
    \end{split}
\end{equation*}
Note that
\begin{equation*}
    \begin{split}
        \frac{\partial v_{\epsilon_{Y}}(\bx)}{\partial\epsilon} & = (1+\gamma)\int_y y h_{Y|\bX, S, T}(y|\bx, 1, t)p(y|\bx,1,t)dy, \\
        \frac{\partial w_{\epsilon_{Y}}(\bx)}{\partial\epsilon} &= [1+\rho e_{1-t}(\bx, 0)]\int_y y h_{Y|\bX, S, T}(y|\bx, 0, t)p(y|\bx,0,t)dy,
    \end{split}
\end{equation*}

and $\left. v_{\epsilon_{Y}}(\bx)\right|_{\epsilon=0} = v(\bx)$, $\left. w_{\epsilon_{Y}}(\bx)\right|_{\epsilon=0} = w(\bx)$. Then,

\begin{equation*}
    \begin{split}
        \nabla_{h_{Y|\bX, S, T}}\theta_2 = &  \int_{\bx} \frac{\partial}{\partial\epsilon} \left.\Bigg[\frac{v_{\epsilon_{Y}}(\bx)\exp\left(\alpha v_{\epsilon_{Y}}(\bx) \right)  +  w_{\epsilon_{Y}}(\bx)\exp\left(\alpha w_{\epsilon_{Y}}(\bx) \right)}{\exp\left(\alpha v_{\epsilon_{Y}}(\bx) \right) + \exp\left(\alpha w_{\epsilon_{Y}}(\bx) \right)}\Bigg]\right|_{\epsilon=0} p(0|\bx)p(x)dx .
    \end{split}
\end{equation*}

Like above, we evaluate the partial derivative piece by piece for clarity. First, the left term of the fraction:
\begin{align*}
        & \frac{\partial}{\partial\epsilon} \left.\frac{v_{\epsilon_{Y}}(\bx)\exp\left(\alpha v_{\epsilon_{Y}}(\bx) \right)}{\exp\left(\alpha v_{\epsilon_{Y}}(\bx) \right) + \exp\left(\alpha w_{\epsilon_{Y}}(\bx) \right)}\right|_{\epsilon=0} \\
        & = \Bigg\{ \left[\exp\left(\alpha v_{\epsilon_{Y}}(\bx) \right) + \exp\left(\alpha w_{\epsilon_{Y}}(\bx) \right) \right] \\
        & \hspace{0.6cm} \times \left[\frac{\partial v_{\epsilon_{Y}}(\bx)}{\partial\epsilon}\exp\left(\alpha v_{\epsilon_{Y}}(\bx) \right) + v_{\epsilon_{Y}}(\bx)\exp\left(\alpha v_{\epsilon_{Y}}(\bx) \right)\times \alpha \frac{\partial v_{\epsilon_{Y}}(\bx)}{\partial\epsilon} \right] \\
        & \hspace{0.6cm} - v_{\epsilon_{Y}}(\bx)\exp\left(\alpha v_{\epsilon_{Y}}(\bx) \right)\left[\exp\left(\alpha v_{\epsilon_{Y}}(\bx) \right)\times \alpha \frac{\partial v_{\epsilon_{Y}}(\bx)}{\partial\epsilon} + \exp\left(\alpha w_{\epsilon_{Y}}(\bx) \right)\times \alpha \frac{\partial w_{\epsilon_{Y}}(\bx)}{\partial\epsilon} \right]\Bigg\} \\
        &\quad \left. \times \left[\exp\left(\alpha v_{\epsilon_{Y}}(\bx) \right) + \exp\left(\alpha w_{\epsilon_{Y}}(\bx) \right) \right]^{-2}\right|_{\epsilon=0} \\
        & = \frac{\frac{\partial v_{\epsilon_{Y}}(\bx)}{\partial \epsilon}\times\Big[\exp\left(\alpha v(\bx) \right) \exp\left(\alpha w(\bx) \right) + \alpha v(\bx) \exp\left(\alpha v(\bx) \right) \exp\left(\alpha w(\bx) \right) + \exp\left(2\alpha v(\bx) \right)\Big]}{\left[\exp\left(\alpha v(\bx)\right) + \exp\left(\alpha w(\bx)\right)\right]^{2}} \\
        &\quad - \frac{\frac{\partial w_{\epsilon_{Y}}(\bx)}{\partial \epsilon}\times \alpha v(\bx) \exp\left(\alpha v(\bx) \right) \exp\left(\alpha w(\bx) \right)}{\left[\exp\left(\alpha v(\bx)\right) + \exp\left(\alpha w(\bx)\right)\right]^{2}}
\end{align*}

The right fraction partial derivative looks similar:
\begin{align*}
        & \frac{\partial}{\partial\epsilon} \left.\frac{w_{\epsilon_{Y}}(\bx)\exp\left(\alpha w_{\epsilon_{Y}}(\bx) \right)}{\exp\left(\alpha v_{\epsilon_{Y}}(\bx) \right) + \exp\left(\alpha w_{\epsilon_{Y}}(\bx) \right)}\right|_{\epsilon=0} \\
        & = \Bigg\{ \left[\exp\left(\alpha v_{\epsilon_{Y}}(\bx) \right) + \exp\left(\alpha w_{\epsilon_{Y}}(\bx) \right) \right] \\
        & \hspace{0.6cm} \times \left[\frac{\partial w_{\epsilon_{Y}}(\bx)}{\partial\epsilon}\exp\left(\alpha w_{\epsilon_{Y}}(\bx) \right) + w_{\epsilon_{Y}}(\bx)\exp\left(\alpha w_{\epsilon_{Y}}(\bx) \right)\times \alpha \frac{\partial w_{\epsilon_{Y}}(\bx)}{\partial\epsilon} \right] \\
        & \hspace{0.6cm} - w_{\epsilon_{Y}}(\bx)\exp\left(\alpha w_{\epsilon_{Y}}(\bx) \right)\left[\exp\left(\alpha v_{\epsilon_{Y}}(\bx) \right)\times \alpha \frac{\partial v_{\epsilon_{Y}}(\bx)}{\partial\epsilon} + \exp\left(\alpha w_{\epsilon_{Y}}(\bx) \right)\times \alpha \frac{\partial w_{\epsilon_{Y}}(\bx)}{\partial\epsilon} \right]\Bigg\} \\
        &\quad \left. \times \left[\exp\left(\alpha v_{\epsilon_{Y}}(\bx) \right) + \exp\left(\alpha w_{\epsilon_{Y}}(\bx) \right) \right]^{-2}\right|_{\epsilon=0} \\
        & = \frac{\frac{\partial w_{\epsilon_{Y}}(\bx)}{\partial \epsilon}\times\Big[\exp\left(\alpha v(\bx) \right) \exp\left(\alpha w(\bx) \right) + \alpha w(\bx) \exp\left(\alpha v(\bx) \right) \exp\left(\alpha w(\bx) \right) + \exp\left(2\alpha w(\bx) \right)\Big]}{\left[\exp\left(\alpha v(\bx)\right) + \exp\left(\alpha w(\bx)\right)\right]^{2}} \\
        &\quad - \frac{\frac{\partial v_{\epsilon_{Y}}(\bx)}{\partial \epsilon}\times \alpha w(\bx) \exp\left(\alpha v(\bx) \right) \exp\left(\alpha w(\bx) \right)}{\left[\exp\left(\alpha v(\bx)\right) + \exp\left(\alpha w(\bx)\right)\right]^{2}}
\end{align*}

Now, we combine terms and group by those multiplied to $\frac{\partial v_{\epsilon_{Y}}(\bx)}{\partial\epsilon}$ and $\frac{\partial w_{\epsilon_{Y}}(\bx)}{\partial\epsilon}$. First, $\frac{\partial v_{\epsilon_{Y}}(\bx)}{\partial\epsilon}$:
\begin{equation*}
    \begin{gathered}
        \frac{\partial v_{\epsilon_{Y}}(\bx)}{\partial \epsilon} \times \Bigg[\frac{\exp\left(\alpha v(\bx) \right) \exp\left(\alpha w(\bx) \right) + \exp\left(2\alpha v(\bx) \right)}{\left[\exp\left(\alpha v(\bx)\right) + \exp\left(\alpha w(\bx)\right)\right]^{2}} + \\
        \frac{\alpha v(\bx) \exp\left(\alpha v(\bx) \right) \exp\left(\alpha w(\bx) \right) -\alpha w(\bx) \exp\left(\alpha v(\bx) \right)\exp\left(\alpha w(\bx) \right)}{\left[\exp\left(\alpha v(\bx)\right) + \exp\left(\alpha w(\bx)\right)\right]^{2}}\Bigg]
        \\
        =
        \\
        \frac{\partial v_{\epsilon_{Y}}(\bx)}{\partial \epsilon} \times \Bigg[\lambda_1(\bx) + \alpha\lambda_1(\bx)\lambda_2(\bx)\left[v(\bx) - w(\bx) \right] \Bigg]
        \\
        =
        \\
        (1+\gamma)\int_y y h_{Y|\bX, S, T}(y|\bx,1,t)p(y|\bx,1,t)dy \times \Bigg[\lambda_1(\bx) + \alpha\lambda_1(\bx)\lambda_2(\bx)\left[v(\bx) - w(\bx) \right] \Bigg].
    \end{gathered}
\end{equation*}
Next, $\frac{\partial w_{\epsilon_{Y}}(\bx)}{\partial\epsilon}$:
\begin{equation*}
    \begin{gathered}
        \frac{\partial w_{\epsilon_{Y}}(\bx)}{\partial \epsilon} \times \Bigg[\frac{\exp\left(\alpha v(\bx) \right) \exp\left(\alpha w(\bx) \right) + \exp\left(2\alpha w(\bx) \right)}{\left[\exp\left(\alpha v(\bx)\right) + \exp\left(\alpha w(\bx)\right)\right]^{2}} + \\
        \frac{\alpha w(\bx) \exp\left(\alpha v(\bx) \right) \exp\left(\alpha w(\bx) \right) -\alpha v(\bx) \exp\left(\alpha v(\bx) \right)\exp\left(\alpha w(\bx) \right)}{\left[\exp\left(\alpha v(\bx)\right) + \exp\left(\alpha w(\bx)\right)\right]^{2}}\Bigg]
        \\
        =
        \\
        \frac{\partial w_{\epsilon_{Y}}(\bx)}{\partial \epsilon} \times \Bigg[\lambda_2(\bx) + \alpha\lambda_1(\bx)\lambda_2(\bx)\left[w(\bx) - v(\bx) \right] \Bigg]
        \\
        =
        \\
        [1+\rho e_{1-t}(\bx, 0)]\int_y y h_{Y|\bX, S, T}(y|\bx,0,t)p(y|\bx,0,t)dy \Bigg[\lambda_2(\bx) + \alpha\lambda_1(\bx)\lambda_2(\bx)\left[w(\bx) - v(\bx) \right] \Bigg].
    \end{gathered}
\end{equation*}

We plug both of these components that make up the partial derivative evaluated at $\epsilon=0$ back into our integral:

\begin{equation*}
    \begin{split}
        \nabla_{h_{Y|\bX, S, T}}\theta_2 = & \int_{\bx} (1+\gamma)\int_y y h_{Y|\bX, S, T}(y|\bx,1,t)p(y|\bx,1,t)dy \\
        & \quad \times \Bigg[\lambda_1(\bx) + \alpha\lambda_1(\bx)\lambda_2(\bx)\left[v(\bx) - w(\bx) \right] \Bigg] p(0|\bx)p(x)dx \\
        & + \int_{\bx} [1+\rho e_{1-t}(\bx, 0)]\int_y y h_{Y|\bX, S, T}(y|\bx,0,t)p(y|\bx,0,t)dy \\
        & \quad \times \Bigg[\lambda_2(\bx) + \alpha\lambda_1(\bx)\lambda_2(\bx)\left[w(\bx) - v(\bx) \right] \Bigg] p(0|\bx)p(x)dx.
    \end{split}
\end{equation*}

Reorganizing terms and writing in expectation form, we get

\begin{equation*}
    \begin{split}
        \nabla_{h_{Y|\bX, S, T}}\theta_2 = & (1 + \gamma)\expe\bigg[g_0(\bX)\left\{\lambda_1(\bX) + \alpha\lambda_1(\bX)\lambda_2(\bX)\left[v(\bX) - w(\bX) \right]\right\} \\
        & \hspace{1.5cm} \times \expe[Y h_{Y|\bX, S, T}(Y|\bX, S, T) | \bX, S=1, T = t]\bigg] \\
        & + \expe\bigg[g_0(\bX)\left[1 + \rho e_{1-t}(\bX, 0)\right]\left\{\lambda_2(\bX) + \alpha\lambda_1(\bX)\lambda_2(\bX)\left[w(\bX) - v(\bX) \right] \right\} \\
        & \hspace{1cm} \times \expe[Y h_{Y|\bX, S, T}(Y|\bX, S, T) | \bX, S=0, T=t]\bigg].
    \end{split}
\end{equation*}

We finish by showing that this equals $\expe[\phi_{\theta_2}h_{Y|\bX, S, T}]$. First, setting it up,

\begin{equation*}
    \begin{split}
        \expe[\phi_{\theta_2}h_{Y|\bX, S, T}] = & (1 + \gamma)\expe[g_0(\bX)\left\{\lambda_1(\bX) + \alpha\lambda_1(\bX)\lambda_2(\bX)\left[v(\bX) - w(\bX) \right]\right\} \\
        & \hspace{1.5cm} \times \expe[(Y - \mu(\bX,1,t))h_{Y|\bX, S, T}(Y|\bX, S, T) | \bX, S=1, T = t]] \\
        & + \expe[g_0(\bX)\left\{\lambda_1(\bX)v(\bX) + \lambda_2(\bX)w(\bX)\right\}\expe[h_{Y|\bX, S, T}(Y|\bX, S, T) | \bX, S=0]] \\
        & + \expe[g_0(\bX)\left[1 + \rho e_{1-t}(\bX, 0)\right]\left\{\lambda_2(\bX) + \alpha\lambda_1(\bX)\lambda_2(\bX)\left[w(\bX) - v(\bX) \right] \right\} \\
        & \hspace{1cm} \times \expe[(Y - \mu(\bX,0,t))h_{Y|\bX, S, T}(Y|\bX, S, T) | \bX, S=0, T=t]] \\
        & + \expe[g_0(\bX)\rho\mu(\bX,0,t)\left\{\lambda_2(\bX) + \alpha\lambda_1(\bX)\lambda_2(\bX)\left[w(\bX) - v(\bX) \right] \right\} \\
        & \hspace{1cm} \times \expe[\left[1 - T - e_{1-t}(\bX, 0) \right]h_{Y|\bX, S, T}(Y|\bX, S, T) | \bX, S = 0]].
    \end{split}
\end{equation*}

Then, we use linearity of expectation to separate terms,

\begin{equation*}
    \begin{split}
        \expe[\phi_{\theta_2}h_{Y|\bX, S, T}] = & (1 + \gamma)\expe\Big[g_0(\bX)\left\{\lambda_1(\bX) + \alpha\lambda_1(\bX)\lambda_2(\bX)\left[v(\bX) - w(\bX) \right]\right\} \\
        & \hspace{1.5cm} \times \expe[Y h_{Y|\bX, S, T}(Y|\bX, S, T) | \bX, S=1, T = t] \\
        & \hspace{1.5cm} - \mu(\bX,1,t)\expe[h_{Y|\bX, S, T}(Y|\bX, S, T) | \bX, S=1, T = t]\Big]
        ] \\
        & + \expe[g_0(\bX)\left\{\lambda_1(\bX)v(\bX) + \lambda_2(\bX)w(\bX)\right\}\expe[h_{Y|\bX, S, T}(Y|\bX, S, T) | \bX, S=0]] \\
        & + \expe\Big[g_0(\bX)\left[1 + \rho e_{1-t}(\bX, 0)\right]\left\{\lambda_2(\bX) + \alpha\lambda_1(\bX)\lambda_2(\bX)\left[w(\bX) - v(\bX) \right] \right\} \\
        & \hspace{1cm} \times \expe[Y h_{Y|\bX, S, T}(Y|\bX, S, T) | \bX, S=0, T=t] \\
        & \hspace{1cm} - \mu(\bX,0,t)\expe[h_{Y|\bX, S, T}(Y|\bX, S, T) | \bX, S=0, T = t]\Big] \\
        & + \expe[g_0(\bX)\rho\mu(\bX,0,t)\left\{\lambda_2(\bX) + \alpha\lambda_1(\bX)\lambda_2(\bX)\left[w(\bX) - v(\bX) \right] \right\} \\
        & \hspace{1cm} \times \expe[\left[1 - T - e_{1-t}(\bX, 0) \right]h_{Y|\bX, S, T}(Y|\bX, S, T) | \bX, S = 0]].
    \end{split}
\end{equation*}

Now, we use the mean zero property of the score function to remove several of the terms,

\begin{equation*}
    \begin{split}
        \expe[\phi_{\theta_2}h_{Y|\bX, S, T}] = & (1 + \gamma)\expe\Big[g_0(\bX)\left\{\lambda_1(\bX) + \alpha\lambda_1(\bX)\lambda_2(\bX)\left[v(\bX) - w(\bX) \right]\right\} \\
        & \hspace{1.5cm} \times \expe[Y h_{Y|\bX, S, T}(Y|\bX, S, T) | \bX, S=1, T = t]\Big] \\
        & + \expe\Big[g_0(\bX)\left[1 + \rho e_{1-t}(\bX, 0)\right]\left\{\lambda_2(\bX) + \alpha\lambda_1(\bX)\lambda_2(\bX)\left[w(\bX) - v(\bX) \right] \right\} \\
        & \hspace{1cm} \times \expe[Y h_{Y|\bX, S, T}(Y|\bX, S, T) | \bX, S=0, T=t] \Big].
    \end{split}
\end{equation*}

This confirms that $\nabla_{h_{Y|\bX, S, T}}\theta_2 = \expe[\phi_{\theta_2}h_{Y|\bX, S, T}]$. The verification of the candidate EIF for $\theta_2$ is complete.

\subsection{EIF for $\theta(t,\rho,\gamma,\alpha)$}
Having derived and validated the candidate EIFs for $\theta_1$ and $\theta_2$, we simply use the linearity property of EIFs to write the full form the EIF for $\theta(t,\rho,\gamma,\alpha)$ as:

\begin{align*}
    \phi(Z; t,\rho,\gamma,\alpha) = & \frac{S\mathbb{I}(T=t)}{e_t(\bX, 1)} \Big[Y - \mu(\bX, 1, t)\Big]
    + S\mu(\bX, 1, t) + \\
    & + S(1 + \gamma)\left\{\lambda_1(\bX) + \alpha\lambda_1(\bX)\lambda_2(\bX)\left[v(\bX) - w(\bX) \right] \right\} \\
    & \hspace{0.5cm} \times \left\{\frac{\mathbb{I}_t(T)}{e_t(\bX, 1)g_1(\bX)}\left[Y - \mu(\bX,1,t) \right] g_0(\bX)\right\} \\        
    & + \left[1 - S\right] \left\{\lambda_1(\bX)v(\bX) + \lambda_2(\bX)w(\bX) \right\} \\
    & + [1-S]\left\{\lambda_2(\bX) + \alpha\lambda_1(\bX)\lambda_2(\bX)\left[w(\bX) - v(\bX) \right] \right\} \\
    & \hspace{0.5cm} \times \Bigg\{\frac{\mathbb{I}_t(T)}{e_t(\bX, 0)}\left[ Y - \mu(\bX, 0, t)\right]\left[1 + \rho e_{1-t}(\bX, 0)\right] \\
    & \hspace{1.5cm} + \rho\mu(\bX,0,t)\left[\mathbb{I}_{1-t}(T) - e_{1-t}(\bX, 0) \right] \Bigg\} \\        
    & - \theta(t,\rho,\gamma,\alpha).
\end{align*}
\section{(In)compatible $\rho$ and $\gamma$}\label{appdx:incompatible}

In Remark~\ref*{rem:incompatible}, we briefly discussed the reasoning behind the Lemma~\ref*{lemma:pot-out-bounds} and Theorem~\ref*{thm:cate_bound} conditions that $v(\bx,t,-\gamma) \leq w(\bx,t,\rho)$ and $w(\bx,t,-\rho) \leq v(\bx,t,\gamma)$. We referred to $(\rho, \gamma)$ pairs that lead to violations in these conditions as incompatible, which we visually represented as one of the four regions in our breakdown frontier plots. In this Appendix section, we aim to provide more context on this concept and discuss how we estimate the (in)compatibility of a given $(\rho, \gamma)$ in practice.

\subsection{Further Discussion of Incompatibility}
We define a pair $(\rho, \gamma)$ as incompatible if they do not sufficiently relax the assumptions to allow overlap between the bounds produced by $\gamma$ and $\rho$. We elaborate on this idea in this subsection.

Incompatibility stems from the fact that there must be some source of unmeasured confounding affecting either study selection or treatment assignment if the observed potential outcomes in the experimental and observational studies are different. In other words, as discussed in Remark~\ref*{rem:incompatible}, if $\exists (\bx, t)$ such that $\left| \expP[Y \mid \bX = \bx, S=1, T=t] - \expP[Y \mid \bX = \bx, S=0, T=t] \right| = \Delta(t) > 0$, then Assumption~\ref*{assum:exchange} and/or Assumption~\ref*{assum:obs-ignorability} must be violated. 

Recall that in our partial identification framework
\[
v(\bx,t,\gamma) := (1+\gamma)\mu(\bx,1,t), \quad
w(\bx,t,\rho) := e_t(\bx,0)\mu(\bx,0,t) + e_{1-t}(\bx,0)(1+\rho)\mu(\bx,0,t),
\]
serve as two upper bounds on $\expP[Y(t) \mid \bX=\bx, S=0]$. Analogously, $v(\bx,t,-\gamma)$ and $w(\bx,t,-\rho)$ serve as two lower bounds. In particular, we showed in the proof of Lemma~\ref*{lemma:pot-out-bounds} in Appendix~\ref{appdx:pid-proofs} that 
\begin{gather*}
    \expP[Y(t) \mid \bX = \bx, S = 0]\in[v(\bx, t, -\gamma), v(\bx, t, \gamma)], \textrm{ and } \\
    \expP[Y(t) \mid \bX = \bx, S = 0]\in[w(\bx, t, -\rho), w(\bx, t, \rho)].
\end{gather*}
We then took the $\max$ over the two lower bounds and the $\min$ over the two upper bounds to get the tightest possible bounds on $\expP[Y(t) \mid \bX = \bx, S = 0]$ that we subsequently use to upper bound $\expP[Y(t) \mid \bX=\bx]$. 

Consider the case that $\rho$ and $\gamma$ are both set to zero and $\Delta(t) > 0$. In this case, we have that the two intervals bounding $\expP[Y(t) \mid \bX = \bx, S = 0]$ are
\begin{gather*}
    \expP[Y(t) \mid \bX = \bx, S = 0]\in[v(\bx, t, -0), v(\bx, t, 0)] = \mu(\bx,1,t), \textrm{ and } \\
    \expP[Y(t) \mid \bX = \bx, S = 0]\in[w(\bx, t, -0), w(\bx, t, 0)] = \mu(\bx,0,t).
\end{gather*}
But, as we established earlier, if $\Delta(t) > 0$, then $\mu(\bx,1,t) \neq \mu(\bx,0,t)$. This leads to a contradiction as the bounds based on $\rho = 0$ and $\gamma = 0$ assume no differences in potential outcomes across studies or treatment groups. But if $\mu(\bx,1,t) \neq \mu(\bx,0,t)$ is observed, then some violation of assumptions must be present. Hence, $(\rho, \gamma) = (0, 0)$ is incompatible with the data.

\begin{figure}[ht]
    \centering
    \includegraphics[width=0.6\textwidth]{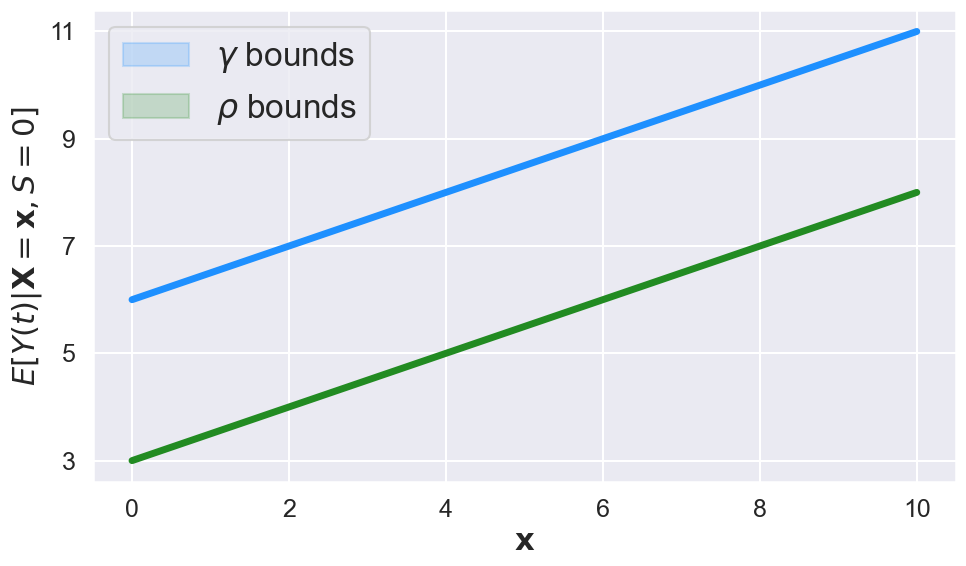}
    \caption{Bounds on $\expP[Y(t) \mid \bX = \bx, S = 0]$ from $[v(\bx, t, -0), v(\bx, t, 0)]$ and $[w(\bx, t, -0), w(\bx, t, 0)]$ when $\mu(\bx,1,t) \neq \mu(\bx,0,t)$.}
    \label{fig:bounds-toy-example-0-0}
\end{figure}

\begin{figure}[ht]
    \centering
    \includegraphics[width=0.6\textwidth]{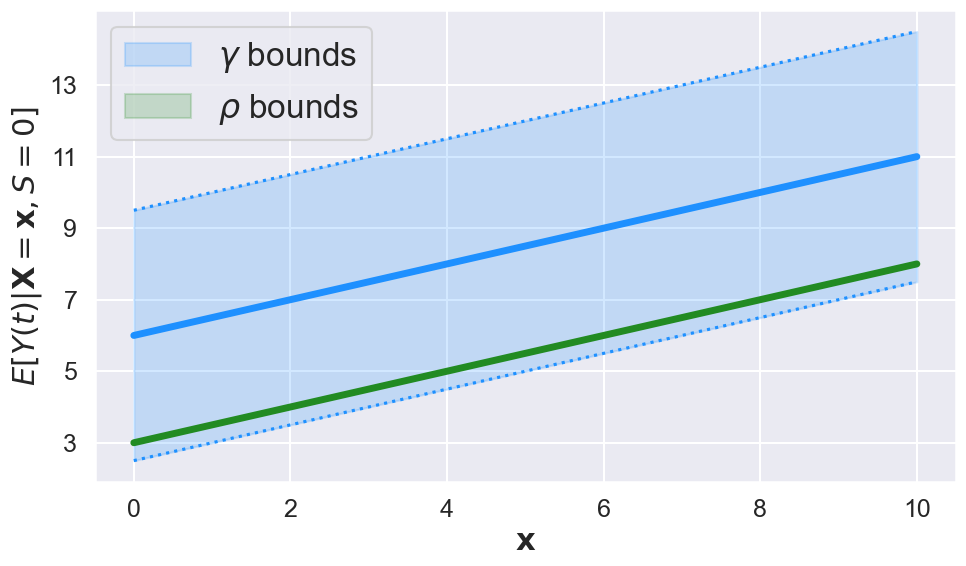}
    \caption{Bounds on $\expP[Y(t) \mid \bX = \bx, S = 0]$ from $[v(\bx, t, -\gamma), v(\bx, t, \gamma)]$ and $[w(\bx, t, -0), w(\bx, t, 0)]$ when $\mu(\bx,1,t) \neq \mu(\bx,0,t)$ and $\gamma$ is large enough that $v(\bx, t, -\gamma) < w(\bx, t, 0)$.}
    \label{fig:bounds-toy-example-big-enough-gamma}
\end{figure}

Towards building up an understanding through a series of visuals, we depict this scenario in a toy example using Figure~\ref{fig:bounds-toy-example-0-0}, where $\mu(\bx,1,t) \neq \mu(\bx,0,t)$. Here, it is clear that the bounds on $\expP[Y(t) \mid \bX = \bx, S = 0]$ do not intersect---leading us to call this an incompatible choice for $\rho$ and $\gamma$. Conversely, consider the scenario where we keep $\rho=0$ but we increase $\gamma$ sufficiently so that $v(\bx, t, -\gamma) = (1-\gamma)\mu(\bx,1,t) < w(\bx, t, 0) = \mu(\bx,0,t)$. This scenario is depicted in Figure~\ref{fig:bounds-toy-example-big-enough-gamma}. Here, we see that the bounds on $\expP[Y(t) \mid \bX = \bx, S = 0]$ do intersect---leading us to call this choice of $\rho$ and $\gamma$ a compatible pair of parameters.

Connecting this back to the conditions in Lemma~\ref*{lemma:pot-out-bounds} and Theorem~\ref*{thm:cate_bound}, in Figure~\ref{fig:bounds-toy-example-0-0} we violated the condition that $v(\bx,t,-\gamma) \leq w(\bx,t,\rho)$ whereas in Figure~\ref{fig:bounds-toy-example-big-enough-gamma} both $v(\bx,t,-\gamma) \leq w(\bx,t,\rho)$ and $w(\bx,t,-\rho) \leq v(\bx,t,\gamma)$.

\begin{figure}[ht]
    \centering
    \includegraphics[width=0.9\textwidth]{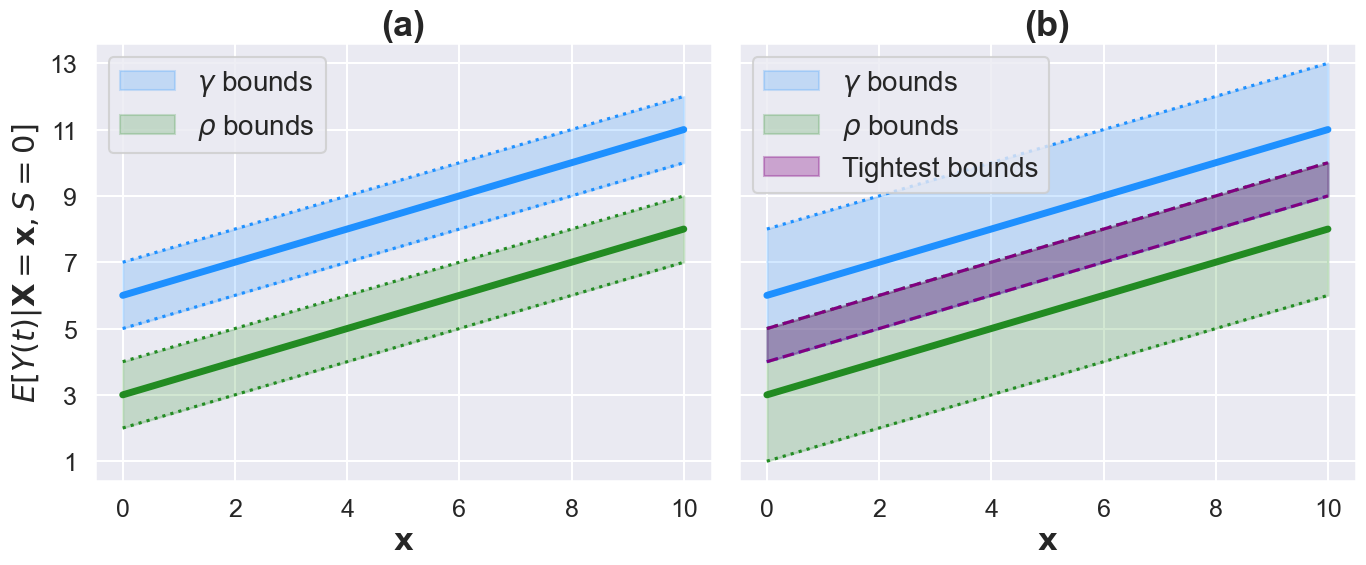}
    \caption{Bounds on $\expP[Y(t) \mid \bX = \bx, S = 0]$ from $[v(\bx, t, -\gamma), v(\bx, t, \gamma)]$ and $[w(\bx, t, -\rho), w(\bx, t, \rho)]$ when $\mu(\bx,1,t) \neq \mu(\bx,0,t)$. In \textbf{(a)}, $\rho$ and $\gamma$ are not large enough for the bounds to intersect. Whereas, in \textbf{(b)}, $\rho$ and $\gamma$ are made large enough for the bounds to intersect.}
    \label{fig:bounds-toy-example-both}
\end{figure}

When $\gamma$ and $\rho$ are both greater than zero, they can still be incompatible if either $v(\bx,t,-\gamma) > w(\bx,t,\rho)$ or $w(\bx,t,-\rho) > v(\bx,t,\gamma)$. Continuing with our example, consider the case depicted in Figure~\ref{fig:bounds-toy-example-both}(a) where the lower bound on the $\gamma$ bound is larger than the upper bound on the $\rho$ bound, i.e. $v(\bx,t,-\gamma) > w(\bx,t,\rho)$. This is an example where $\Delta(t) > 0$ and the values of $\rho$ and $\gamma$ are not sufficiently large to explain this discrepency. Conversely, in Figure~\ref{fig:bounds-toy-example-both}(b), the values of $\rho$ and $\gamma$ are increased, making $v(\bx,t,-\gamma) \leq w(\bx,t,\rho)$, and leading to compatible parameter values. In this plot, we also show how we take the $\max$ of the two lower bounds and the $\min$ of the two upper bounds to get the tightest bounds---corresponding to how we constructed the tightest bounds in Lemma~\ref*{lemma:pot-out-bounds} and Theorem~\ref*{thm:cate_bound}.

\subsection{Estimating Incompatibility}
In practice, we do not observe the true values of $v(\bx,t,\gamma)w(\bx,t,\rho)$ and $w(\bx,t,\rho)$ and must estimate them from data. As a result, determining whether a given $(\rho, \gamma)$ pair is incompatible must account for estimation uncertainty. 

Ideally, for any given $(\rho, \gamma)$, we would test at each $(\bx, t)$ whether the estimated bounds based on $\rho$ and $\gamma$ intersect, indicating compatibility. However, this is intractable with finite samples in high-dimensional covariate spaces. To address this challenge, we assess compatibility at the level of the target estimand, rather than at individual covariate profiles. Specifically, we evaluate whether the bounds intersect on average over $\bX$, separately for each treatment arm $t\in\{0, 1\}$. When focusing on the CATE, this expectation is taken over the subpopulation of interest, whereas for the ATE, this expectation is taken over the full population.

As shown in the proof of Lemma~\ref*{lemma:pot-out-bounds} in Appendix~\ref{appdx:pid-proofs}, the condition $\max\{v(\bx, t, -\gamma), w(\bx, t, -\rho)\} \leq \min\{v(\bx, t, \gamma), w(\bx, t, \rho)\}$ is equivalent to checking that both $v(\bx,t,-\gamma) \leq w(\bx,t,\rho)$ and $w(\bx,t,-\rho) \leq v(\bx,t,\gamma)$. Both conditions ensure that the $\rho$ and $\gamma$ intervals overlap. To make this condition testable, we assess whether the average size of the overlap region is nonnegative. Specifically, we check whether the expected difference of $\min\{v(\bx, t, \gamma), w(\bx, t, \rho)\} - \max\{v(\bx, t, -\gamma), w(\bx, t, -\rho)\} \geq 0$. We define $G(t) = \mathbbm{E}[\min\{v(\bx, t, \gamma), w(\bx, t, \rho)\} - \max\{v(\bx, t, -\gamma), w(\bx, t, -\rho)\}]$. This expectation is taken over the relevant covariate distribution and is evaluated separately for each treatment arm $t\in\{0,1\}$. 

Because generating resamples that both satisfy the null and re-estimate the nuisance functions is non-trivial, we adopt a simplified resampling approach that treats these functions as fixed. We describe the full procedure below and return to its limitations at the end of this section.

For each treatment arm $t$, we conduct a one-sided hypothesis test where the null hypothesis is that the average size of the overlap region is nonnegative, $H_0: G(t) \geq 0$, and the alternative is that the average size is negative, $H_A: G(t) < 0$. Rejecting the null corresponds to concluding that the specified parameter pair is incompatible. Given a dataset $\mathcal{D}_n$, parameter values ($\rho'$,$\gamma'$) and treatment arm $t'$, we construct a resampling-based test as follows. We first compute the observed test statistic as the sample mean of overlap sizes: 
\[T_{obs} = \frac{1}{n} \sum_{i=1}^n \min\{v(\bx_i, t', \gamma'), w(\bx_i, t', \rho')\} - \max\{v(\bx_i, t', -\gamma'), w(\bx_i, t', -\rho')\}.\]
To simulate the null distribution where $G(t')=0$, we subtract $T_{obs}$ from each individual overlap value, producing a centered dataset that has mean zero. We denote these centered values as $o_i = \min\{v(\bx_i, t', \gamma'), w(\bx_i, t', \rho')\} - \max\{v(\bx_i, t', -\gamma'), w(\bx_i, t', -\rho')\} - T_{obs}$. We then perform $R$ resampling iterations. In each iteration, $r$, we randomly draw $n$ values with replacement from the centered overlap size set $\{o_i,...,o_n\}$, and compute the mean of the resampled values $T_r = \frac{1}{n} \sum_{j\in J_r} o_j$, where $J_r$ is the set of indices sampled in iteration $r$. Finally, we compute the one-sided p-value as the proportion of $T_r$ values that are greater than or equal to $T_{obs}$: $p = \frac{1}{R}\sum_{r=1}^R \mathbbm{I}[T_r \geq T_{obs}]$.

We write this testing procedure as an algorithm in Appendix~\ref{appdx:compatible-test-algo}. We also note two primary limitations of this approach. 

First, as stated above, we do not test this condition at each $(\bx, t)$ pair. This is primarily a practical limitation, as checking compatibility pointwise across the full covariate-treatment domain is infeasible in high dimensions or with continuous covariates. However, in settings with only discrete covariates and sufficiently large datasets, such a pointwise test could be applied to each covariate profile.

Second, as briefly mentioned above, our resampling-based procedure operates on estimated overlap values rather than resampling at the dataset level. This differs from a formal bootstrap test, which would involve resampling full observations and re-estimating the functions $e_t(\bx,0)$ and $\mu(\bx,s,t)$ each time, propagating the uncertainty in estimating these quantities. In our setting it is extremely difficult to generate new datasets that satisfy the null $G(t')=0$ while preserving the original data generation process. We therefore center the empirical distribution of the overlap values to satisfy the null and perform hypothesis testing relative to that. 

While our approach captures part of the sampling uncertainty---namely, the variance in the derived overlap values---it treats the estimates of $e_t(\bx,0)$ and $\mu(\bx,s,t)$ for $s\in\{0,1\}$ as fixed. Ignoring the uncertainty in these nuisance functions understates the true variability of $G(t)$. Consequently, the reference distribution of $G(t)$ under the null is too narrow, causing the test to reject the null, and classify $(\rho, \gamma)$ pairs as incompatible, more often than a fully bootstrap-based procedure that re-estimates $e_t(\bx,0)$ and $\mu(\bx,s,t)$ in every resample. For our sensitivity analysis framework, this potential for an inflated incompatible region is an acceptable (albeit not desirable) behavior. We prefer to label $(\rho, \gamma)$ pairs as incompatible that may, in fact, be compatible rather than risk retaining ones that do not sufficiently explain differences between the study types. Nevertheless, developing a testing scheme that fully accounts for nuisance-estimation uncertainty is an important direction for future work and would strengthen the overall framework.

\section{Algorithms}\label{appdx:algorithms}
In this section, we include algorithms for the various components of our paper. We begin with the bias corrected estimator from Section~\ref*{sec:estimation}, followed be procedures for constructing breakdown frontier plots, like those in Section~\ref*{sec:experiments}, and for estimating the (in)compatibility of parameter pairs, described in Appendix~\ref{appdx:incompatible}. For each algorithm, we also discuss hyperparameter choice considerations and note relevant computational considerations.

\subsection{Bias corrected Estimator}
Algorithm~\ref{alg:bias_corrected_bounds} outlines the cross-fitting procedure used to implement the bias-corrected estimators defined in Section~\ref*{sec:estimation}. When estimating population-level ATE, the algorithm is applied to the full dataset. For CATE estimation, the dataset is first filtered to include only those samples whose covariate profiles $\bX_i$ fall within the subgroup of interest.

There are two hyperparameters beyond our sensitivity parameters $\rho$ and $\gamma$, namely $\alpha$ and the number of folds $k$. Larger values of $\alpha$ will more closely approximate the hard minimum and maximum operators. However, especially in small sample sizes, large $\alpha$ values can lead to instability in the bias-correction term. This occurs in regions where the maximum or minimum function switches between its arguments, as the Boltzmann operator exhibits steep gradients at these transitions. For this paper, we use a moderate value of $\alpha=10$, which balances stability and approximation quality. This choice may be increased in larger datasets.

The value of $k$ controls the number of splits to use for the cross-fitting procedure. Larger $k$ values provide more training data per fold but also increase computational demand. A suitable choice should consider the type of model used to estimate the nuisance functions, the dataset size, and the available computational resources.

Lastly, the choice of model used to estimate the nuisance functions $\hat{\eta}^{(j)}$ is another implicit hyperparameter. Flexible machine learning models are commonly used, though choices should consider sample size and computational constraints.

\begin{algorithm}[t]
    \caption{Bias-corrected estimators for lower and upper bounds}
    \label{alg:bias_corrected_bounds}
    \begin{algorithmic}[1]
        \Require Dataset $\mathcal{D}_n$ with $n$ samples, sensitivity parameters $\rho, \gamma \geq 0$, Boltzmann operator hyperparameter $\alpha \geq 0$, number of folds $k\in\mathbbm{N}$, with $(2\leq k \leq n)$.
        \State Split data into $k$ folds $\{\mathcal{E}_j\}_{j=1}^{k}$, where $\mathcal{E}_j$ is the set of indices for the samples in the hold-out set for fold $j$, and $\mathcal{T}_j = \{1, \dots, n\} \setminus \mathcal{E}_j$ is the corresponding training set indices.
        \For{$j = 1, \dots, k$}
            \State Estimate the set of nuisance function, $\hat{\eta}^{(j)} = (\hat{g}^{(j)}, \hat{e}^{(j)}_t, \hat{\mu}^{(j)})$, using training data $\mathcal{T}_j$:
            \State \hspace{0.5cm} $\hat{g}^{(j)}_s(\bx)$: study propensity function
            \State \hspace{0.5cm} $\hat{e}^{(j)}_t(\bx, s)$: treatment propensity functions for $s \in \{0,1\}$
            \State \hspace{0.5cm} $\hat{\mu}^{(j)}(\bx, s, t)$: expected outcome functions for $(s,t) \in \{0,1\}^2$
            \For{$i \in \mathcal{E}_j$}
                \For{$(t', \rho', \gamma', \alpha') \in \{(1,-\rho,-\gamma,\alpha), (0,\rho,\gamma,-\alpha), (1,\rho,\gamma,-\alpha), (0,-\rho,-\gamma,\alpha)\}$}
                    \State Calculate the following values at $\bx_i$:
                    \begin{align*}
                        \hat{v}^{(j)}(\bx_i, t', \gamma') & = (1+\gamma')\hat{\mu}^{(j)}(\bx_i, 1, t'), \\
                        \hat{w}^{(j)}(\bx_i, t', \rho') & = \hat{e}^{(j)}_{t'}(\bx, 0)\hat{\mu}^{(j)}(\bx, 0, t') + (1 - \hat{e}^{(j)}_{t'}(\bx, 0))(1 + \rho')\hat{\mu}^{(j)}(\bx, 0, t'), \\
                        \hat{\lambda}^{(j)}_1(\bx_i, t', \rho', \gamma', \alpha') & = \frac{\exp(\alpha'\hat{v}^{(j)}(\bx_i, t', \gamma'))}{\exp(\alpha'\hat{v}^{(j)}(\bx_i, t', \gamma')) + \exp(\alpha'\hat{w}^{(j)}(\bx_i, t', \rho'))}, \\
                        \hat{\lambda}^{(j)}_2(\bx_i, t', \rho', \gamma', \alpha') & = \frac{\exp(\alpha'\hat{w}^{(j)}(\bx_i, t', \rho'))}{\exp(\alpha'\hat{v}^{(j)}(\bx_i, t', \gamma')) + \exp(\alpha'\hat{w}^{(j)}(\bx_i, t', \rho'))}.
                    \end{align*}
                
                    \State Compute and store the plug-in term value, 
                    $\hat{b}_{i, t', \rho', \gamma', \alpha'} =  \hat{g}^{(j)}_1(\bx_i) \hat{\mu}^{(j)}(\bx_i,1,t) + \hat{g}^{(j)}_0(\bx_i) \Big\{\hat{\lambda}^{(j)}_1(\bx_i, t', \rho', \gamma', \alpha')\hat{v}^{(j)}(\bx_i, t', \gamma') + \hat{\lambda}^{(j)}_2(\bx_i, t', \rho', \gamma', \alpha')\hat{w}^{(j)}(\bx_i, t', \rho') \Big\}$.
                    \State Compute and store the (uncentered) EIF value $\hat{\phi}_{i, t', \rho', \gamma', \alpha'}$ (see Equation~\ref{eq:eif-alg-form}).
                \EndFor
            \EndFor
        \EndFor
        \State Compute the lower and upper bound plug-in estimates:
        \begin{align*}
            \hat{\theta}_{LB}^{plugin}(\rho,\gamma,\alpha; \hat{\eta}) = \frac{1}{n}\sum_{i} \hat{b}_{i, 1, -\rho, -\gamma, \alpha} - \hat{b}_{i, 0, \rho, \gamma, -\alpha}, \quad
            \hat{\theta}_{UB}^{plugin}(\rho,\gamma,\alpha; \hat{\eta}) = \hat{b}_{i, 1, \rho, \gamma, -\alpha} - \hat{b}_{i, 0, -\rho, -\gamma, \alpha}.
        \end{align*}
        \State Compute and store the lower and upper bound (centered) EIF values for each unit $i$:
        \begin{align*}
            \hat{\phi}_{i, LB, \rho, \gamma, \alpha} & = \big(\hat{\phi}_{i, 1, -\rho, -\gamma, \alpha} - \hat{\phi}_{i, 0, \rho, \gamma, -\alpha}\big) - \hat{\theta}_{LB}^{plugin}(\rho,\gamma,\alpha; \hat{\eta}), \\
            \hat{\phi}_{i, UB, \rho, \gamma, \alpha} & = \big(\hat{\phi}_{i, 1, \rho, \gamma, -\alpha} - \hat{\phi}_{i, 0, -\rho, -\gamma, \alpha}\big) - \hat{\theta}_{UB}^{plugin}(\rho,\gamma,\alpha; \hat{\eta}).
        \end{align*}              
        \State Compute the lower and upper bound bias corrected estimates:
        \begin{align*}
            \hat{\theta}_{LB}^{bc}(\rho,\gamma,\alpha; \hat{\eta}) = & \hat{\theta}_{LB}^{plugin}(\rho,\gamma,\alpha; \hat{\eta}) + \frac{1}{n}\sum_{i}^n \hat{\phi}_{i, LB, \rho, \gamma, \alpha}, \\
            \hat{\theta}_{UB}^{bc}(\rho,\gamma,\alpha; \hat{\eta}) = & \hat{\theta}_{UB}^{plugin}(\rho,\gamma,\alpha; \hat{\eta}) + \frac{1}{n}\sum_{i}^n \hat{\phi}_{i, UB, \rho, \gamma, \alpha}
        \end{align*}
        \State Compute bound variance estimates via the sample variance of the EIFs:
        \begin{align*}
            \hat{\sigma}_{LB}^2 = & \frac{1}{n} \sum_i^n \Big[\hat{\phi}_{i, LB, \rho, \gamma, \alpha} - \frac{1}{n}\sum_j^n\hat{\phi}_{j, LB, \rho, \gamma, \alpha}  \Big]^2, \\
            \hat{\sigma}_{UB}^2 = & \frac{1}{n} \sum_i^n \Big[\hat{\phi}_{i, UB, \rho, \gamma, \alpha} - \frac{1}{n}\sum_j^n\hat{\phi}_{j, UB, \rho, \gamma, \alpha}  \Big]^2.
        \end{align*}        
        \State \Return $\hat{\theta}_{LB}^{bc}(\rho,\gamma,\alpha; \hat{\eta}), \hat{\theta}_{UB}^{bc}(\rho,\gamma,\alpha; \hat{\eta}), \hat{\sigma}_{LB}^2, \hat{\sigma}_{UB}^2$.
    \end{algorithmic}
\end{algorithm}

In Algorithm~\ref{alg:bias_corrected_bounds}, we omit the specifics of calculating the uncentered EIF value $\hat{\phi}_{i, t', \rho', \gamma', \alpha'}$ for a given sample $i$, treatment indicator $t'$ and parameter values $\rho',\gamma'$ and $\alpha'$ given the length and complexity of the term. We include that full form in Equation~\ref{eq:eif-alg-form}, where we dropped the arguments from $\hat{\lambda}_1^{(j)}$, $\hat{\lambda}_2^{(j)}$, $\hat{v}^{(j)}$, and $\hat{w}^{(j)}$ for brevity, but note that they correspond to those functions evaluated at $\bx_i$ and parameter values $(t', \rho', \gamma', \alpha')$.

\begin{equation}\label{eq:eif-alg-form}
\begin{aligned}
\hat{\phi}_{i, t', \rho', \gamma', \alpha'} =\ & 
\frac{s_i \cdot \mathbb{I}(t_i = t')}{\hat{e}_{t'}^{(j)}(\bx_i, 1)}\left[y_i - \hat{\mu}^{(j)}(\bx_i, 1, t')\right] + s_i \cdot \hat{\mu}^{(j)}(\bx_i, 1, t') \\
& + s_i(1 + \gamma')\left\{\hat{\lambda}_1^{(j)} + \alpha' \hat{\lambda}_1^{(j)} \hat{\lambda}_2^{(j)}\left[\hat{v}^{(j)} - \hat{w}^{(j)}\right]\right\} \\
& \quad \times \left\{\frac{\mathbb{I}(t_i = t')}{\hat{e}_{t'}^{(j)}(\bx_i, 1)\hat{g}_1^{(j)}(\bx_i)}\left[y_i - \hat{\mu}^{(j)}(\bx_i, 1, t')\right] \hat{g}_0^{(j)}(\bx_i)\right\} \\
& + (1 - s_i)\left\{\hat{\lambda}_1^{(j)} \hat{v}^{(j)} + \hat{\lambda}_2^{(j)} \hat{w}^{(j)}\right\} \\
& + (1 - s_i)\left\{\hat{\lambda}_2^{(j)} + \alpha' \hat{\lambda}_1^{(j)} \hat{\lambda}_2^{(j)}\left[\hat{w}^{(j)} - \hat{v}^{(j)}\right]\right\} \\
& \quad \times \Bigg\{ \frac{\mathbb{I}(t_i = t')}{\hat{e}_{t'}^{(j)}(\bx_i, 0)} \left[y_i - \hat{\mu}^{(j)}(\bx_i, 0, t')\right] \left[1 + \rho' \cdot \hat{e}_{1 - t'}^{(j)}(\bx_i, 0)\right] \\
& \qquad + \rho' \cdot \hat{\mu}^{(j)}(\bx_i, 0, t') \left[\mathbb{I}(t_i = 1 - t') - \hat{e}_{1 - t'}^{(j)}(\bx_i, 0)\right] \Bigg\}
\end{aligned}
\end{equation}

\subsection{Breakdown Fontier Plot}\label{appdx:bf-algo}
Algorithm~\ref{alg:breakdown-frontier} outlines the procedure for constructing a breakdown frontier plot like those in Section~\ref*{sec:experiments}. As with the algorithm for the bias-corrected estimator, when estimating population-level ATE, the algorithm is applied to the full dataset. Whereas, for CATE estimation, the dataset is first filtered to include only those samples whose covariate profiles $\bX_i$ fall within the subgroup of interest. 

The hyperparameters used are as follows: $\alpha$ and $k$ are inputs to the bias-corrected estimator, $R$ specifies the number of resampling iterations in the incompatibility test, and $c$ denotes the confidence level used throughout the algorithm. The most critical design choice is the grid of $(\rho, \gamma)$ values. Large grids that include many $(\rho, \gamma)$ pairs yield more detailed breakdown frontiers, but increase computational cost. A reasonable approach is to select maximum values of $\rho$ and $\gamma$ based on what a domain expert deems plausible, and then construct a grid of evenly spaced pairs of $(\rho, \gamma)$ between $(0,0)$ and those maximum values. In our experiments, we consistently cap $\rho$ and $\gamma$ at 0.2, corresponding to a $20\%$ relative violation. While this choice reflects a substantial but interpretable level of assumption violation, selecting a plausible range involves subjective judgment based on the context and domain knowledge.

\begin{algorithm}[H]
\caption{Breakdown Frontier Plot Construction}
\label{alg:breakdown-frontier}
\begin{algorithmic}[1]
\Require Dataset $\mathcal{D}_n$, grid of $(\rho, \gamma)$ values, confidence level $c \in (0,1)$, Boltzmann smoothing parameter $\alpha \geq 0$, number of cross-fitting folds $k$, number of resampling iterations $R$
\Ensure Heatmap assigning each $(\rho, \gamma)$ pair to one of four regions

\State \textbf{Estimate nuisance functions once} via cross-fitting (see Algorithm~\ref{alg:bias_corrected_bounds}, Steps 1–7) to obtain $\hat{\eta} = (\hat{g}, \hat{e}_t, \hat{\mu})$
\For{each $(\rho, \gamma)$ in the grid}
    \State Compute bias-corrected bounds $(\hat{\theta}^{bc}_{LB}, \hat{\theta}^{bc}_{UB})$ and variances $(\hat{\sigma}^2_{LB}, \hat{\sigma}^2_{UB})$ using Algorithm~\ref{alg:bias_corrected_bounds}
    \State Construct $100(1{-}c)\%$ confidence intervals:
    \[
        \text{CI}_{LB} = \left[\hat{\theta}^{bc}_{LB} - z_{1-c/2} \cdot \frac{\hat{\sigma}_{LB}}{\sqrt{n}},\ 
                               \hat{\theta}^{bc}_{LB} + z_{1-c/2} \cdot \frac{\hat{\sigma}_{LB}}{\sqrt{n}} \right]
    \]
    \[
        \text{CI}_{UB} = \left[\hat{\theta}^{bc}_{UB} - z_{1-c/2} \cdot \frac{\hat{\sigma}_{UB}}{\sqrt{n}},\ 
                               \hat{\theta}^{bc}_{UB} + z_{1-c/2} \cdot \frac{\hat{\sigma}_{UB}}{\sqrt{n}} \right]
    \]
    \State Run incompatibility test (Algorithm~\ref{alg:compatibility-test}) for $t = 0$ and $t = 1$ using $R$ resamples
    \If{p-value $< c$ for either $t$}
        \State Assign region: \texttt{Incompatible}
    \ElsIf{both bounds are strictly $> 0$ or strictly $< 0$, and both intervals exclude 0}
        \State Assign region: \texttt{Conclusive}
    \ElsIf{both point estimates have the same sign, but at least one CI includes 0}
        \State Assign region: \texttt{Tentative}
    \ElsIf{point estimates have opposite signs}
        \State Assign region: \texttt{Inconclusive}
    \EndIf
\EndFor
\State Render heatmap over the grid with regions color-coded
\end{algorithmic}
\end{algorithm}

\subsection{(In)compatible $\rho$ and $\gamma$ Test}\label{appdx:compatible-test-algo}
Finally, we include include Algorithm~\ref{alg:compatibility-test} to outline the steps of the (in)compatibility test. We refer to Appendix~\ref{appdx:incompatible} for a detailed explanation and discussion of this procedure.

\begin{algorithm}[H]
\caption{Resampling-based Test for Parameter Compatibility}
\label{alg:compatibility-test}
\begin{algorithmic}[1]
\Require Dataset $\mathcal{D}_n = \{(X_i, T_i, S_i, Y_i)\}_{i=1}^n$, parameter values $(\rho', \gamma')$, treatment arm $t'$, number of resamples $R$
\Ensure p-value for testing $H_0: G(t') \geq 0$ vs. $H_A: G(t') < 0$

\State \textbf{Compute observed test statistic:}
\[
T_{\text{obs}} = \frac{1}{n} \sum_{i=1}^n \left[\min\{v(X_i, t', \gamma'), w(X_i, t', \rho')\} - \max\{v(X_i, t', -\gamma'), w(X_i, t', -\rho')\}\right]
\]

\State \textbf{Center overlap values to simulate null:}
\[
o_i = \min\{v(X_i, t', \gamma'), w(X_i, t', \rho')\} - \max\{v(X_i, t', -\gamma'), w(X_i, t', -\rho')\} - T_{\text{obs}} \quad \text{for } i = 1,\ldots,n
\]

\For{$k = 1$ to $R$}
    \State Sample $n$ indices $J_r = \{j_1, \ldots, j_n\}$ with replacement from $\{1, \ldots, n\}$
    \State Compute resampled statistic: 
    \[
    T_r = \frac{1}{n} \sum_{j \in J_r} o_j
    \]
\EndFor

\State \textbf{Compute p-value:}
\[
p = \frac{1}{R} \sum_{r=1}^R \mathbbm{1}[T_r \geq T_{\text{obs}}]
\]

\State \Return $p$
\end{algorithmic}
\end{algorithm}

\section{Simulation Setup}\label{appdx:sim-setup}
Each synthetic dataset in Section~\ref*{sec:sim-study} is composed of 2500 i.i.d. samples where for each sample $i$, we start by generating three observable covariates and one unobserved confounder,
\begin{gather*}
    X_{i,1}, X_{i, 2}, X_{i,3} \sim \mathcal{N}(1,1), \\
    U_i \sim \mathcal{N}(1,1).
\end{gather*}

We then define $C_i = (1 - \beta)X_{i,1} + \beta U_i$, where $\beta$ is a hyperparameter passed to the data generating process (DGP). This parameter controls the level of unobserved confounding, with larger values of $\beta$ corresponding to more unobserved confounding. Using this, we generate the study and treatment indicators as
\begin{gather*}
    S_i \sim \textrm{Bernoulli}\Big(\text{expit}(-C_i)\Big), \textrm{ and } \\
    T_i \sim \textrm{Bernoulli}\Big(S_i \times 0.5 + (1 - S_i) \times \text{expit}(C_i)\Big),
\end{gather*}
where expit is the logistic sigmoid: \(\text{expit}(x) = \frac{1}{1 + e^{-x}}\). The potential outcomes and observed outcome are then generated as
\begin{gather*}
    Y_i(0) = 100 + X_{i, 2}, \\
    Y_i(1) = Y_i(0) + 12C_i - 10X_{i,3} + \tau, \textrm{ and } \\
    Y_i = T_i \times Y_i(1) + (1 - T_i) \times Y_i(0) + \epsilon_i,
\end{gather*}
where $\epsilon_i \sim \mathcal{N}(0,1)$ and $\tau$ is another hyperparameter passed to the DGP that controls the size of the constant treatment effect. Therefore, a larger $\tau$ corresponds to a larger treatment effect.

There are five different breakdown frontier plots in Section~\ref*{sec:sim-study}. Each plot is generated from a different dataset that is created from the above DGP by varying the two hyperparameters, $\beta$ and $\tau$. In particular,
\begin{itemize}
    \item Base: $\beta=0.4$ and $\tau=5$.
    \item Larger $\tau$: $\beta=0.4$ and $\tau=8$.
    \item Smaller $\tau$: $\beta=0.4$ and $\tau=2$.
    \item Larger $U$: $\beta=0.6$ and $\tau=5$.
    \item Smaller $U$: $\beta=0.2$ and $\tau=5$.
\end{itemize}

Note that in the plots, "Larger $U$" and "Smaller $U$" refer to higher a lower levels of unobserved confounding induced by the hyperparameter $\beta$, not to the values of the unobserved confounder $U$ itself. With full knowledge of the DGP, more accurate labels would be "Larger $\beta$" and "Smaller $\beta$". But given that the full DGP was not introduced in the main text, we use "Larger $U$" and "Smaller $U$" for simplicity and accessibility.
\section{Experimental Details}\label{appdx:exp-details}
This section provides implementation specifics for the results presented in Section~\ref*{sec:experiments}. We outline the hyperparameters and other relevant experimental settings used to generate each of the breakdown frontier plots shown in Section~\ref*{sec:experiments}. Appendix~\ref{appdx:algorithms} includes the algorithm for constructing breakdown frontier plots (Algorithm~\ref{alg:breakdown-frontier}), as well as algorithms for the double machine learning estimator (Algorithm~\ref{alg:bias_corrected_bounds}) and the procedure for determining (in)compatible sensitivity parameters (Algorithm~\ref{alg:compatibility-test}). These components together form the full procedure used to construct a breakdown frontier plot.

All experiments where conducted in Python. We reference relevant packages and classes were necessary. See accompanying code for additional details.

\paragraph{Datasets.} For details on the data-generating process used to create the synthetic datasets in Section~\ref*{sec:sim-study}, see Appendix~\ref{appdx:sim-setup}. For the Project STAR dataset, the outcome of interest was defined as the average score across standardized tests on math, reading, language, and social science. The objective was to assess the effect of small class sizes from kindergarten through third grade on this outcome. Measured covariates included gender, race, and age at the start of kindergarten.

For the ATE plot in Figure~\ref*{fig:star}(a), the analysis was conducted on the full dataset. For the CATE plots in Figure~\ref*{fig:star}(b), we restricted the dataset to two subgroups: students who began kindergarten before age six, and students who were at least six years old at the start of kindergarten. 

Further details on the Project STAR dataset, including the raw dataset files and cleaning scripts, can be found in the accompanying code.

\paragraph{Hyperparameter Settings.} Several hyperparameters are held constant across all breakdown frontier plots. Specifically, the following settings are used throughout:
\begin{itemize}
    \item Grid of $(\rho, \gamma)$: We construct a grid over the \((\rho, \gamma)\) parameter space by taking all pairwise combinations of values sampled uniformly from the intervals \([0, 0.2]\). Specifically, we define:
            \[
            \gamma \in \text{linspace}(0, 0.2, 50), \quad \rho \in \text{linspace}(0, 0.2, 50)
            \]
            where \(\text{linspace}(a, b, n)\) denotes a sequence of \(n\) evenly spaced values from \(a\) to \(b\), inclusive. The resulting grid contains \(50 \times 50 = 2500\) \((\rho, \gamma)\) pairs.
            
    \item Confidence level $c$: 0.95 
    \item Boltzmann smoothing parameter $\alpha$: 10
    \item Models used to estimate the nuisance functions:
    \begin{itemize}
        \item $\hat{g}^{(j)}_s(\bx)$: \texttt{sklearn.linear\_model.LogisticRegressionCV(n\_jobs=1)}
        \item $\hat{e}^{(j)}_t(\bx, s)$: \texttt{sklearn.linear\_model.LogisticRegressionCV(n\_jobs=1)}
        \item $\hat{\mu}^{(j)}(\bx, s, t)$: \texttt{sklearn.linear\_model.RidgeCV()}
    \end{itemize}
\end{itemize}

The remaining hyperparameters---namely the number of cross-fitting folds to use for the double machine learning estimators, $k$, and the number of resampling iterations to use for the (in)compatible test, $R$---are set according to the data type. For all breakdown frontier plots based on simulated (Section~\ref*{sec:sim-study}), we use $k=2$ and $R=100$. For plots based on the Project STAR data (Section~\ref*{sec:star-results}), we use $k=5$ and $R=1000$. The larger values for the Project STAR dataset were chosen to generate more precise results. 

\paragraph{Uncertainty Estimation.} As discussed at the end of Section~\ref*{sec:estimation} and in the double machine learning estimator Algorithm~\ref{alg:bias_corrected_bounds}, the variance of the bounds can be estimated either using the sample variance of the estimated efficient influence functions (EIF) or through resampling methods. For the simulated datasets, we used the sample variance of the EIFs. For the Project STAR dataset, however, we used a bootstrap resampling procedure with 1000 iterations. 

The bootstrap approach was chosen for Project STAR due to the relatively small dataset size, specifically in the observational study arm, and the extreme nuisance function estimates it produced---which led to large and unstable variance estimates when relying solely on the EIFs. In contrast, the bootstrap procedure produced more consistent and stable variance estimates.

\paragraph{Computational Setting and Details.} All experiments were run on a Slurm-managed cluster using VMware virtual machines, each equipped with an Intel(R) Xeon(R) CPU E5-2699 v4 @ 2.20GHz. No GPU or specialized hardware was used.

For the simulated datasets, we constructed breakdown frontier results by submitting a single Slurm job with 1 CPU core and 32 GB of RAM. Each breakdown frontier plot dataset was generated in a just a few minutes, and compatible region tests (Algorithm~\ref{alg:compatibility-test}) were run within the same job. 

For the Project STAR dataset, where we estimated uncertainty using 1000 bootstrap resamples, we distributed the work across 20 Slurm jobs, each allocated the same compute resources as the simulated dataset setup. Each job ran 50 iterations with distinct random seeds. The compatible region tests for Project STAR were run in a separate Slurm job using the same resources. 

Each job completed in approximately 1-2 minutes, and the overall compute cost was low. The primary limiting factor was the number of bootstrap iterations. All scripts can be run on a local machine with sufficient memory to store the breakdown frontier grids.

\end{document}